\pdfoutput=1 
\documentclass[12pt]{article}        %%%%%%%%%%%%%%%%%%
\usepackage{amsmath}
\usepackage{amssymb}
\usepackage{euscript}

\usepackage{inputenc}

\usepackage{amsmath,amssymb,bbm,color}
\usepackage{amsbsy}
\usepackage{euscript}
\usepackage{graphicx}
\usepackage{hyperref}
\usepackage{cite}
\usepackage{enumerate}
\numberwithin{equation}{section}

\usepackage{alltt} %--- teletype verbatim
\usepackage{color}

\newcommand{\kkmc}{{\tt   KKMC}}

\newcommand{\bhlumi}{{\tt BHLUMI}}
\newcommand{\bhwide}{{\tt BHWIDE}}
\newcommand{\koralz}{{\tt KORALZ}}
\newcommand{\koralw}{{\tt KORALW}}

\newcommand{\yfsww}{{\tt  YFSWW}}

\newcommand{\kandy}{{\tt  KandY}}
\newcommand{\racoon}{{\tt RACOONWW}}

\newcommand{\winhac}{{\tt WINHAC}}
\newcommand{\photos}{{\tt PHOTOS}}

\def\Order#1{${\cal O}(#1)$}
\def\Ordex#1{${\cal O}(#1)_{exp}$}
\def\Oeex#1{${\cal O}(#1)_{\rm EEX}$}
\def\Oceex#1{${\cal O}(#1)_{\rm CEEX}$}

\def\bbeta{\bar{\beta}}
\def\hbeta{\hat{\beta}}
\def\tbeta{\tilde{\beta}}

\newcommand{\order}[1]{${\cal O}(#1)$}

\newcommand{\Meu}{\EuScript{M}}

\newcommand{\Mmf}{\mathfrak{M}}

\newcommand{\eqop}{\operatornamewithlimits{=}}

\begin{document}                     %%%%%%%%%%%%%%%%%%%%%%

\allowdisplaybreaks

%%%%%%%%%%%%%%%%%%%%%%%%%%%%%%%%%%%%%%%%%%%%%%%%%%%%%%%%%%%%%%%%%%%%%%%%%%%%%%%%%
%%%%%%%%%%%%%%%%%%%%%%%%%%%%%%%%%%%%%%%%%%%%%%%%%%%%%%%%%%%%%%%%%%%%%%%%%%%%%%%%%
\begin{titlepage}

\begin{flushright}
{\bf  IFJPAN-IV-2019-5}
\end{flushright}

\vspace{1mm}
\begin{center}{\bf\Large QED Exponentiation for quasi-stable charged particles:
                         the $e^-e^+\to W^-W^+$ process${^\dag}$}
\end{center}

%\vspace{1mm}
\begin{center}
  {\bf   S. Jadach$^{a}$,}
  {\bf   W. P\l{}aczek$^{b}$}
  {\em and}
  {\bf   M. Skrzypek$^{a}$}\\
{\em $^a$ Institute of Nuclear Physics, Polish Academy of Sciences,\\
  ul.\ Radzikowskiego 152, 31-342 Krak\'ow, Poland}\\
{\em $^b$ Institute of Applied Computer Science, Jagiellonian University,\\
ul.\ \L{}ojasiewicza 11, 30-348 Krak\'ow, Poland}
\end{center}

\vspace{1mm}
\begin{abstract}
All real and virtual infrared singularities in the standard analysis
of the perturbative Quantum Electrodynamics (like that of Yennie--Frautschi--Suura)
are associated with photon emissions from the external legs in the scattering process.
External particles are stable, with the zero decay width.
Such singularities are well understood at any perturbative order and are resummed.
The case of production and decay of the semi-stable {\em neutral} particles,
like the $Z$-boson or the $\tau$-lepton, 
with the narrow decay width, $\Gamma/M \ll 1$, is also well understood at any
perturbative order and soft-photon resummation can be done.
For an absent or loose upper cut-off on the total photon energy $\omega$,
production and decay processes of the semi-stable (neutral) particles decouple
approximately and can be considered quasi-independently.
In particular, the soft-photon resummation can be done
separately for the production and the decay, treating
a semi-stable (neutral) particle as stable.
QED interference contributions between the production and decay stages 
are suppressed by the $\Gamma/M$ factor. 
If experimental precision
$\omega$ is comparable with or better than $\Gamma/M$,
these interferences have to be included.
In the case of $\omega \ll \Gamma$ decoupling of production and decay does not work
any more and the role of semi-stable particles is 
reduced to the same role as that of other internal off-shell particles.
So far, consistent treatment of the soft photon resummation for semi-stable
{\em charged} particles like the $W^\pm$ bosons is not available in the literature, and
the aim of this work is to present a solution to this problem.
Generally, this should be feasible because
the underlying physics is the same as in the case of neutral semi-stable resonances --
in the limit of $\Gamma/M \ll 1$ the production and decay processes 
for charged particles also necessarily decouple due to long lifetime of intermediate particles.
Technical problems to be solved in this work
are related to the fact that semi-stable charged particle are able to emit photons.
Practical importance of the presented technique to the $e^+e^-\to W^+W^-$ process
at the Future electron--positron Circular Collider (FCC-ee) is underlined.
\end{abstract}

\footnoterule
\noindent
{\footnotesize
\begin{itemize}
\item[${\dag}$]
This work is partly supported by
%%%%% NCN-Jadach
 the Polish National Science Center grant 2016/23/B/ST2/03927
% the Citadel Foundation 
 and the CERN FCC Design Study Programme.
\end{itemize}
}
\vspace{1mm}
\begin{flushleft}
{\bf 
  IFJPAN-IV-2019-5
%  {\small (Last correction \today)}
}\end{flushleft}

\end{titlepage}

%%%%%%%%%%%%%%%%%%%%%%%%%%%%%%%%%%%%%%%%%%%%%%%%%%%%%%%%%%%%%%%%%%%%%%%%%%%%%%%%%%%%%%%%%%%%%
\tableofcontents
\newpage

%%%%%%%%%%%%%%%%%%%%%%%%%%%%%%%%%%%%%%%%%%%%%%%%%%%%%%%%%%%%%%%%%%%
\section{Introduction}
%%%%%%%%%%%%%%%%%%%%%%%%%%%%%%%%%%%%%%%%%%%%%%%%%%%%%%%%%%%%%%%%%%%

The Standard Model Electroweak (EW) Field Theory was confirmed 
as the correct physics theory of electromagnetic and weak interactions 
between elementary particles by precision measurements 
of the LEP experiments~\cite{Alcaraz:2007ri,Schael:2013ita}.
The LEP data were precise enough to test
all important dynamical properties of the EW theory, 
such as quantum loop effects, consequences of the renormalisation,
multiple photon emission, etc.
In particular, EW gauge cancellations and quantum loop effects
were verified experimentally at LEP in the $e^+e^- \to W^+W^-$ process
at the precision tag for the total cross section at the level of $0.3$--$05\%$.
The mass of the $W$-boson was also measured directly with the precision of $33\,$MeV.

The electron-positron Future Circular Collider 
(FCC-ee)~\cite{Abada:2019lih,Abada:2019zxq},
considered as the future project at CERN, 
will be able to produce the number of $W$-boson pairs by a factor of $10^3$ higher than at 
LEP.
This will serve to determine total $WW$ cross section, the mass and width of $W$
with the unprecedented precision
and search for any anomalous phenomena beyond the Standard Model (SM)
of the EW and strong interactions.
Obviously, analysing FCC-ee data will also require new
SM perturbative calculations for the $e^+e^- \to W^+W^-$ process,
much more precise than these available at the LEP era
\cite{Blondel:2018mad,Blondel:2019qlh}  
The precision tag expected in FCC-ee experiments
is at the level of about $0.01\%$, a factor of $10$ better than at LEP.
This will require to go beyond the state of the art of the LEP era 
in the calculations of the SM predictions for the $e^+e^- \to W^+W^-$ or
$e^+e^- \to 4f$ processes.

For general discussion of the theoretical issues in the $W$-pair production process
the reader should consult the reviews of refs.~\cite{Grunewald:2000ju,Placzek:2002ft}.
In particular, the delicate question of the EW gauge invariance for
the Dyson summation leading to imaginary part of the $W$ and $Z$ 
propagators is covered there.

Here we shall focus on the important QED part of the EW/SM perturbative corrections to 
the $W$-pair production process.
More precisely, on this part of the QED corrections which is related
to soft and collinear (SC) singularities for real and virtual photon emissions
on the external legs%
\footnote{It is tempting to call them ``universal'' but, in fact,
  non-soft subleading collinear perturbative corrections are process-dependent,
  hence non-universal,
  while all soft corrections are universal.
}.
According to the accumulated knowledge on the SC photonic contribution,
it is quite clear that they factorise either at the amplitude level,
or for the differential distributions and can be calculated separately
to a much higher order than the remaining genuine EW corrections%
\footnote{The genuine EW part of the SM perturbative corrections include non-soft,
  non-collinear remnants of the QED origin.}.
This is very convenient, because SC contributions are
much bigger numerically than genuine EW corrections,
simpler to calculate, and can be resummed to the infinite order.
Once separation of the QED and EW parts is established,
resummation of some higher-order contributions in each of these two classes
can be done independently.
The important nontrivial final step is then
merging/matching them in the final results.

There is little doubt that the
factorisation and resummation of the QED soft/collinear corrections 
is the key to the success in the high-precision calculations of the SM predictions 
for the $W$-pair production process at FCC-ee.

There are four classes of QED corrections to the $W$-pair production and decay process: 
initial-state corrections (ISR), 
final-state corrections (FSR) in the decays of two $W^\pm$,
final-state Coulomb corrections (FSC) and
the so-called non-factorisable interferences (NFI)
between the production and the decays (IFI) and between two $W^\pm$ decays (FFI).
The IFI corrections are suppressed due to relatively long lifetime of $W$'s
and FFI due to large space separation.
The effects due to ISR are numerically the biggest but also easier to control,
while the FSR effects can be also quite sizeable for typical experimental cut-offs.

The IFI and FFI interferences are small, 
suppressed by the factor $\Gamma_W/M_W$ away from the $WW$-production threshold, 
strongly cut-off dependent and algebraically most complicated.
At LEP they could be neglected but for the FCC-ee precision they 
have to be handled with great care!
%{\bf \\ 
%WP: Ponizszy fragment bym usunal! Niefaktoryzowalnosc w odniesieniu do tych 
%interferencji oznacza, 
%ze nie da sie w nich rozdzielic procesow produkcji i rozpadow na poziomie 
%przekroju czynnego.
%\\}
%{\em 
%In reality, IFI and FFI contributions to differential cross sections
%do factorise -- for the real photon emission at the amplitude level
%and for the virtual ones at the loop integrand level.
%Hence naming them as non-factorisable is somewhat misleading.
%Nevertheless, we shall stick to this traditional naming,
%NFI in short.
%}

The relative narrowness of the $W$ boson resonance not only causes
suppression of the QED interferences, but also provides for
the expansion in terms of $\Gamma_W/M_W\sim$ \order{\alpha}
of the matrix element of the $e^+e^- \to 4f$ process
into the numerically biggest and physically most interesting 
double-resonant $e^+e^- \to W^+W^-$ part,
and less important single-resonant and non-resonant background parts.
In the following we shall refer to them as the double-pole (DP),
single-pole (SP) and non-pole (NP) contributions, as it was common in the LEP-era literature.

The above pole expansion (POE) in the powers of $\Gamma_W/M_W$, disentangling 
the DP, SP and NP components at the scattering amplitude-level
is very useful because it allows for each of these three components 
to calculate the genuine EW corrections at a different perturbative order
and to perform resummation of the QED soft/collinear contributions
at a different sophistication level.
In the final stage of the calculation, 
the best way is to sum POE contributions coherently at the amplitudes level,
before summing over spin and taking modulus squared,
rather than for differential cross sections,
thus avoiding proliferation of many interference terms.

At the time of LEP experiments,
two solutions based on the pole expansion were worked out,
in which the \order{\alpha^1} EW corrections were complete only
for the DP component $e^+e^- \to W^+W^-$ of the $e^+e^- \to 4f$ process.
One of them, nicknamed \kandy~\cite{Jadach:2001mp,Jadach:2000kw},
was based on the combination of \yfsww3\cite{Jadach:1996hi,Jadach:2001uu} Monte Carlo (MC)%
\footnote{Including EW \order{\alpha^1} corrections 
   of refs.~Refs.~\cite{Fleischer:1988kj, Kolodziej:1991pk,Fleischer:1991nw, Fleischer:1994sq}.
}
for the $e^+e^- \to W^+W^-$ and $W^\pm$-decay processes
with another MC program \koralw\cite{Jadach:1998gi} for the remaining background.
The multiphoton emission for ISR, including higher orders, was implemented using
the soft-photon resummation inspired by the Yennie--Frautschi--Suura (YFS) work\cite{Yennie:1961ad}.
The QED FSR was added in $W$ decays using the \photos\ program \cite{Jadach:1993hs,Barberio:1990ms}.
Another POE-based solution was that of \racoon~\cite{Denner:2000bj,Denner:2002cg},
also with the complete EW \order{\alpha^1} corrections implemented 
only for the signal  $e^+e^- \to W^+W^-$ process and not for the background 
part.

Implementation of QED corrections in \racoon\ was very different from that in \kandy.
On the one hand, \racoon\ was using exact matrix element for the entire $e^+e^- \to 4f\gamma$ process
but it was lacking sophisticated soft photon resummation of the \kandy.
For more detailed comparison of the two approaches
see the review of ref.~\cite{Placzek:2002ft} 
or more recent of ref.\cite{Jadach:2019bye}.
Both approaches were instrumental in the analysis of the LEP data
for the $e^+e^- \to W^+W^-$ process~\cite{Schael:2013ita},
where the gauge cancellations and the quantum effects of the EW theory 
were tested experimentally for the first time.

Both approaches, \kandy\ and \racoon,
neglect terms of \order{\alpha\Gamma_W/M_W}.
The QED NFI interferences between $W$ production and decays were either neglected completely
(\kandy) or included in the soft-photon approximation (\racoon) without resummation.
The overall precision of these calculations was about $0.3$--$0.5\%$.
The FCC-ee experiments will require new calculations with the precision tag below $0.1\%$,
thus adding missing $\alpha\Gamma_W/M_W$ corrections, 
\Order{\alpha^2} electroweak corrections to the DP component,
a more advanced QED factorisation/resummation scheme,
subleading QED \Order{\alpha^2} 
corrections and more will be needed \cite{Blondel:2018mad,Blondel:2019qlh}.
In particular, inclusion the QED NFI corrections
in the fully exclusive way%
\footnote{They depend strongly on experimental cut-offs.},
taking into account the $\Gamma_W/M_W$ suppression, will be necessary.

The aim of the present work is to work out a new methodology of
the soft photon resummation including NFI corrections
for charged unstable particles, similarly as it was done
for the production and decay of the narrow neutral $Z$-boson 
in the process $e^+e^-\to f\bar{f}+n\gamma$
with a built-in  $\Gamma_Z/M_Z$ suppression for the 
QED initial-final interferences (IFI) 
at any perturbative order~\cite{Jadach:1999vf,Jadach:2000ir}.
This method was already tested for the $Z$ resonance
in the Monte Carlo event generator \kkmc\cite{Jadach:1999vf}.
Its matrix element is built according to
the so-called coherent exclusive exponentiation (CEEX) scheme,
in which factorisation of the infrared (IR) divergences
is done entirely at the amplitude level (before squaring and spin-summing).
The older version of the exclusive exponentiation (EEX) of refs.\cite{Jadach:1988gb,Jadach:1991dm}
was done at the level  of differential distributions
for the same $e^+e^-\to f\bar{f}+n\gamma$ process and features multiphoton resummation
of ISR and FSR.
Both approaches, CEEX and EEX, are inspired by the pioneering work of
Yennie--Frautschi--Suura~\cite{Yennie:1961ad}.

In the present work we shall generalise the CEEX scheme 
to the case of any number of narrow {\em charged} intermediate resonances, 
like the $W$-boson --
the scheme is however quite general and applies to any charged resonance of any spin.
The new  CEEX scheme provides exclusive (unintegrated) description
for multiple real photons of any energy, for
$E_\gamma\sim \Gamma_W$, $E_\gamma \ll \Gamma_W$ and $E_\gamma\sim \sqrt{s}$,
with  all QED interferences between production and decays properly accounted for.
Multiple real and virtual photon emission from all external stable particles
and the intermediate  {\em semi-stable} charged resonance will be described correctly
in the soft photon limit and summed up to the infinite order.
As in the case of CEEX of refs.~\cite{Jadach:1999vf,Jadach:2000ir},
its present extension will provide for a well-defined methodology of
incorporating {\em non-soft} contributions%
\footnote{This will be done without introducing any parameter in the photon energy
 distinguishing between soft and hard photons. 
 Minimum photon energy in the Monte Carlo implementation 
 can be set to an arbitrarily low value without any effect on the physical results.}
(including the genuine EW corrections)
calculated up to a finite perturbative order into multiphoton amplitudes
of the soft-photon resummation scheme.
In particular, sizeable but easier to calculate QED non-soft collinear
contributions can also be included easily up to an arbitrarily high order.

The consistent resummation of the apparently IR-divergent contribution due to photon
emissions from the semi-stable intermediate charged particle (narrow 
resonances) 
in the perturbative expansion is a non-trivial issue.
Let us first consider $\Gamma\to 0$ limit.
The best illustrative example is that of the $\tau^\pm$-pair production and decay
in the $e^+e^-$ annihilation where a time scale of the $\tau$-pair
formation (production process) is shorter than the $\tau$ lifetime by 
at least a factor of $\Gamma_\tau/m_\tau \simeq 3 \cdot 10^{-12}$, 
hence photons emitted in these two stages get completely decoupled
and the QED effects in the production and the decay 
can be implemented separately~\cite{Jadach:1984iy,Jadach:1991ws,Jadach:1999vf}.

The situation in the $W$-pair production is similar but the suppression
factor $\Gamma_W/M_W\simeq 0.026$ is not that small.
The QED interferences are therefore expected to be of the order of
$\alpha\Gamma_W/M_W\simeq 2\cdot 10^{-4}$.
In LEP experiment this size could be neglected, but for the FCC-ee precision,
effects of this size have to be calculated and taken into account.
Moreover, such interferences depend on kinematical cut-offs --
from the experience with the $Z$-boson case we know that they may
grow by a factor of $2$--$5$ even for relatively mild cut-offs on photon energies.
Also, in the case when photon energy resolution $\omega$
of the detector approaches $2\,$GeV,
which is the case for FCC-ee detectors,
photon emission from FSR in the production process
and from $W$ decays cannot be separated
and treated in the soft photon approximation,
consequently the off-shell $W$'s have to be treated
the same way as other internal exchanges
in the $e^+e^- \to 4f$ process.

Our aim is to construct a variant of CEEX spin amplitudes
in which we profit as much as possible from the smallness
of $\Gamma_W/M_W$ and the classic YFS soft-photon limit
for the entire $e^+e^- \to 4f$ process is correctly reproduced
for $\omega \ll \Gamma_W$.
The basic technical problem will be that if we want to treat
$W$'s as stable particles in the $W$-pair production process with the zero width,
then amplitudes of photon emission from $W$ must be IR-singular,
while for the semi-stable $W$'s they are not (the $W$ width acts as a IR regulator).
Our aim is to reconcile these two contradictory situations in
a single algebraic framework.

In the \yfsww3\ program, photon emission in the $e^+e^- \to W^+W^- \to 4f$ process 
was treated in a similar way as
in the above $\tau$-pair production and decay,
except that $W$ invariant masses were not fixed but modelled 
according to the Breit--Wigner shape.
The QED matrix element in \yfsww3\ for $e^+e^-\to W^+W^-$ 
with the soft photon resummation is of the EEX type,
including ISR, FSR and IFI.
Decays of $W^\pm$s are supplemented with additional photons using \photos.
However, it could be easily replaced with the multiphoton MC implementation
of the EEX of the \winhac\ program~\cite{Placzek:2003zg}.
Once EEX implementation is available in the $W$-pair process
for the production and decays, the new CEEX matrix element 
developed in the present work can be introduced using 
an additional multiplicative MC weight%
\footnote{The same way as in \kkmc.},
without any change in the underlying MC program.
The above would be the solution for the resummed QED corrections
of the DP part of the $e^+e^- \to 4f$ process.
The \order{\alpha^1} and \order{\alpha^2} genuine EW corrections
can be added in the on-shell approximation within the CEEX matrix
element in the similar way as it was done for the $e^+e^-\to 2f$ process
in refs.~\cite{Jadach:1999vf,Jadach:2000ir}.
So far only \order{\alpha^1} EW corrections are available.
In order to exploit fully  FCC-ee data, the \order{\alpha^2} EW corrections
will be needed.
As pointed out in ref.~\cite{Blondel:2018mad}, the clear and clean separation
of the QED and the genuine EW correction at any perturbative order
is a useful built-in feature of the CEEX factorisation/resummation scheme.

The single-pole group of diagrams of the $e^+e^- \to 4f$ process
process is separated at the amplitude level in the CEEX scheme.
It would be enough to include the genuine EW corrections to the SP part
at \order{\alpha^1}.
They are in principle known, because they are part
of the \order{\alpha^1} corrections to $e^+e^- \to 4f$ process
in refs.~\cite{Denner:2005es, Denner:2005fg}, 
although it may be not simple to disentangle them 
from the rest of the existing calculations.
For the non-pole part of the $e^+e^- \to 4f$ process
it would probably be enough to take it at the tree-level
as far as the genuine EW corrections are concerned and take care of the QED corrections only,
either in the CEEX or EEX scheme.

In this work, the CEEX scheme will be defined only for the DP part
leaving the easier SP and NP variants for the future development.
On the other hand,
we shall also discuss in a more detail the explicit algebraic relation
between the CEEX scheme and the EEX scheme of the \yfsww3.
This will provide better understanding of the theoretical foundation of the existing
EEX scheme of the \yfsww3.
The main result of this work is, however, that it provides an important building
block for the future high-precision calculations for the $W$ pair production process,
and also for any other process with narrow charged resonances.

Close to the $WW$ threshold, where the $W$ mass
is planned in the FCC-ee experiment to be measured with the $\leq 0.5\,$MeV precision
(using the total cross section~\cite{Abada:2019lih,Abada:2019zxq}),
the problem is that
the pole expansion for the non-QED part of the scattering matrix element
is not efficient any more.
The partial suppression of the QED IFI and FFI corrections
will still work close to threshold as long as resonant curves of $W$'s
are not fully ``distroyed'' by the threshold cut-off.
However, as shown in works based 
on the effective field theory (EFT)~\cite{Beneke:2007zg,Actis:2008rb},
near the threshold one may exploit expansion in the Lorentz velocity 
$\beta=\sqrt{s-4M_W^2}/2M_W \ll 1$ of the $W$'s in order to reduce substantially a number
of diagrams, such that higher-order EW and QED corrections
are again within the reach of practical evaluations.
This kind of expansion should be exploited in the standard diagrammatic approach as well.

Summarising,
a combination of the pole expansion  and of the QED exclusive exponentiation
has already proven to be an economical solution for precision calculations of 
the SM prediction for the $W$-pair production process at LEP
and is the best candidate for the further development
in future electron-positron collider projects, especially for FCC-ee.
The inclusion of the QED interferences between the $W$ production and decays,
and of other missing corrections of the order of $\alpha\Gamma_W/M_W$
will require applying a more sophisticated 
soft/collinear photon factorisation and resummation scheme, combined with POE.
We propose here a new solution based on the coherent exclusive exponentiation, CEEX,
in which resummation of the infrared (IR) divergences 
is done entirely at the amplitude level.
The interesting feature of this new scheme is that the
$\Gamma_W/M_W$ suppression of the QED interferences between production and decay
is a built in feature valid in any order and at any photon energy scale/resolution,
all over the entire multiphoton phase space.
The new scheme is similar to the CEEX scheme previously formulated and successfully applied
to the case of the neutral intermediate resonances (the $Z$-boson).

One should not give up on the more traditional EEX schemes, however.
We shall discuss briefly alternative solutions within the traditional EEX schemes
(extensions of EEX of \yfsww3).
We shall also examine approximations or simplification done in the transition from
the CEEX to EEX schemes, and between various variants of them.

Concluding, this work provides an important building block for the future
high-precision Standard Model calculations 
for the $W$-pair production process at the future $e^+e^-$ colliders.

The paper is organised as follows. 
In Section~2 we describe the pole expansion for the $W$-pair production process. 
Section~3 is devoted to a general discussion of various kinds of the exclusive QED exponentiation 
and a problem of photon emission from an intermediate semi-stable charged particle.
In Section~4 we present details on the CEEX scheme for the process $e^+e^- \to 4f$ 
involving intermediate resonant $W$-bosons. Relations between the CEEX and EEX schemes are
discussed in Section~5. Section~6 contains summary and outlook of our work. 
Finally, detailed derivations of factoring multiphoton radiation from an intermediate semi-stable charged particle, 
resummation of real-photon emissions 
and the virtual YFS form-factor 
for the pertinent process are given in Appendices A, B and C, respectively. 

Shorter version of this work was reported 
in the conference materials of Ref.~\cite{Jadach:2019yhw}. 

%\newpage
%%%%%%%%%%%%%%%%%%%%%%%%%%%%%%%%%%%%%%%%%%%%%%%%%%%%%%%%%%%%%%%%%%%
\section{Pole expansion for $W$-pair production}
%%(based on the original text of Wiesiek).

As pointed out by R.~Stuart~\cite{Stuart:1995zr}, it is always possible to
decompose the matrix element into a combination of Lorentz covariant 
tensors and Lorentz invariant functions. If unstable particles are
involved in a process, one can then perform a Laurent expansion about
complex poles corresponding to those unstable particles. However, only
the Lorentz invariant functions (mathematically, analytic functions of complex
variables) are subject to this expansion, while the Lorentz covariant
and spinor structure of the matrix element should remain untouched. 
In the so-called leading-pole approximation (LPA) one retains only
the leading terms in the above expansion, neglecting the rest of the
Laurent series. As discussed in Ref.~\cite{Stuart:1995zr}, the whole 
procedure does not violate gauge invariance of the matrix element.
This is guaranteed by the fact that all terms in the pole expansion
are independent of each other, e.g. in the case of two unstable
particles, the doubly-resonant terms are independent of
the singly-resonant and non-resonant ones, therefore there cannot be
gauge cancellations between those terms. 
In Ref.~\cite{Stuart:1995zr}, the process of $Z$-pair production and
decay was presented as an example. 

Here, we discuss the process of $W$-pair production and decay: 
%%%%%%%%%%%%%%%%%%%%%%%%%%%%%%%%%%%%%%%%%%%%%%%%%%%%%%%%%%%%%%%%%%%%%%
\begin{equation}
e^-(p_1) + e^+(p_2) \longrightarrow W^-(Q_1) + W^+(Q_2)
\longrightarrow f_1(q_1) + \bar{f}_2(q_2) + f_3(q_3) + \bar{f}_4(q_4),
\label{process}
\end{equation}
%%%%%%%%%%%%%%%%%%%%%%%%%%%%%%%%%%%%%%%%%%%%%%%%%%%%%%%%%%%%%%%%%%%%%%
where $W^-$ decays into $f_1,\bar{f}_2$ and $W^+$ into $f_3,\bar{f}_4$.
At the lowest order, the minimum
gauge invariant subset of Feynman diagrams needed for this process
is the so-called CC11-class of graphs. It includes apart from doubly-resonant
$WW$ graphs (the so-called CC03) also singly-resonant $W$ graphs.   
Below we discuss how to apply the pole expansion this process.

Since we are interested only in LPA (a double-pole approximation in this case)
we start from extracting a part of the full matrix element that can give
rise to doubly-resonant contributions (the rest will drop in LPA anyway). 
It can be written as follows:
%%%%%%%%%%%%%%%%%%%%%%%%%%%%%%%%%%%%%%%%%%%%%%%%%%%%%%%%%%%%%%%%%%%%%%
\begin{eqnarray}
{\cal M} &=& \sum_i \left[\bar{v}_e(p_2) T^i_{\mu\nu} u_e(p_1) \right]
             M_i(s,t,s_1,s_2) \nonumber \\
         & & \times D_W^{-1}(s_2) 
             \left[\bar{u}_{f_3}(q_3) \gamma^{\mu} V_{Wf}(s_2) 
             \omega_- v_{f_4}(q_4) \right] \nonumber \\
         & & \times D_W^{-1}(s_1) 
             \left[\bar{u}_{f_1}(q_1) \gamma^{\nu} V_{Wf}(s_1)
             \omega_- v_{f_1}(q_1) \right] \;, 
\label{2res-me}
\end{eqnarray}
%%%%%%%%%%%%%%%%%%%%%%%%%%%%%%%%%%%%%%%%%%%%%%%%%%%%%%%%%%%%%%%%%%%%%%
where 
%%%%%%%%%%%%%%%%%%%%%%%%%%%%%%%%%%%%%%%%%%%%%%%%%%%%%%%%%%%%%%%%%%%%%%
\begin{equation}
D_W(s) = s - M_W^2 + \Pi_W(s)
\label{Wpropag}
\end{equation}
%%%%%%%%%%%%%%%%%%%%%%%%%%%%%%%%%%%%%%%%%%%%%%%%%%%%%%%%%%%%%%%%%%%%%%
is a Dyson-resumed $W$ propagator with $\Pi_W(s)$ being the $W$ self-energy
correction. In the above we have used the following notation:
%%%%%%%%%%%%%%%%%%%%%%%%%%%%%%%%%%%%%%%%%%%%%%%%%%%%%%%%%%%%%%%%%%%%%%
\begin{eqnarray}
 & & \omega_- = \frac{1}{2}(1 - \gamma_5), \nonumber \\
 & & s_1 = Q_1^2,\; s_2 = Q_2^2,
   \label{notat1} \\
 & & Q_1 = q_1 + q_2, \; Q_2 = q_3 + q_4. \nonumber
\end{eqnarray}
%%%%%%%%%%%%%%%%%%%%%%%%%%%%%%%%%%%%%%%%%%%%%%%%%%%%%%%%%%%%%%%%%%%%%%
$T^i_{\mu\nu}$ are the Lorentz covariant tensors spanning the tensor
structure of the matrix element, while $M_i,\,\Pi_W,\,V_{Wf}$ are
Lorentz scalars that are analytic functions of independent Lorentz
invariants of the process. These functions then undergo the Laurent
expansion about the complex poles corresponding to a finite-range propagation
of two $W$'s. Keeping only the leading terms in the above expansion,
we end up with the LPA matrix element~\cite{Jadach:2000kw,Jadach:2001cz}
%%%%%%%%%%%%%%%%%%%%%%%%%%%%%%%%%%%%%%%%%%%%%%%%%%%%%%%%%%%%%%%%%%%%%%
\begin{eqnarray}
{\cal M}_{\rm LPA} &=& \sum_i \left[\bar{v}_e(p_2) T^i_{\mu\nu} u_e(p_1) \right]
             M_i(s,t,s_p,s_p) \nonumber \\
         & & \times \frac{F_W(s_p)}{s_2-s_p} \left[\bar{u}_{f_3}(q_3) 
             \gamma^{\mu} V_{Wf}(s_p) \omega_- v_{f_4}(q_4)\right]\nonumber \\
         & & \times \frac{F_W(s_p)}{s_1-s_p} \left[\bar{u}_{f_1}(q_1) 
             \gamma^{\nu} V_{Wf}(s_p) \omega_- v_{f_1}(q_1)\right]\;, 
\label{LPA-me}
\end{eqnarray}
%%%%%%%%%%%%%%%%%%%%%%%%%%%%%%%%%%%%%%%%%%%%%%%%%%%%%%%%%%%%%%%%%%%%%%
where the pole position $s_p$ is a solution to the equation
%%%%%%%%%%%%%%%%%%%%%%%%%%%%%%%%%%%%%%%%%%%%%%%%%%%%%%%%%%%%%%%%%%%%%%
\begin{equation}
s - M_W^2 + \Pi_W(s) = 0,\quad F_W(s_p) = [1 + \Pi^{'}_W(s_p)]^{-1}.
\label{spequ}
\end{equation}
%%%%%%%%%%%%%%%%%%%%%%%%%%%%%%%%%%%%%%%%%%%%%%%%%%%%%%%%%%%%%%%%%%%%%%

At the lowest order the Lorentz tensors read
%%%%%%%%%%%%%%%%%%%%%%%%%%%%%%%%%%%%%%%%%%%%%%%%%%%%%%%%%%%%%%%%%%%%%%
\begin{eqnarray}
T_{\mu\nu}^{1,2} &=& \gamma^{\lambda}\Gamma_{\lambda\mu\nu}(Q,Q_1,Q_2),
\label{T12} \\
T_{\mu\nu}^3 &=& \gamma_{\mu}(\not\!p_2 - \not\!Q_2) \gamma_{\nu},
\label{T3} 
\end{eqnarray}
%%%%%%%%%%%%%%%%%%%%%%%%%%%%%%%%%%%%%%%%%%%%%%%%%%%%%%%%%%%%%%%%%%%%%%
and the Lorentz scalars are
%%%%%%%%%%%%%%%%%%%%%%%%%%%%%%%%%%%%%%%%%%%%%%%%%%%%%%%%%%%%%%%%%%%%%%
\begin{eqnarray}
M_1 &=& e^2 \frac{1}{s}, 
\label{M1} \\
M_2 &=&-e^2 \frac{s_W}{c_W}[v_e - a_e\gamma_5] 
            \frac{1}{s-M_Z^2 + iM_Z\Gamma_Z}, 
\label{M2} \\
M_3 &=& \frac{e^2}{s_W^2}\frac{1}{t} ,
\label{M3} \\
V_{Wf} &=& \frac{eU_{ij}\sqrt{N_c}}{2s_W^2},
\label{Vwf} 
\end{eqnarray}
%%%%%%%%%%%%%%%%%%%%%%%%%%%%%%%%%%%%%%%%%%%%%%%%%%%%%%%%%%%%%%%%%%%%%%
where $Q = p_1+p_2,\; s=Q^2,\;t=(p_2-Q_2)^2$, $U_{ij}$ is the CKM matrix
element, $N_c$ is the QCD colour factor, $s_W = \sin\theta_W$,
$c_W = \cos\theta_W$, $v_e$ and $a_e$ are the vector and axial
couplings of a $Z$ boson to electrons,  $\Gamma_{\lambda\mu\nu}$
is the $VWW$ coupling ($V=\gamma,Z$):
%%%%%%%%%%%%%%%%%%%%%%%%%%%%%%%%%%%%%%%%%%%%%%%%%%%%%%%%%%%%%%%%%%%%%%
\begin{equation}
\Gamma_{\lambda\mu\nu}(Q,Q_1,Q_2) = (Q+Q_1)_{\nu}g_{\lambda\mu} 
                                  + (Q_2-Q_1)_{\lambda}g_{\mu\nu}
                                  - (Q+Q_2)_{\mu}g_{\nu\lambda}.
\label{TGC}
\end{equation}
%%%%%%%%%%%%%%%%%%%%%%%%%%%%%%%%%%%%%%%%%%%%%%%%%%%%%%%%%%%%%%%%%%%%%%
In the scalar function $M_2$ we have also applied LPA to the
intermediate $Z$-boson. It is done in a similar way as for $W$'s.
The $W$-pole position, up to ${\cal O}(\alpha^2)$, is given by
%%%%%%%%%%%%%%%%%%%%%%%%%%%%%%%%%%%%%%%%%%%%%%%%%%%%%%%%%%%%%%%%%%%%%%
\begin{equation}
s_p = M_W^2 - iM_W\Gamma_W + {\cal O}(\alpha^2),
\label{spappr}
\end{equation}
%%%%%%%%%%%%%%%%%%%%%%%%%%%%%%%%%%%%%%%%%%%%%%%%%%%%%%%%%%%%%%%%%%%%%%
where $M_W,\,\Gamma_W$ are the usual on-shell scheme $W$ mass and width,
and $F_W=1$. One can easily check that at the lowest order this LPA
matrix element has the same form as the CC03 matrix element calculated
in 't Hooft--Feynman gauge and in the constant $W$-width scheme. 
It was noticed in Ref.~\cite{Beenakker:1994vn} that when 
the CC03
matrix element is calculated in the axial gauge also singly-resonant
terms appear. This indicates that the singly-resonant graphs
are needed to guarantee gauge invariance of the matrix element, i.e.
that CC03 itself is not gauge-invariant, but one has to take at least
CC11 for hadronic, CC10 for semi-leptonic and CC09 for leptonic final
states. In the LPA approach described above it does not matter what
gauge is used in the calculations. We start from the gauge-invariant
matrix element and then apply the pole expansion. In the resulting
LPA matrix element all non-double-pole terms drop out.     

One of the complications that arises when going to higher orders is
the fact that $W$'s are electrically charged and therefore radiate photons.
When a real or virtual photon is emitted from the $W$ one has more
than just two $W$ propagators in the matrix element and the question 
is how to apply the pole expansion in such a case.   
Here, however, one can exploit a partial-fraction 
decomposition of a product of two propagators, namely:
%%%%%%%%%%%%%%%%%%%%%%%%%%%%%%%%%%%%%%%%%%%%%%%%%%%%%%%%%%%%%%%%%%%%%%
\begin{equation}
\frac{1}{Q^2 - M^2}\frac{1}{{Q'}^2 - M^2}
\equiv \frac{1}{2kQ' + k^2}\frac{1}{{Q'}^2 - M^2}
- \frac{1}{Q^2  - M^2}\frac{1}{2kQ  - k^2},
\label{pfd-prop}
\end{equation}
%%%%%%%%%%%%%%%%%%%%%%%%%%%%%%%%%%%%%%%%%%%%%%%%%%%%%%%%%%%%%%%%%%%%%%
where $M^2=M^2_W+i\Gamma_W M_W$, $Q,\,Q'=Q-k$ 
are the $W$ four-momenta before and after
radiation of a photon of the four-momentum $k$, respectively%
\footnote{See also Appandix \ref{appendixA}.}. 
So, a product of two propagators can be replaced by a sum
of single propagators multiplied by eikonal factors. This corresponds
to splitting the photon radiation into the radiation in the $W$-production
stage and  the radiation in the $W$-decay stage. These two stages
are separated by the finite-range $W$ propagation. 
The above decomposition can be applied both to the real and virtual
photon emissions. In the case of the real photons the radiation
amplitude splits into the sum of the amplitudes corresponding
to photon emission in the $WW$-production and two $W$-decays.
At the level of the cross section this results in the sum
of contributions corresponding to the photon radiation at each
stage of the process -- the factorisable corrections, 
and the contributions corresponding to interferences between various stages 
-- the non-factorisable corrections. Similarly, for the virtual corrections, 
the contributions with photons attached to the same stage
give rise to the factorisable corrections, while the ones
where photons interconnect different stages of the process contribute to
the non-factorisable corrections. In this way all radiative corrections
can be split in a gauge-invariant way into the factorisable and
non-factorisable ones. 

Since the non-factorisable corrections were
negligible for the main LEP2 observables one could drop them%
\footnote{In fact, we use an approximation for the non-factorisable
          corrections in terms of the so-called screened Coulomb ansatz 
          of Ref.~\cite{Chapovsky:1999kv}.}
and concentrate only on the factorisable ones. 
For factorisable corrections one can employ
the existing calculation for the on-shell $WW$-production and the
on-shell $W$-decay. Our aim is to treat the QED corrections according
to the YFS exclusive exponentiation procedure and also apply the LPA, 
described above, in order to obtain the gauge-invariant formulation. 
How to do this?
Extraction of infra-red (IR) contributions for both real and
virtual photons can be done in a gauge-invariant way according to
the YFS theory for each of the stages separately. These contributions
are then sum up to infinite order and result in the so-called 
YFS form-factor. 
This means that the YFS form-factors and the IR real-photon  
$\tilde{S}$-factors  involving $W$'s do not have to be taken on-pole but 
can be calculated like for stable particles. 
After having done this we can apply the pole expansion
to the IR-residuals -- the YFS $\bar{\beta}$-functions. 
We proceed in the way described at the beginning of this section
and retain only the leading-pole (double-pole) terms. 
The ${\cal O}(\alpha)$ LPA matrix element for the real photon 
contribution reduces, similarly to the lowest order, to the form
that can be obtained from the doubly-resonant Feynman graphs with
single-photon emission in the 't Hooft--Feynman gauge.  
The ${\cal O}(\alpha)$ virtual correction form-factors should,
in principle, be evaluated on the complex pole.
This would require an analytic continuation of the usual one-loop 
results to the second Riemann sheet (this may be a technical problem).
However, for the aimed
LPA accuracy, it is sufficient to use the approximation $s_p \simeq M_W^2$.
This would correspond to neglecting terms of  
${\cal O}(\frac{\alpha}{\pi}\frac{\Gamma_W}{M_W})$. 
More details about implementation of the ${\cal O}(\alpha)$ corrections
in the $WW$-production process in the MC event generator \yfsww3 can
be found in Ref.~\cite{Jadach:1996hi}.

%\newpage
%%%%%%%%%%%%%%%%%%%%%%%%%%%%%%%%%%%%%%%%%%%%%%%%%%%%%%%%%%%%%%%%%%%
%%%%%%%%%%%%%%%%%%%%%%%%%%%%%%%%%%%%%%%%%%%%%%%%%%%%%%%%%%%%%%%%%%%
\section{General discussion}
\label{sec:GenDis}

In this section we collect discussion on various aspects of the photon radiation
in the $W$ pair production process, in particular we discuss
various exponentiation schemes
preparing grounds for defining them explicitly
in the following sections. 
We define more precisely our aims, discuss various constraints,
introduce notation and terminology.

The fact that $W$'s are narrow resonances and behave like almost stable particles
is of great practical importance for the evaluation of the radiative corrections,
because it provides an additional small parameter $\Gamma_W/M_W$ which
can be used as an expansion parameter, leading to
reduction of the complexity of calculations of radiative corrections.
As a result, the dominant double-resonant part of the process (\ref{eeWW4f})
can be well approximated as three independent processes:
one production process and two decay processes.
For the double resonant part it is possible to use simpler on-shell radiative corrections,
while for the single-resonant part we may stay at the Born-level or use some crude
leading-order (LO) approximations for the radiative corrections.
Of course, we have to have at our disposal a method of splitting the Born 
amplitude and the amplitude with the radiative corrections into the double- and single-resonant parts,
without breaking gauge invariance and other elementary principles.
The pole expansion (POE) seems to be the best method available.
Once POE is used for $W$-pair production process to isolate 
the double-pole (DP), single-pole (SP) and  non-pole (NP) parts,
photon emission from the intermediate unstable $W$'s
has to be reorganised in a consistent way.
In addition, it would be desirable to sum up photon emission from $W$'s
to infinite order (exponentiate),
for instance using one of EEX or CEEX schemes.

In the following we shall characterise various methods of the known
soft photon resummation and then characterise problems
related to soft-photon emission from charged semi-stable intermediate
particles (resonances), like the $W$-bosons.

%%%%%%%%%%%%%%%%%%%%%%%%%%%%%%%%%%%%%%%%%%%%%%%%%%%%%%%%%%%%%%%%%%%
\subsection{Various kinds of exclusive exponentiation}
%%%%%%%%%%%%%%%%%%%%%%%%%%%%%%%%%%%%%%%%%%%%%%%%%%%%%%%%%%%%%%%%%%%
\label{subsec:VarEex}

%%%%%%%%%%%%%%%%%%%%%%%%%%%%%%%%%%%%%%%%%%%%%%%%%%%%%%%%%%%%%%%%%%%%%%%%%%%%%%
%%%%%%%%%%%%%%%%%%%%%%%%%%%%%%%%%%%%%%%%%%%%%%%%%%%%%%%%%%%%%%%%%%%%%%%%%%%%%%%
\begin{table}
\centering
\begin{tabular}{|c|c|c|c|c|}
\hline
 Resummation & Formalism                      & NFI interf. & Implementations & Order  \\
\hline
  \multicolumn{5}{|c|}{ No resonances} \\
\hline
EEX$_{B}$  &\cite{Jadach:1988rr,Yennie:1961ad} & --          & {\tt YFS1}   & \order{\alpha^1} \\
CEEX$_{B}$ & None                              & --          & None         & -- \\
\hline
  \multicolumn{5}{|c|}{ Neutral semi-stable intermediate particles} \\
\hline
EEX$_{R}$  &\cite{Jadach:1991dm} & No          & {\tt YFS3, KORALZ}         & \order{\alpha^3}\\
CEEX$_{R}$ &\cite{Jadach:1999vf,Jadach:2000ir} & Yes         & \kkmc\       & \order{\alpha^2} \\
\hline
  \multicolumn{5}{|c|}{ Charged semi-stable intermediate particles} \\
\hline
EEX$_{R}$  &\cite{Jadach:1996hi,Jadach:2001uu} & No          & {\tt YFSWW3} & \order{\alpha^3}\\
CEEX$_{R}$ & This work                         & Yes         & None         & -- \\
\hline
\end{tabular}
\caption{\sf
 The list of the exclusive soft photon resummation schemes and their implementations.
 The 2nd column indicates the primary reference for the formalism.
 Inclusion of the non-factorisable interference is marked in the 3rd column.
 Practical implementations in the MC codes are listed in the 4th column.
 The maximum (LO) order of the complete non-soft QED corrections is indicated in the last column.
}
\label{tab:table1}
\end{table}

Generally, there are two kinds of exclusive exponentiation schemes: 
(1) the older one, which we call EEX, in which isolation of IR
singularities due to {\em real} photons is done for differential distributions (probabilities),
as in the classic work of Yennie--Frautschi--Suura (YFS)~\cite{Yennie:1961ad},
and (2) the newer one of 
refs.~\cite{Jadach:1998jb,Jadach:2000ir,Jadach:1999vf}, referred to as CEEX,
in which the same isolation of the real photon IR singularities is done 
for the amplitudes themselves, that is before squaring and spin-summing them.
CEEX has a number of advantages over EEX. The price to pay is that it can be more complicated
in the implementation and slower in the numerical evaluation.

Since we are interested mainly in the exclusive exponentiation
for the processes with the narrow resonances,
it is worth to note that, within EEX and CEEX families, there are
two distinct subgroups of implementations which differ rather strongly 
in the treatment of the narrow resonances (or of sharp $t$-channel peaks).
The key difference is in the treatment of the shift of the energy-momentum
in the propagator of the resonance due to emission of the real or virtual photons.
Let us, for the purpose of this work, call this effect a ``recoil effect'' or shortly a ``recoil''.
%{\footnotesize [If somebody has better name than recoil please propose something]}.

Within the EEX family there is 
a baseline variant based on the original YFS work~\cite{Yennie:1961ad},
in which the recoil is realised in an order-by-order way.
Let us denote them with EEX$_{B}$.
Examples of the EEX$_{B}$ variants are: 
the unpublished MC code {\tt YFS1} described in ref.~\cite{Jadach:1988rr} 
and \bhlumi\ 1.x of ref.~\cite{Jadach:1988ec}%
\footnote{In the case of the sharp $t$-channel exchange singularity in the low-angle Bhabha
   scattering, the analog of the recoil effect between the electron and positron lines is also worth
   to take into account in a better way than in EEX$_{B}$.
}.
In  EEX$_{B}$ the recoil is absent completely at the level of \Oeex{\alpha^0}.
Then, it is gradually introduced in an order-by-order manner, 
through the so-called IR-finite $\bbeta$-functions.
For instance, in \Oeex{\alpha^2} the exact recoil in the differential distribution is realised
due to two hard real photons -- if there is a third ``spectator'' hard photon,
then its contribution to resonance propagator is simply ignored.
The problem is that,
from the point of view of the strong variation of the resonance propagator,
a photon with the energy of the order of the resonance width $\Gamma$ is already hard!
This is why EEX$_{B}$ can be disastrous for narrow resonances, 
where in order to realise the recoil, it would be mandatory to jump immediately 
to very high perturbative orders,
otherwise the perturbative convergence for the QED corrections would be  miserable.
EEX$_{B}$ can be a convenient and natural choice if there are no resonances at all.

In the second class of the EEX scenarios, the recoil in the resonance propagator 
(or sharp $t$-channel exchange) is a built-in feature of the scheme,
which is present already in \Oceex{\alpha^0}.
Let us call such a scheme EEX$_{R}$. 
It is realised for the first time
in the {\tt YFS3} event generator~\cite{Jadach:1991dm} and later on
included in the \koralz~\cite{Jadach:1993yv}, \kkmc~\cite{Jadach:1999vf}
programs and finally in the \yfsww3 program~\cite{Jadach:1996hi,Jadach:2001uu}.
The analogous scheme for a process dominated by the $t$-channel was implemented
in the \bhlumi\ MC program \cite{Jadach:1991by,Jadach:1996is}.
In EEX$_{R}$, the total energy-momentum in the resonance propagator 
(or $t$-channel exchange) includes the contribution from all 
real photons emitted prior to resonance formation ($t$-channel exchange).
This means that for each photon we have to know whether 
it belongs to resonance production or decay process (ISR or FSR).
This is possible because in this scenario one always neglects completely and
irreversibly the QED interferences between the ISR and FSR%
\footnote{ In the case of the low angle Bhabha process neglected 
  are interferences between the electron and positron lines in the Feynman diagram.}.
Neglecting these interferences may be not so harmful as compared to experimental precision,
because they are suppressed by the $\Gamma/M$ factor.
The EEX$_{R}$ is obviously very well suited for narrow resonances, 
as long as we can afford neglecting \Order{\alpha\Gamma/M} interference corrections,
and we do not attempt to examine experimentally spectra of photons 
with energies $E_\gamma\simeq \Gamma$.

In the CEEX family of exponentiations there are analogous two sub-classes:
either the recoil is implemented in the infinite order (CEEX$_{R}$) 
or in the order-by-order manner (CEEX$_{B}$).
One great advantage of CEEX is that, 
in the process with the resonant component and the non-resonant background, 
one may apply CEEX$_{R}$ to the resonant part of the amplitude and CEEX$_{B}$ 
to the background and add the two coherently afterwards.

Let us comment on the relation of the above schemes to the classic YFS work
and the relation of EEX$_{R}$ to other ones.
All the above exponentiation schemes are inspired by the classic YFS work~\cite{Yennie:1961ad}
in one way or another.
However, it is in fact only the EEX$_{B}$ scheme
which was formulated explicitly in the original YFS work.
CEEX is a non-trivial extension of the YFS exponentiation scheme,
see ref.~\cite{Jadach:2000ir} for more discussion.
So far, there is no implementation of the CEEX$_{B}$ scheme,
while more sophisticated CEEX$_{R}$ is successfully
implemented in \kkmc~\cite{Jadach:1999vf} program
for the neutral semi-stable $Z$ boson production and decay
in the electron--positron annihilation 
and recently in the proton--proton collision~\cite{Jadach:2017sqv}.

The above inventory of all schemes of the exclusive QED exponentiations
and their implementations are summarised in table~\ref{tab:table1}.

Finally, let us note that  there is another variant of the EEX$_{R}$ scheme implemented
in the \bhwide\ program of ref.~\cite{Jadach:1995nk},
featuring partial implementation of the QED NFI interferences for semi-stable
neutral boson exchanges.
It will be discussed in the following whether this kind of scheme
could be extended to include the QED NFI interferences for the charged semi-stable $W$-boson.

%%%%%%%%%%%%%%%%%%%%%%%%%%%%%%%%%%%%%%%%%%%%%%%%%%%%%%%%%%%%%%%%%%%%%%%%
%%%%%%%%%%%%%%%%%%%%%%%%%%%%%%%%%%%%%%%%%%%%%%%%%%%%%%%%%%%%%%%%%%%%%%%%
\subsection{Photons from intermediate semi-stable charged particle}
\label{subsec:PhoInChPar}

Let us present an introductory discussion on the photon emission from
the intermediate charged unstable $W$'s.

In order to better grasp physics of the photon emission from unstable
charged particles, let us consider one more time the case of
$e^+e^-\to \tau^+\tau^- +n\gamma,\;\; \tau^\pm \to X^\pm$ process.
In this case, with $\Gamma_\tau/m_\tau= 2.27\cdot 10^{-12}$, 
the production and decay processes are well separated in time due to this factor.
For instance, the formation time of the $\tau$-pair at $\sqrt{s}=100\,$GeV is $\sim 10^{-24}\,$sec
while the $\tau$ lifetime is much longer, $2.9\cdot 10^{-13}\,$sec.
This is why the ISR photons emitted from initial beams have no chance 
to interfere with these of the $\tau$ decays.
The FSR photons emitted from the outgoing ultrarelativistic $\tau$'s are quite copiously,
because $\ln(s/m_\tau^2)=8.06$, 
but still, the emissions of the FSR photons and photons in the decays
are time-separated by the  factor of $\Gamma_\tau/m_\tau= 2.27\cdot 10^{-12}$%
\footnote{At LEP energies $\tau$ decays 
 are separated from the production by the giant 2 millimeter distance.}.
The suppression of the interferences between photon emission from two decays is even stronger,
by the factor $\Gamma_\tau/\sqrt{s} \sim 10^{-14}$.
Consequently, all practical calculation for QED effects 
in the $\tau$-pair production and decay process
from the production threshold onwards were implemented in
the Monte Carlo programs independently for the production and decay 
parts~\cite{Jadach:1984iy,Jadach:1991ws,Jadach:1999vf}.
The $\tau$-leptons in the production process are treated
in the perturbative/diagrammatic QED calculations and in the phase-space integration
as stable particles {\em with the fixed mass and the zero decay width}.
Photon emission from the unstable intermediate $\tau$'s is of course 
exponentiated
-- the same way in the decay parts.
Can the above production-decay separation break down?
Yes, if the energy resolution in the photon energy
(a cut on photon energies) is smaller than the $\tau$ width,
that is below $0.003\,$eV, which is experimentally unfeasible.

%/////////////////////////////////////////////////////////////////////////////////////////
\begin{figure}[!ht]
\centering
\setlength{\unitlength}{0.1mm}
\includegraphics[height=70mm,width=110mm]{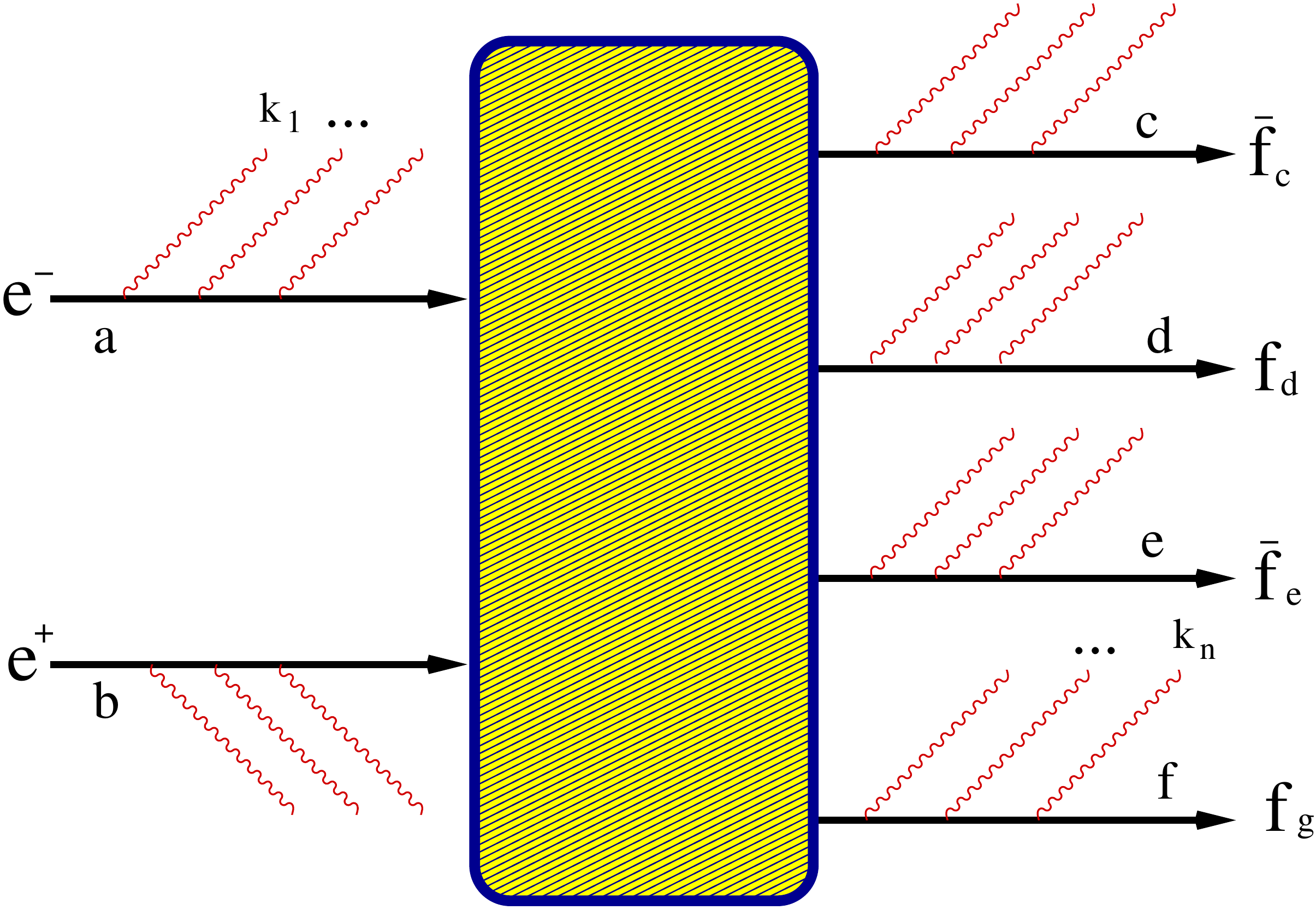} % was 40 mm
\caption{\small\sf
  Kinematics of the four-fermion production process with multiple photons.
}
\label{fig:ee4f}
\end{figure}
%-----------------------------------------------------------------------------------------

In order to see that the problem of the
photon emission from the unstable intermediate $W$'s is not a completely 
trivial,
let us recall a well-known elementary fact~\cite{Yennie:1961ad}:
the emission of photons from the stable initial beams
and four final fermions can be factorised 
into a product of the soft factors  $\prod_i J_{6f}^{\mu_i}(k_i)$
with the total electric current for all six {\em external} particles:
\begin{equation}
  \label{4fcurrent}
  J_{6f}^\mu(k) = \hat{J}_{a}^\mu(k) +\hat{J}_{b}^\mu(k) +\hat{J}_{c}^\mu(k) 
                 +\hat{J}_{d}^\mu(k) +\hat{J}_{e}^\mu(k) +\hat{J}_{f}^\mu(k),
\end{equation}
where
\begin{equation}
  \hat{J}_{x}^\mu(k) = \theta_x Q_x
  \frac{2p_x^\mu\theta_x +k}{k^2+2k\cdot p_x\theta_x +i\varepsilon},
\end{equation}
$p_x$ and $Q_x$ are the momentum and charge (in the units of positron charge) of the
emitter particle $x$, and $\theta_x=+1,-1$
for the initial- and final-state particle, respectively.
For the virtual photons there might be contractions among the
pairs of the currents $J_{6f}^{\mu_i}(k_i)$ and $J_{6f}^{\mu_j}(k_j)$,
see next sections for the explicit formulation.
Fig.~\ref{fig:ee4f} provides a visual representation
of the process of four-fermion production in electron--positron collisions.
All possible contractions (loops) for the virtual photons are not explicitly marked there.

Strictly speaking, in the orthodox YFS scheme~\cite{Yennie:1961ad},
the emissions from the intermediate $W$'s {\em should not}
be included in the IR soft factors, because $W$'s are internal exchanges
and the corresponding emission  does not contribute any IR singularity.
This is true, not only because each $W$ resonance is off-shell ($p_W^2\neq M_W^2$),
but also because photons with energy below $W$ width, $E_\gamma \ll \Gamma_W$, 
emitted according to the above $J_{6f}^\mu$, ``know nothing'' about $W$'s%
\footnote{Finite $W$ width acts as IR regulator.}.
The reason is that, $W$'s live too shortly to affect the distributions of such
very soft (long-wavelength) photons.

%/////////////////////////////////////////////////////////////////////////////////////////
\begin{figure}[!ht]
\centering
\setlength{\unitlength}{0.1mm}
\includegraphics[height=80mm,width=130mm]{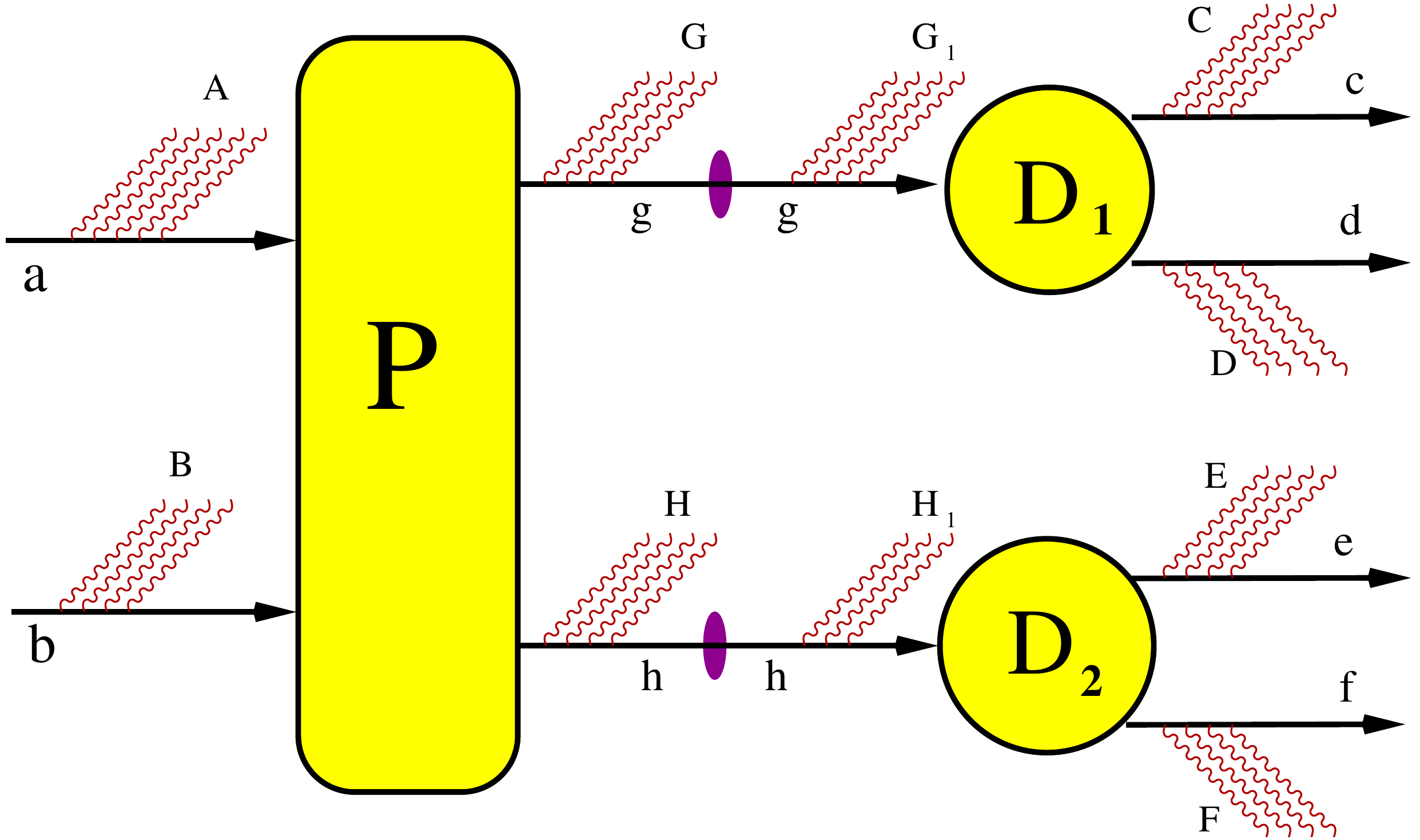} % was 40 mm
\caption{\small\sf
  Kinematics of the double-resonant process.
}
\label{fig:eeWW4f}
\end{figure}
%-----------------------------------------------------------------------------------------
On the other hand, looking into the example of the $\tau$-pair production and decay,
the emission of soft and hard photons out of $W$'s definitely
makes a lot of sense.
However, in the case of the $W$-pair, the time separation of the production and decay stages
is not that extremely long --
this is why it is desirable to implement {\em smooth} analytical transition
from the situation in which emission of photons with $E_\gamma < \Gamma_W$ is governed
solely by the $J_{6f}^\mu$ currents to a situation in which the emission of photons with
$E_\gamma > \Gamma_W$ gets a well-defined contribution from the intermediate $W$'s.
The above situation is visualised in fig.~\ref{fig:eeWW4f}
which describes a double-resonant process
\begin{equation}
  \label{eeWW4f}
  \begin{split}
   &e^-(p_{a}) +e^+(p_{b}) \to W^-(p_{g}) +W^+(p_{h})+n\gamma,\\
   &W^-(p_{g}) \to f_{c}(p_c) +\bar{f}_{d}(p_d)+n\gamma,\quad
    W^+(p_{h}) \to f_{e}(p_e) +\bar{f}_{f}(p_f)+n\gamma,
  \end{split}
\end{equation}
where we understand again that we may also contract any pair of the photon lines
into a virtual photon exchange (loop).
Here and in the following we use the following short-hand notation:
%//////////////////////////////////////////////////
\begin{equation}
  p_{ab}=p_a+p_b,\quad p_{cd}=p_c+p_d,\quad p_{ef}=p_e+p_f.
\end{equation}

The key point is a very special way in which
the recoil is implemented in the resonance propagators.
To understand this problem better,
let us consider first the case with one real photon $n=1$ in the two soft limit regimes:
(i) semi-soft, $k^0 \sim \Gamma \ll \sqrt{s}$ and (ii) true-soft, $k^0\ll \Gamma_W \ll\sqrt{s}$.
The true-soft case (ii) is the case of the standard YFS, in which we have
%//////////////////////////////////////////////////
\begin{equation}
  \label{eq:current-yfs61}
  \begin{aligned}
  &{\Meu^{(0)}}^{\mu_1}(k_1) \simeq
  Const\; \frac{1}{p_{cd}^2-M^2}\; \frac{1}{p_{ef}^2-M^2}
  \\
  & \times \Bigg\{
   Q_a \frac{2p_a^\mu}{2p_a k_1}
  +Q_b \frac{2p_b^\mu}{2p_b k_1}
  -Q_c \frac{2p_c^\mu}{2p_c k_1}
  -Q_d \frac{2p_d^\mu}{2p_d k_1}
  -Q_e \frac{2p_e^\mu}{2p_e k_1}
  -Q_f \frac{2p_d^\mu}{2p_f k_1}
    \Bigg\}.
  \end{aligned}
\end{equation}

In eq.~(\ref{eq:current-yfs61}) there is
no emission from any internal $W$ line and no dependence in the resonance propagators
due to photon emission. 
In the semi-soft regime (i) we have to restore such a dependence in the resonance propagators,
that is take into account the recoil.
This {\em cannot} be done without introducing
photon emission from the intermediate charged resonance into the total electromagnetic current
(unless we drop the NFI corrections altogether, as we already discussed).
In order to see this point more clearly, let us write down
a naive extension of the formula of eq.~(\ref{eq:current-yfs61})
in the {\em complete analogy}
with the CEEX for the neutral resonances, like the $Z$-boson:
%//////////////////////////////////////////////////
\begin{equation}
  \begin{aligned}
    {\Meu^{(0)}_1}^{\mu}(k_1) &\simeq
%%  \\&
  \frac{1}{p_{cd}^2-M^2}\; \frac{1}{p_{ef}^2-M^2}
  \Bigg\{  Q_a \frac{2p_a^\mu}{2p_a k_1} +Q_b \frac{2p_b^\mu}{2p_b k_1} \Bigg\}
  \\&
 +\frac{1}{(p_{cd}+k_1)^2-M^2}\; \frac{1}{p_{ef}^2-M^2}
  \Bigg\{ -Q_c \frac{2p_c^\mu}{2p_c k_1} -Q_d \frac{2p_d^\mu}{2p_d k_1} \Bigg\}
  \\&
 +\frac{1}{p_{cd}^2-M^2}\; \frac{1}{(p_{ef}+k_1)^2-M^2}
  \Bigg\{ -Q_e \frac{2p_e^\mu}{2p_e k_1} -Q_f \frac{2p_d^\mu}{2p_f k_1} \Bigg\}.
  \end{aligned}
\end{equation}
The above extension is, however, {\em useless}, because it is not gauge invariant.
We have to restore emission from the internal $W$ in order to 
cure the gauge invariance, while maintaining recoil in the resonance propagator!

We therefore {\em restore} photon emission from the internal $W$ in the soft photon approximation
(starting from Feynman diagrams)
and next, factorise it into the product of the emission factors
using the identity (\ref{eq:ident1phot}) given in Appendix A.
This identity also shows why it is necessary to sum up {\em coherently} over two photon
assignments, either to $W$ in the production or to $W$ in the decay. 

For the single real semi-soft photon under consideration,
we obtain immediately the following gauge-invariant amplitude
being the sum of three parts, each of them gauge invariant by itself%
\footnote{The gauge invariance is manifest:
  $j^\mu_P k_{1\mu} =  j^\mu_{D_1}k_{1\mu} =  j^\mu_{D_2}k_{1\mu} =0.$
}:
%//////////////////////////////////////////////////
\begin{equation}
  \label{eq:real-single}
  \begin{aligned}
   &{\Meu^{(0)}_1}^{\mu}(k_1) \simeq
  \\&
  \frac{1}{p_{cd}^2-M^2}\; \frac{1}{p_{ef}^2-M^2}
  \Bigg\{  Q_a \frac{2p_a^\mu}{2p_a k_1} +Q_b \frac{2p_b^\mu}{2p_b k_1}
          -Q_g \frac{2p_g^\mu}{2p_g k_1} -Q_h \frac{2p_h^\mu}{2p_h k_1}
  \Bigg\}
  \\&
  +\frac{1}{(p_{cd}+k_1)^2-M^2}\; \frac{1}{p_{ef}^2-M^2}
  \Bigg\{ Q_g \frac{2p_g^\mu}{2p_g k_1}
         -Q_c \frac{2p_c^\mu}{2p_c k_1} -Q_d \frac{2p_d^\mu}{2p_d k_1} \Bigg\}
  \\&
  +\frac{1}{p_{cd}^2-M^2}\; \frac{1}{(p_{ef}+k_1)^2-M^2}
  \Bigg\{ Q_h \frac{2p_h^\mu}{2p_h k_1}
          -Q_e \frac{2p_e^\mu}{2p_e k_1} -Q_f \frac{2p_f^\mu}{2p_f k_1} \Bigg\}
  \\&
  = \frac{1}{p_{cd}^2-M^2}\; \frac{1}{p_{ef}^2-M^2}
    \Bigg\{ j^\mu_P 
           +\frac{p_{cd}^2-M^2 }{(p_{cd}+k_1)^2-M^2}\; j^\mu_{D_1} 
           +\frac{p_{ef}^2-M^2 }{(p_{ef}+k_1)^2-M^2}\; j^\mu_{D_2}  \Bigg\}
  \\&
  =   \sum_{\wp=(P,D_1,D_2)}^{3}  
  \frac{1}{p_{G}^2-M_W^2}
  \frac{1}{p_{H}^2-M_W^2}
   j_{\wp}^{\mu},
  \end{aligned}
\end{equation}
where $p_g=p_c+p_d+k_1$ and $p_h=p_e+p_f+k_1$. In the last line we have used
\begin{equation}
  \label{eq:current}
  \begin{aligned}
& p_G=p_c+p_d+K_{D_1},~~p_H=p_e+p_f+K_{D_2},~~K_X =\sum_{i\in X}k_i
\\ &  j_{P}^{\mu_i}=
  \frac{2 p_{a}^{\mu_i}}{2p_{a} k_i}
  +\frac{2 p_{b}^{\mu_i}}{2p_{b} k_i}
     -\frac{2 p_{G}^{\mu_{i}}}{2p_{G} k_i}
     -\frac{2 p_{H}^{\mu_{i}}}{2p_{H} k_i},
\\ &
   j_{D_1}^{\mu_i}=
  \frac{2 p_{G}^{\mu_{i}}}
           {2p_{G} k_i}
     -\frac{2 p_{c}^{\mu_i}}{2p_{c} k_i}
     -\frac{2 p_{d}^{\mu_i}}{2p_{d} k_i},
   ~~j_{D_2}^{\mu_i}=
  \frac{2 p_{H}^{\mu_{i}}}
           {2p_{H} k_i}
     -\frac{2 p_{e}^{\mu_i}}{2p_{e} k_i}
     -\frac{2 p_{f}^{\mu_i}}{2p_{f} k_i}.
  \end{aligned}
\end{equation}
We keep in mind that in general $p_{g,h}^2\neq M^2$.
The strange looking notation in the last line with the sum over partitions
assigning photon to production or decays is done for the purpose of easy
generalisation to the $n$-emissions case.

The single-photon amplitude of eq.~(\ref{eq:real-single}) coincides precisely
(up to fermion spinors) with the $n=1$ case
of the multiphoton \Ordex{\alpha^0} amplitude
of eq.~(\ref{eq:ccex-alf0}) in the next section.
It features a proper dependence of the resonance propagators on the photon momentum
in the entire photon energy region $k^0\ll \sqrt{s}$,
including $k^0 \sim \Gamma_W$, and interpolates smoothly with the classic
YFS formula of eq.~(\ref{eq:current-yfs61}), in the limit $k^0 \ll \Gamma_W$.
The same will be true for the amplitude of eq.~(\ref{eq:ccex-alf0})
in a more general case of $n>1$.

Let us close this section with the multiple-photon extension of the formula
(\ref{eq:real-single}) with the notation of (\ref{eq:current})  
(details of its derivation can be found in Appendix \ref{app:b}):
\begin{equation}
  \label{eq:multi-WW-real4}
  \begin{aligned}
   &{\Meu^{(0)}_N}^{\mu_1,\dots,\mu_N}(k_1,\dots,k_N)
  \simeq
  \sum_{\wp=(P,D_1,D_2)^N}^{3^N}
%  \frac{-g^{\lambda\sigma}+Q_{g}^\lambda Q_{g}^{\sigma}/M_W^2}
  \frac{1}{p_{G}^2-M_W^2}
  \frac{1}{p_{H}^2-M_W^2}
%\\ &
  \prod_{i=1}^{N}
  j_{\wp_i}^{\mu_i}.
  \end{aligned}
\end{equation}

%\newpage
%%%%%%%%%%%%%%%%%%%%%%%%%%%%%%%%%%%%%%%%%%%%%%%%%%%%%%%%%%%%%%%%%%%%%%%%%%%%%%%
%%%%%%%%%%%%%%%%%%%%%%%%%%%%%%%%%%%%%%%%%%%%%%%%%%%%%%%%%%%%%%%%%%%%%%%%%%%%%%%
\section{CEEX scheme for charged unstable emitters}

In the following we shall implicitly assume that IR-singularities are regularised
with the photon mass $m_\gamma$. The exact IR cancellations
between the real photons phase-space integrals $\int_{m_\gamma} d\Phi$ 
and the virtual form-factor $\alpha B(m_\gamma)$ work very schematically as follows:
\begin{equation}
  \sigma = \sum_{n=0}^\infty \frac{1}{n!}
           \int_{m_\gamma} d\Phi_{4+n}(k_1\dots k_n)
           \sum_{spin} |e^{\alpha B(m_\gamma)} \Mmf(k_1\dots k_n)|^2.
\end{equation}
One may, of course, introduce the traditional IR-cut $E_\gamma > E_{\min}$ for all real photons,
see refs.~\cite{Yennie:1961ad,Jadach:2000ir} for details.
This we shall not do in the following, because it would obscure
notation and is in fact unnecessary 
(even in the MC realisation we could stick to the $m_\gamma$ regulator).

In the following we shall present the formalism of CEEX for
$e^-e^+\to W^+W^-,\;\; W^\pm \to X^\pm$. However, 
this formalism is quite general and applies also
to the single-$W^\pm$ production and decay (also in hadron--hadron collisions)
and also to any other process with any unstable intermediate 
charged particles of arbitrary spin.

%%%%%%%%%%%%%%%%%%%%%%%%%%%%%%%%%%%%%%%%%%%%%%%%%%%%%%%%%%%%%%%%%%%%%%%%%%%%%%%
%%%%%%%%%%%%%%%%%%%%%%%%%%%%%%%%%%%%%%%%%%%%%%%%%%%%%%%%%%%%%%%%%%%%%%%%%%%%%%%
\subsection{ Non-resonant variant of \Order{\alpha^1} CEEX for $e^-e^+\to 4f$}
Let us start from defining CEEX for the $e^-e^+\to 4f$ process with the simplest possible
variant of \Order{\alpha^1} CEEX,
in which the exponentiation procedure is
{\em not} influenced by the presence of any narrow charged resonances in the
Born matrix element $M^{(0)}$.
This CEEX$_{B}$ scheme (according to the notation introduced in Subsection~\ref{subsec:VarEex})
can be used for the non-resonant background in the $e^-e^+\to 4f$ process.
It is a kind of warm-up example in which we introduce
some notation and terminology employed in the following.

Suppressing momenta and spin indices of the fermions,
the \Ordex{\alpha^0} and \Ordex{\alpha^1} 
$n$-photon spin amplitudes can be written in a straightforward way
%///////////////////////////////////////////////////
%     CEEX  O(alf1)
%///////////////////////////////////////////////////
\begin{equation}
  \label{eq:ceexA-master}
  \begin{aligned}
  &{\Meu^{(0)}_n}^{\mu_1,\mu_2,...\mu_n}(k_1,k_2,...k_n)=
  \frac{1}{n!} e^{\alpha B_6^{\rm YFS}}\;
  \hbeta_0^{(0)}  \prod_{i=1}^n j^{\mu_i}(k_i),\qquad \hbeta_0^{(0)}=M^{(0)},
  \\
  &{\Meu^{(1)}_n}^{\mu_1,\mu_2,...\mu_n}(k_1,k_2,...k_n)=
   \frac{1}{n!} e^{\alpha B_6^{\rm YFS}} \Bigg\{
   \hbeta_0^{(1)}  \prod_{i=1}^n j^{\mu_i}(k_i)
  +\sum_{j=1}^n  \hbeta_1^{(1)}{}^{\mu_j}(k_j) \prod_{i\neq j} j^{\mu_i}(k_i)
   \Bigg\},
  \end{aligned}
\end{equation}
where the total electric current 
\begin{equation}
\label{eq:real-curr}  
  j^{\mu}(k_i) = ie \sum_{X=a,b,c,d,e,f} \hat{j}_X^\mu(k_i),\qquad
  \hat{j}_X^\mu(k_i) \equiv Q_X \theta_X \frac{2p_X^\mu}{2p_X k_i},
\end{equation}
sums contributions from all six external fermions $X=a,b,...f$, see fig.~\ref{fig:ee4f},
and $\theta_X=+1$ for the incoming particle $X$ (in the initial state),
    $\theta_X=-1$ for the outgoing particle $X$  (in the final state).
No emission from $W$'s is seen in $j^{\mu}$.
The IF-finite $\beta$-functions are defined in the usual way
%//////////////////////////////////////////////////
\begin{equation}
\begin{aligned}
  &\hbeta^{(1)}_0
  = \left[ e^{-\alpha B_6^{\rm YFS}} M^{(0)}_0 \right]_{ {\cal O}(\alpha^1)},
  \\
  & {\hbeta^{(1)}_1}{}^\mu(k)
  = {M^{(1)}_1}^\mu(k) -j^\mu(k)  M^{(0)}_0.
  \end{aligned}
\end{equation}

The UV-finite, IR-divergent, gauge-invariant
YFS form-factor is defined in the standard way, see also Appendix C:
%//////////////////////////////////////////////////
\begin{equation}
\label{eq:B6YFS}  
\begin{aligned}
&B_6^{\rm YFS} = \int \frac{i}{(2\pi)^3}
\frac{d^4 k}{ k^2-m_\gamma^2 +i\varepsilon}\; 
J^\mu(k) \circ J_\mu(k),
\\&
J^\mu(k)= \sum_{X=a,b,c,d,e,f} \hat{J}_{X}^\mu(k),\qquad
\hat{J}_{X}^\mu(k) \equiv
 Q_X \theta_X\; 
   \frac{ 2p_X^\mu\theta_X +k^\mu}{k^2+ 2p_X k\theta_X +i\varepsilon},
\end{aligned}
\end{equation}
where $\theta_X$ is defined as above, and we use the following short-hand notation:
%%%%%%%%%%%%%%%%%%%%%%%%%%%%
\begin{equation}
\label{eq:virt-curr}  
\begin{aligned}
&S(k)= J(k) \circ J(k) =\sum_{ {X=a,b,c,d,e,f}\atop{Y=a,b,c,d,e,f} } J_X(k) \circ J_Y(k),
\\&
  J_X(k) \circ J_Y(k) \equiv J_X(k) \cdot J_Y(-k),\; {\rm for}\; X\neq Y,\quad
\\&
  J_X(k) \circ J_X(k) \equiv J_X(k) \cdot J_X(k).
\end{aligned}
\end{equation}
As we see, $B_6^{\rm YFS}$ sums up the contributions from all six external fermions.
IR-cancellations occur after squaring, spin-summing and integrating over the phase space,
in a way which was shown using several methods in 
refs.~\cite{Yennie:1961ad,Jadach:2000ir}

%%%%%%%%%%%%%%%%%%%%%%%%%%%%%%%%%%%%%%%%%%%%%%%%%%%%%%%%%%%%%%%%%%%%%%%%%%%%%%%%%%%%%%%
%%%%%%%%%%%%%%%%%%%%%%%%%%%%%%%%%%%%%%%%%%%%%%%%%%%%%%%%%%%%%%%%%%%%%%%%%%%%%%%%%%%%%%%
\subsection{ Resonant variant of CEEX \Order{\alpha^1} for $e^-e^+\to 4f$}
In the following we shall discuss the  \Order{\alpha^1} variant of CEEX for $e^-e^+\to 4f$
in which the recoil in resonance propagators is realised at any perturbative order
and the $\Gamma_W/M_W$ suppression of the NFI contributions 
is a natural, built-in feature, valid in every perturbative 
order \Ordex{\alpha^r}, $r=0,1,2,...$.
In order to formulate such a scheme completely, 
one has to re-consider the isolation of IR-singular
photon-emission factor to infinite order from the internal $W$ lines,
{\em going beyond the scope of the classic scheme of YFS~\cite{Yennie:1961ad}}.
The important element of the isolation of 
the apparent IR-singularities due to emission of photons
from the resonant charged particles
is the reorganisation of the product of the internal
propagators, derived in Appendix A.
The virtual exponential form-factor has also a more complicated structure 
and is re-derived in Appendix C.
Our derivation of the CEEX amplitudes
is based on rearrangement of the infinite
perturbative expansion in terms of Feynman diagrams,
as in refs~\cite{Yennie:1961ad,Jadach:2000ir} and the use of the pole-expansion%
\footnote{We hope that the mathematical {\em rigour} of this proof
  will be improved in the future works.}.
Although our aim are the \Oceex{\alpha^1} amplitudes,
the main features of the scheme can already be defined and discussed for
the simpler \Oceex{\alpha^0} case, which will be discussed first.
The extension of the presented technique to \Oceex{\alpha^2} with the complete
non-soft second-order photonic corrections and the genuine EW
corrections is straightforward.

%%%%%%%%%%%%%%%%%%%%%%%%%%%%%%%%%%%%%%%%%%%%%%%%%%%%%%%%%%
\subsubsection{ Introductory double-pole \Order{\alpha^0} CEEX}
%%%%%%%%%%%%%%%%%%%%%%%%%%%%%%%%%%%%%%%%%%%%%%%%%%%%%%%%%%
\label{sssect:ceex0}
Let us assume that for the $e^-e^+\to 4f$ process depicted in fig.~\ref{fig:ee4f}
we have at our disposal the Born matrix element ${\cal M}^{(0)}_0$ which
we expand into the non-pole part ${M}^{(0)}_0()$,
the single-pole part $M^{(0)}_0(Q)$ and the double-pole part $M^{(0)}_0(Q,R)$,
where $Q$ and $R$ are four-momenta in the $W$ propagators
\begin{equation}
 {{\cal M}^{(0)}_0}^\mu()={M^{(0)}_0}^\mu() +{M^{(0)}_0}^\mu(Q)+{M^{(0)}_0}^\mu(Q,R),
\end{equation}
The same pole-expansion is done for the  exact single-photon spin amplitudes
\begin{equation}
 {{\cal M}^{(1)}_1}^\mu(k)={M^{(1)}_1}^\mu(k)+{M^{(1)}_1}^\mu(Q,k)+{M^{(1)}_1}^\mu(Q,R,k),
\end{equation}
where $k$ is the photon four-momentum and the index $\mu$ is understood to be contracted
with the photon polarisation vector.
The one-loop corrected complete \Order{\alpha^1} spin amplitudes in the POE
we denote as $M^{(1)}_0()$, $M^{(1)}_0(Q^2)$ and $M^{(1)}_0(Q,R)$:
\begin{equation}
 {{\cal M}^{(1)}_0}^\mu={M^{(1)}_0}^\mu() +{M^{(1)}_0}^\mu(Q)+{M^{(1)}_0}^\mu(Q,R).
\end{equation}

Let us focus now on the double-resonant part of the amplitudes
${M^{(0)}_0}^\mu(Q,R)$ and ${M^{(1)}_0}^\mu(Q,R)$.
The single-resonant part is completely analogous (we shall list the differences)
and the non-resonant case has already been discussed
in the previous subsection.

The CEEX \Order{\alpha^0} spin amplitudes for $n$ photons can be derived 
as the following gauge-invariant subset of the
complete perturbative series
%///////////////////////////////////////////////////
%     CEEX  O(alf0)
%///////////////////////////////////////////////////
\begin{equation}
  \label{eq:ccex-alf0}
  \begin{aligned}
  &{\Meu^{(0)}_n}^{\mu_1,\mu_2,...,\mu_n}(k_1,k_2,...,k_n)
   =\sum_{\wp\in\{P,D_1,D_2\}^n}\!\!\!\!\!\!
     e^{\alpha B_{10}(U_\wp,V_\wp)}\;
     \hbeta^{(0)}_0\left(U_\wp,V_\wp\right)
     \prod_{i=1}^n \; j_{\{\wp_i\}}^{\mu_i}(k_i),\\
  & U_\wp= p_c+p_d+\sum_{\wp_i=D_1}k_i,\quad
    V_\wp= p_e+p_f+\sum_{\wp_i=D_2}k_i.
  \end{aligned}
\end{equation}
Here, the fermion four-momenta $p_A$ and helicities $\lambda_A,A=a,b,c,d,e,f$ are suppressed.
Photons are grouped into three sets: production, first decay and second decay, denoted as
$P,D_1,D_2$.
The {\em coherent} sum is taken over all
$3^n$ assignments of a photon to 3 stages of the process.
Each assignment is represented by the vector $(\wp_1,...,\wp_n)$
whose components are taking three possible values $\wp_j= P,D_1,D_2$.
The cornerstone of this construction are three gauge invariant
electric currents
%//////////////////////////////////////////////////
\begin{equation}
  \begin{aligned}
    &j^{\mu}_P(k_i)     = ie \sum_{X=a,b,g,h} \hat{j}^{\mu}_X(k),\quad
     j^{\mu}_{D_1}(k_i) = ie \sum_{X=g,c,d}   \hat{j}^{\mu}_X(k),\quad
     j^{\mu}_{D_2}(k_i) = ie \sum_{X=h,e,f}   \hat{j}^{\mu}_X(k),\quad\\
    &p_g=U_\wp,\quad p_h=V_\wp, 
  \end{aligned}
\end{equation}
defined in terms of elementary currents $\hat{j}_X(k)$ of eq.~(\ref{eq:real-curr}).
They include also $\hat{j}$'s for two $W$'s, see eq.~(\ref{eq:real-curr}).
The essential steps in derivation of the CEEX formula of eq.~(\ref{eq:ccex-alf0})
are given in Appendices A, B and C.

The dependence of the amplitude in
eq.~(\ref{eq:ccex-alf0}) on the four-momenta was already analysed
in the case of the single real photon in the previous section.
The case of many real photons is completely analogous.
Let us turn now our attention to a more interesting case of multiple virtual
photos which contribute to the virtual form-factor $\exp(B_{10})$.

The virtual IR-singularities factorise off in eq.~(\ref{eq:ccex-alf0}) into
the factor $\exp(B_{10})$.
Let us recall that our aim is to reproduce
the $\Gamma/M$ suppression of the NFI corrections already 
at the \Ordex{\alpha^0} level.
It would be incorrect to employ here
the classic YFS form-factor $B_6^{\rm YFS}$ of eq.~(\ref{eq:B6YFS}).
This choice would render eq.~(\ref{eq:ccex-alf0}) IR-finite, however,
it would fail to resum the $\alpha\ln(\Gamma/M)$ contributions
and miss the $\Gamma/M$ suppression of NFI corrections,  at the \Ordex{\alpha^0} level.
How to see it? 
One may check it by explicit analytical calculation, similar to the one performed
in ref.~\cite{Jadach:2000ir}, or numerically.
Quite generally, the reason for the above failure is that
the effective energy scale for NFI is not $\sqrt{s}$ but $\Gamma_W$.
The NFI contributions for the real photon energies above $\Gamma_W$
are suppressed strongly by the resonance propagator.
However, this works for the real but not for virtual photons in  $B_6^{YFS}$, 
hence the energy scale for virtual photons is necessarily $\sqrt{s}$.
The mismatch between the scale for real and virtual photon will cause the
NFI contribution to blow up at the \Ordex{\alpha^0} by orders of magnitude,
and even for \Ordex{\alpha^1} they may be far from the reality.

The remedy for the above problem is well known for the neutral 
resonances~\cite{Greco:1975rm,Greco:1980mh,Jadach:2000ir}
and also can be deduced from the \Order{\alpha^1} calculation (without exponentiation) 
of the NFI term for the charged resonance of $W$,
see refs.~\cite{Fadin:1993dz,Chapovsky:1999kv,Dittmaier:1999mb}.
The modified CEEX form-factor which should be used in eq.~(\ref{eq:ccex-alf0}) 
is the following:
%//////////////////////////////////////////////////
\begin{equation}
\label{eq:B10}
  \begin{aligned}
    &B_{10}(p_{cd},p_{ef})= 
      \int  \frac{i}{(2\pi)^3}\; \frac{d^4k}{k^2-\lambda^2+i\varepsilon}\;
\\&\qquad
    \bigg\{  J_P(k) \circ J_P(k)   +J_{D_1}(k)\circ J_{D_1}(k)   +J_{D_2}(k) \circ J_{D_2}(k)
\\&\qquad
          +\frac{p_{cd}^2-M^2 }{(p_{cd}-k)^2-M^2}\; 2 J_P(k)\circ J_{D_1}(k)
          +\frac{p_{ef}^2-M^2 }{(p_{ef}-k)^2-M^2}\; 2 J_P(k)\circ J_{D_2}(k)  
\\&\qquad
          +\frac{p_{cd}^2-M^2 }{(p_{cd}+k)^2-M^2}
         \;\frac{p_{ef}^2-M^2 }{(p_{ef}-k)^2-M^2}\;
          2 J_{D_1}(k)\circ J_{D_2}(k)
\bigg\},
  \end{aligned}
\end{equation}
where
%//////////////////////////////////////////////////
\begin{equation}
  \begin{aligned}
    &J^{\mu}_P(k)     = \sum_{X=a,b,g,h} \hat{J}_X^\mu(k),\qquad
     J^{\mu}_{D_1}(k) = \sum_{X=g,c,d}   \hat{J}_X^\mu(k),\qquad
     J^{\mu}_{D_2}(k) = \sum_{X=h,e,f}   \hat{J}_X^\mu(k),\\
    &p_g = p_{cd}+K_1,\quad p_h = p_{ef}+K_2,
  \end{aligned}
\end{equation}
see eq.~(\ref{eq:virt-curr}) for definition of elementary virtual current $\hat{J}_X$
and of its circle-products.
In eq.~(\ref{eq:ccex-alf0}) the four-momenta $U_\wp,V_\wp$ in  $B_{10}(U_\wp,V_\wp)$
should be identified with $p_{cd}+K_1$ and $p_{ef}+K_2$ in eq.~(\ref{eq:B10}),
where $K_1$ and $K_2$ are total four momenta of all {\em real} 
photons in the two decay processes.
Note that the above form-factor is gauge invariant and UV-finite.
Moreover, each of its six components is also separately gauge invariant and
UV-finite.
Almost all its components are already available in the literature.
We have omitted from discussion the important Coulomb effect,
see ref.~\cite{Chapovsky:1999kv,Actis:2008rb} for more details.

The index $10$ in $B_{10}$ reflects the fact that we have $10$ emission currents
in $B_{10}$: $6$ for fermions and $4$ for $W$'s 
-- that is $2$ for $W$'s in the production and $2$ for $W$'s in the decays.

Heuristic derivation of
the above CEEX form-factor, directly from the Feynman diagram, is done in
Appendix C using similar techniques as in Subsection~3.2.2 of 
ref.~\cite{Jadach:2000ir}.
In this derivation one may see explicitly why the first three components
for the production and decays are exactly like in the standard YFS scheme, while three
interferences are modified.

%%%%%%%%%%%%%%%%%%%%%%%%%%%%%%%%%%%%%%%%%%%%%%%%%%%%%%%%%%%%%%%
\subsubsection{ The \Order{\alpha^1} CEEX for  double-pole component}
%%%%%%%%%%%%%%%%%%%%%%%%%%%%%%%%%%%%%%%%%%%%%%%%%%%%%%%%%%%%%%%
\label{sssect:dpceex1}
The construction of  \Ordex{\alpha^0} for the $e^+e^-\to 4f$ process
of the previous subsection was based,
on one hand, on the gauge invariant POE of the Born spin amplitudes into
the double-, single- and non-pole parts
and, on the other hand, on the soft photon approximation in which
real and virtual photon emission/absorption
is represented as a product of the universal (spin-independent) factors,
taking care of the recoil in all resonance propagators.

We intend now to extend the above scheme in such a way that
the complete \Order{\alpha^1} to the $e^+e^-\to 4f$ process
are or can be included.
The immediate question is to what extent POE into the double-, single- and non-pole parts
can be kept at all at \Order{\alpha^1}?

Concerning POE at \Order{\alpha^1},
we assume that both the \Order{\alpha^1}
amplitudes:  $M_1^{(1)\mu}(k)$ with the emission of an additional single photon
and $M_0^{(1)}$ with the complete one-loop corrections can be pole-expanded
into the double-, single- and non-pole parts%
\footnote{The ultimate proof will be provided by someone who will do it in practice.}.
Obviously,  this can be done in many ways.
Essentially it can be done (in principle) because
the two propagators for the {\em internal} $W$ line due to photon emission
can always be replaced by a sum of ``two poles'' using the identity of eq.~(\ref{eq:ident1phot}).
Each of these terms can be made gauge invariant by taking a residue value
for the entire expression multiplying the pole term, or more selectively, in its scalar part.
This can be done (in principle) for both the amplitudes $M_1^{(1)\mu}(k)$ and $M_0^{(1)}$
representing the exact results of the Feynman diagrams at \Order{\alpha^1}.
The soft-photon-approximated universal part is already included in the calculation 
due to the exponentiation, in the same way as at \Order{\alpha^0}.

The double-pole \Order{\alpha^1} CEEX amplitude, including
terms of \Order{\frac{\alpha}{\pi}\frac{\Gamma}{M}} 
due to the NFI interferences, reads as follows:
%///////////////////////////////////////////////////
%     CEEX  O(alf1)
%///////////////////////////////////////////////////
\begin{equation} 
\label{eq:ceex1b-resonanat}
  \begin{aligned}
 &{\Meu^{(1)}_n}^{\mu_1,\mu_2,...,\mu_n}(k_1,k_2,...,k_n)_{\rm DP}
   =\sum_{\wp\in\{P,D_1,D_2\}^n} \!\!\!\!
     e^{\alpha B_{10}(U_\wp,V_\wp)}
     \hbeta^{(1)}_0\left(U_\wp,V_\wp\right)
     \prod_{i=1}^n \; j_{\{\wp_i\}}^{\mu_i}(k_i)\; 
\\&\qquad
   + \sum_{j=1}^n\;\;
     \sum_{\wp\in\{P,D_1,D_2\}^{n-1}} \!\!\!\!
     e^{\alpha B_{10}(U_\wp,V_\wp)}
     \hbeta^{(1)\mu_j}_{1\{\wp_j\}}\left(U_\wp,V_\wp,k_j\right)
     \prod_{i\neq j}\; j_{\{\wp_i\}}^{\mu_i}(k_i),
  \end{aligned}
\end{equation}
%%------------
where
%//////////////////////////////////////////////////
\begin{equation}
  \hbeta^{(1)\mu}_{1}(U,V,k)
  =M^{(1)\mu}_{1}(U,V,k)
    -\sum_{\wp=P,D1,D2} j_{\wp}^\mu(k) M^{(0)}_0(U_{\wp},V_{\wp}).
\end{equation}
The IR-finite $\hbeta_0$-functions is here defined as follows:
%//////////////////////////////////////////////////
\begin{equation}
  \hbeta^{(1)}_0(U,V)
  = \left[ e^{-\alpha B_{10}(U,V)} M^{(1)}_0(U,V)
    \right]_{ {\cal O}(\alpha^1)} = M^{(1)}_0(U,V) - B_{10}(U,V) M^{(0)}_0(U,V),
\end{equation}
where $B_{10}(U,V)$ is the complete variant of eq.~(\ref{eq:B10})
and the one-loop corrections in 
the double-pole $M^{(1)}_0(U,V)$ have to be complete at the \Order{\alpha^1},
including terms of \Order{\frac{\alpha}{\pi}\frac{\Gamma}{M}}.
Special care should be taken in order to preserve gauge invariance.
Infrared regulation using $m_\gamma$ or any other method may be employed
in the intermediate steps, but the final $B_{10}(U,V)$ will be IR-finite.

Needless to say that in the above expressions, as usual in all resummation schemes,
one has to provide a recipe for extrapolating the \Order{\alpha^1} results,
originally defined in the phase space with zero or one real photon,
to the phase space enriched with many additional ``spectator'' photons%
\footnote{It is typically done using some kinematic manipulations
  on the four-momenta which are fed into \Order{\alpha^1} formulae
  or using Mandelstam variables -- they are less sensitive to the presence
  of spectators.}.
The uncertainty due to freedom in this extrapolation is of
the \Order{\alpha^2} class.

%%%%%%%%%%%%%%%%%%%%%%%%%%%%%%%%%%%%%%%%%%%%%%%%%%%%%%%%%%%%%%%
\subsubsection{\Order{\alpha^1} CEEX for single-pole component}
%%%%%%%%%%%%%%%%%%%%%%%%%%%%%%%%%%%%%%%%%%%%%%%%%%%%%%%%%%%%%%%

The above implementation of \Order{\alpha^1} CEEX for the DP component
of the QED \Order{\alpha^1} corrections are complete including 
\Order{\frac{\alpha}{\pi}\frac{\Gamma}{M}} corrections due
the NFI interferences.
However, the \Order{\frac{\alpha}{\pi}\frac{\Gamma}{M}} corrections
arise also from the entire QED \Order{\alpha^1} correction
to a single-pole component (which by itself is of \Order{\frac{\Gamma}{M}}).
It is therefore necessary to define \Order{\alpha^1} CEEX for the SP part.
In addition, CEEX for the SP process is also of the vital importance
for the $q\bar{q}\to W \to f\bar{f}$ process at hadron colliders, such as the LHC.

On the other hand, the non-pole (background) part,
which is of \Order{(\frac{\Gamma}{M})^2}),
may included without QED corrections or any kind of implementation
of QED corrections, for instance using the simple baseline \Order{\alpha^0} CEEX
version of Subsection~\ref{sssect:ceex0}.

The CEEX \Ordex{\alpha^1} single-pole and double-pole spin amplitudes 
will be combined {\em additively} as follows%
\footnote{%
  In some four-fermion channels there is no possibility to form a single-resonant $W$.
}:
%//////////////////////////////////////////////////
\begin{equation}
   {\Meu^{(1),\mu_1,\dots,\mu_n}_{n}}(k_1,k_2,...,k_n)_{\rm DSP}
  ={\Meu^{(1),\mu_1,\dots,\mu_n}_{n}}(k_1,k_2,...,k_n)_{\rm SP}
  +{\Meu^{(1),\mu_1,\dots,\mu_n}_{n}}(k_1,k_2,...,k_n)_{\rm DP}.
\end{equation}

The single-pole ${\Meu^{(1)}_{n}(\ldots)_{\rm SP}}$ amplitude is constructed analogously
as in eq.~(\ref{eq:ccex-alf0}). 
The differences are that:
(i) the current $j^\mu_P$ in the production process $e^+e^-\to f_c+\bar{f}_d +W^+$
has five components instead of four,
(ii) the function $B_8$ replaces $B_{10}$,
the $B_8$ has less components, in particular
one interference term instead of three,
(ii) the sum over photon assignment is reduced to the sum over the set 
$\{\wp\}=(P,D_1)^n$ corresponding to $2^n$ assignments:
%///////////////////////////////////////////////////
%     CEEX  O(alf1)
%///////////////////////////////////////////////////
\begin{equation} 
\label{eq:ceex1c-resonanat}
  \begin{aligned}
 &{\Meu^{(1)}_n}^{\mu_1,\mu_2,...,\mu_n}(k_1,k_2,...,k_n)_{SP}
   =\sum_{\wp\in\{P,D_1\}^n} \!\!\!\!
     e^{\alpha B_{8}(U_\wp)}
     \hbeta^{(1)}_0\left(U_\wp\right)
     \prod_{i=1}^n \; j_{\{\wp_i\}}^{\mu_i}(k_i)\; 
\\&\qquad\qquad\qquad
   + \sum_{j=1}^n\;
     \sum_{\wp\in\{P,D_1\}^{n-1}} \!\!\!\!
     e^{\alpha B_{8}(U_\wp)}
     \hbeta^{(1)\mu_j}_{1\{\wp_j\}}\left(U_\wp,k_j\right)
     \prod_{i\neq j}\; j_{\{\wp_i\}}^{\mu_i}(k_i),
  \end{aligned}
\end{equation}
%%------------
where
%//////////////////////////////////////////////////
\begin{equation}
  \hbeta^{(1)\mu}_{1}(U,k)
  =M^{(1)\mu}_{1}(U,k) -\sum_{\wp=P,D1} j_{\wp}^\mu(k) M^{(0)}_0(U_{\wp}).
\end{equation}
The IR-finite $\hbeta_0$-functions is defined here as follows:
%//////////////////////////////////////////////////
\begin{equation}
  \hbeta^{(1)}_0(U)
  = \left[ e^{-\alpha B_{8}(U)} M^{(1)}_0(U)
    \right]_{ {\cal O}(\alpha^1)} = M^{(1)}_0(U) - B_{8}(U) M^{(0)}_0(U),
\end{equation}
where $M^{(1)}_0(U)$ is the single-pole part in the Born amplitude 
of the $e^+e^-\to 4f$ process
and the one-loop corrected single-pole $M^{(1)}_0(U)$ amplitude
is complete at \Order{\alpha^1}.
The $B_{8}(U)$ function is the following variant of that in eq.~(\ref{eq:B10}):
%//////////////////////////////////////////////////
\begin{equation}
\label{eq:B8}
  \begin{aligned}
    &B_{8}(p_{cd})= 
     ie \int  \frac{i}{(2\pi)^3}\; \frac{d^4k}{k^2-\lambda^2+i\varepsilon}\;
\\&\qquad
    \bigg\{  J_P(k) \circ J_P(k)   +J_{D_1}(k)\circ J_{D_1}(k)
%\\&\qquad
          +\frac{p_{cd}^2-M^2 }{(p_{cd}+k)^2-M^2}\; 2 J_P(k)\circ J_{D_1}(k)
\bigg\}.
  \end{aligned}
\end{equation}

The above \Order{\alpha^1} CEEX for the single-pole part of the $e^+e^-\to 4f$ process
implemented in $\Meu^{(1)}_{n}(\ldots)_{\rm SP}$
provides, together with the double-pole CEEX amplitude 
$\Meu^{(1)}_{n}(\ldots)_{\rm DP}$ of the previous section,
the complete QED corrections at the order of \Order{\alpha^1}, 
\Order{\frac{\Gamma}{M}} and
\Order{\frac{\alpha}{\pi}\frac{\Gamma}{M}}
for the $e^+e^-\to 4f$ process.
Let us keep in mind that the definition of the \Order{\frac{\alpha}{\pi}\frac{\Gamma}{M}}
terms in $\Meu^{(1)}_{n}(\ldots)_{\rm SP}$ and $\Meu^{(1)}_{n}(\ldots)_{\rm DP}$
depends on the exact definition of the SP and DP components in POE. 
Only the sum of them is uniquely defined -- more precisely up to the terms
of \Order{\frac{\alpha}{\pi}(\frac{\Gamma}{M})^2}.

In the above formalism, the fermions labeled $e$ and $f$ do not form the resonance.
In the case of the single-$W$ production in the quark--antiquark annihilation
in hadron--hadron collision, the same formalism applies but the particles
$e$ and $f$ are just absent.

%%%%%%%%%%%%%%%%%%%%%%%%%%%%%%%%%%%%%%%%%%%%%%%%%%%%%%%%%%%%%%%
\subsubsection{ Approximate version of \Order{\alpha^1} CEEX}
%%%%%%%%%%%%%%%%%%%%%%%%%%%%%%%%%%%%%%%%%%%%%%%%%%%%%%%%%%%%%%%

Let us also consider one simpler case of the CEEX matrix element,
with the incomplete \Order{\frac{\alpha}{\pi}\frac{\Gamma}{M}} corrections.
It may be of some practical significance for applications
with limited precision and will be described for the DP part only.

In this alternative scheme, the \Order{\alpha^0} part is kept the same
as in the full version of the CEEX scheme for the DP part of Subsection~\ref{sssect:dpceex1}.
The main difference is in the simplification
of the non-soft \Order{\alpha^1} remnants, 
in which the non-factorisable QED interferences between the production and the decays
are downgraded to the soft-photon approximation.

In such an approximation, the \Order{\alpha^1} non-soft corrections
are calculated separately for the production and two decay processes,
and they contribute separately and additively 
to both real $\hbeta^{(1)\mu_1}$ and virtual $\hbeta^{(1)\mu}_0$:
\begin{equation}
 \hbeta^{(1)\mu}_1(U,V,k) = \sum_{X=P,D_1,D_2} \hbeta^{(1)\mu}_{1,X}(U,V,k),\quad
 \hbeta^{(1)}_0(U,V) = \sum_{X=P,D_1,D_2} \hbeta^{(1)}_{0,X}(U,V),
\end{equation}
where $U=p_{cd}$, $V=p_{ef}$.
For instance, the non-soft
contributions from penta-box diagrams in the NFI class
are neglected completely in the $\hbeta^{(1)}_0(U,V)$, 
because their soft part (including resonance effects) 
is already included in the $B_{10}(U,V)$ function.
The single real photon emission spin amplitudes factorise into the production
and decay parts
\begin{equation}
\begin{aligned}
&\Meu_1^{(1)\mu} (k)
    =\Meu_{1,P}^{(1)\mu}(k)\;  \Meu_{0,D_1}^{(0)}\;       \Meu_{0,D_2}^{(0)}
    +\Meu_{1,P}^{(0)}\;        \Meu_{1,D_1}^{(1)\mu}(k)\; \Meu_{0,D_2}^{(0)}
    +\Meu_{1,P}^{(0)}\;        \Meu_{0,D_1}^{(0)}\;       \Meu_{1,D_2}^{(1)\mu}(k)\;
\\&
= \Meu_{0,P}^{(0)} \big[  j_{P}^\mu(k) \Meu_{0,D_1}^{(0)}  \Meu_{0,D_2}^{(0)} 
                   + \Meu_{0,D_1}^{(0)}(k) j_{D_1}^\mu(k)  \Meu_{0,D_2}^{(0)}
                   + \Meu_{0,D_1}^{(0)}  \Meu_{0,D_2}^{(0)}(k) j_{D_2}^\mu(k))\big]
\\&~~~
+ \tbeta^{(1)\mu}_P(k) \Meu_{0,D_1}^{(0)}  \Meu_{0,D_2}^{(0)}
+ \Meu_{1,P}^{(0)}\;   \tbeta^{(1)\mu}_{D_1}(k)   \Meu_{0,D_2}^{(0)}
+ \Meu_{1,P}^{(0)}\;   \Meu_{0,D_1}^{(0)} \tbeta^{(1)\mu}_{D_2}(k)
\\&
= \hbeta^{(0)}_0\left(U,  V \right) j_{P}^\mu(k)
 +\hbeta^{(0)}_0\left(U+k,V \right) j_{D_1}^\mu(k)
 +\hbeta^{(0)}_0\left(U,V+k \right) j_{D_2}^\mu(k)
\\&~~~
 +\hbeta^{(1)\mu}_{1P}(k) +\hbeta^{(1)\mu}_{1D_1}(k)+\hbeta^{(1)\mu}_{1D_2}(k),
\end{aligned}
\end{equation}
where $\tbeta^{(1)\mu}_X(k),\;\; X=P,D_1,D_2$ 
are the CEEX elements for the production and the decays separately,
and we have adopted a convention that the $W$ propagator is included in the 
lowest order decay amplitude $\Meu_{0,D_i}^{(0)}$.
An additional argument $(k)$ in $\Meu_{0,D_i}^{(0)}(k)$
marks that this $W$ propagator includes the momentum 
$k$ of the photon emitted in the decay.

The resulting variant of the \Order{\alpha^1} CEEX amplitude reads as follows:
%///////////////////////////////////////////////////
%     CEEX  O(alf1)
%///////////////////////////////////////////////////
\begin{equation} 
\label{eq:ceex1-approx}
  \begin{aligned}
 &{\Meu^{(1)}_n}^{\mu_1,\mu_2,...,\mu_n}(k_1,k_2,...,k_n)
\\&
   =\!\!\! \sum_{\wp\in\{P,D_1,D_2\}^n} \!\!\!\!\!
     e^{\alpha B_{10}(U_\wp,V_\wp)}
     \Bigg\{ 
     \hbeta^{(1)}_0\left(U_\wp,V_\wp\right)
     \prod_{i=1}^n \; j_{\{\wp_i\}}^{\mu_i}(k_i)\; 
     +\sum_{j=1}^n 
     \hbeta^{(1)\mu_j}_{1\{\wp_j\}}\left(U_\wp,V_\wp,k_j\right)
     \prod_{i\neq j}\; j_{\{\wp_i\}}^{\mu_i}(k_i)\;
    \Bigg\}.
  \end{aligned}
\end{equation}
%%------------
The important difference with respect to the previous case
is that due to the splitting of $\hbeta^{(1)}$ into the production and decay parts,
the photon $k_j$ entering $\hbeta^{(1)}$ is included into the sum over
the photon assignments.

%%%%%%%%%%%%%%%%%%%%%%%%%%%%%%%%%%%%%%%%%%%%%%%%%%%%%%%%%%%%%%%%%%%%%%%%%%%%%%%%%%%%%%%
\subsubsection{ Higher order upgrades and inclusion of genuine electroweak corrections}
%%%%%%%%%%%%%%%%%%%%%%%%%%%%%%%%%%%%%%%%%%%%%%%%%%%%%%%%%%%%%%%%%%%%%%%%%%%%%%%%%%%%%%%

The upgrade of the CEEX amplitudes from \Order{\alpha^1}
to \Order{\alpha^2} is straightforward, following the same
path as in the analogous case of the QED \Order{\alpha^2} CEEX scheme
implemented in the \kkmc\ project\cite{Jadach:1999vf,Jadach:2000ir}.
The CEEX scheme offers great flexibility, 
allowing to truncate a perturbative series at a different order for ISR, FSR,
IFI and IFF.
This may be exploited in a convenient
staging of construction of the respective numerical Monte Carlo program.
In particular,
for the ISR corrections it would be good to include the LO \Order{\alpha^3} corrections.
From the experience of the \kkmc\ project we know that calculations of
the CEEX \Order{\alpha^2} matrix element may be slow, due to the need of
summations over the assignments of photons among production and decays.
However, most of numerical contributions from these 
photon assignments are numerically
negligible and one may invent methods of the effective forecasting which assignments
can be omitted from the evaluation.
This would speed up significantly numerical MC calculations%
\footnote{This will be mandatory for the LO \Order{\alpha^3} corrections.}.

In the present work we concentrate on the QED part of the SM calculations
for the $e^+e^-\to W^+W^-$ process.
Is it possible to factorise and treat separately the QED part from the rest
of the SM corrections, the genuine EW corrections?
The answer is positive because the soft-photon factorisation for both the real
and virtual photons is well established in the 
framework of perturbative calculations~\cite{Yennie:1961ad}.
The remaining genuine EW \Order{\alpha^r} $r=1,2$
corrections are located in the IR-finite remnants 
$\hbeta^{(r)\mu}_0$, $\hbeta^{(r)\mu}_1(k), \hbeta^{(r)\mu_1\mu_2}_2(k_1,k_2)$.
It is only important to remember that the CEEX scheme works at the amplitude
level and in the calculation of the loop corrections leading to
$\hbeta^{(r)\mu}_0$ or $\hbeta^{(r)\mu}_1(k)$, all the IR divergences
are removed by means of subtracting the $B_{10}$ function --
adding the real emissions \`a la Bloch--Norsieck in order to obtain finite
results is a methodological mistake!
Because of that it is much easier to manage the genuine EW corrections
in the CEEX scheme of any perturbative order than in any other scheme,
especially beyond \Order{\alpha^1}.

In the \kandy\ (\yfsww3) calculations of the LEP era, the \Order{\alpha^1} genuine 
EW corrections were included in $\hbeta^{(1)\mu}_0$ 
for the DP production part of the process (similarly as in \racoon).
In order to match a very high precision of the FCC-ee experiments,
it will be necessary to introduce  the \Order{\alpha^2} corrections
in $\hbeta^{(2)\mu}_0$ and $\hbeta^{(1)\mu}_1(k)$ of the DP component.
They are not available yet.
In addition, it will be needed to introduce
the \Order{\alpha^1} EW corrections in $\hbeta^{(1)\mu}_0$ of the SP component.
This subgroup of corrections can, in principle, be extracted from
the existing EW \Order{\alpha^1} calculations for the entire $e^+e^-\to 4f$ process
of ref.\cite{Denner:2005es,Denner:2005fg}.

%%%%%%%%%%%%%%%%%%%%%%%%%%%%%%%%%%%%%%%%%%%%%%%%%%%%%%%%%%%%%%%%%%%%%%%%%%%
%%%%%%%%%%%%%%%%%%%%%%%%%%%%%%%%%%%%%%%%%%%%%%%%%%%%%%%%%%%%%%%%%%%%%%%%%%%
%\newpage
\section{Relations between CEEX and EEX schemes}

Tracing exact relations between various CEEX and EEX schemes 
is quite important for at least two reasons.
The EEX implementation of the exclusive exponentiation in \yfsww3
is the only existing one for the $e^+e^-\to 4f$ process, so it is desirable
to show that it can be embedded in the CEEX scheme as a kind
of a well-defined approximation.
It will also help to better understand the physics of photon emission 
from unstable charged intermediate particles and the inherent limitations 
of the EEX exponentiation scheme in \yfsww3,
in particular clarifying the question: 
what is exactly the mechanism of neglecting the NFI interferences 
in EEX of \yfsww3?

Another important reason is that it would be desirable
to implement the CEEX matrix element using a MC correction weight
on top of the same baseline MC distributions, which is implemented
in the MC event generator for the EEX matrix element.
This strategy was successfully exploited in the \kkmc\ program
and also in the \kandy\ hybrid Monte Carlo.
For these reasons it is interesting to establish the relation between the CEEX and EEX
distributions all over the entire multiphoton phase space.

%%%%%%%%%%%%%%%%%%%%%%%%%%%%%%%%%%%%%%%%%%%%%%%%%%%%%%%%%%%%%%%%%%%%%%%%%%%
\subsection{From CEEX$_R$ to EEX$_R$ algebraically}

As we have already indicated in the introduction,
the EEX differential distributions
for the process $e^-e^+\to W^-W^+,\; W^\pm\to f\bar{f}$,
can be obtained as a limiting case of the CEEX
scheme for the process $e^-e^+\to 4f$, defined in this paper.
Let us do it in the following.
This is analogous to the derivation of EEX of \koralz\
out of the CEEX amplitudes given in Section~4 of ref.~\cite{Jadach:2000ir}%
\footnote{The analogy is however incomplete,
  because here we take into account photon emission from the
  intermediate charged $W$ boson, while in ref.~\cite{Jadach:2000ir} 
  neutral resonance $Z$ was considered.}.
The transition to EEX of \yfsww3 requires
a few additional steps described in the next subsection.

As a starting point we take 
an approximate variant of CEEX of eq.~(\ref{eq:ceex1-approx}),
which is obtained from the exact one of eq.~(\ref{eq:ceex1b-resonanat})
by means of neglecting some non-IR interference NFI terms:
%///////////////////////////////////////////////////
%     sigma
%///////////////////////////////////////////////////
\begin{equation}
  \begin{aligned}
 &\sigma = \frac{1}{flux}\; \sum_{n=0}^\infty \frac{1}{n!}
  \int dLips_{4+n}(p_a+p_b;p_c,p_d,p_e,p_f,k_1...k_n)
\\&\times
    \!\!\! \sum_{\wp\in\{P,D_1,D_2\}^n} \!\!\!\!\!
     e^{\alpha B_{10}(U_\wp,V_\wp)}
     \Bigg\{ 
     \hbeta^{(1)}_0\left(U_\wp,V_\wp\right)
     \prod_{i=1}^n \; j_{\{\wp_i\}}^{\mu_i}(k_i)\; 
     +\sum_{j=1}^n 
     \hbeta^{(1)\mu_j}_{1\{\wp_j\}}\left(U_\wp,V_\wp,k_j\right)
     \prod_{i\neq j}\; j_{\{\wp_i\}}^{\mu_i}(k_i)\;
    \Bigg\}
\\&\times
     \!\!\!\sum_{\wp'\in\{P,D_1,D_2\}^n} \!\!\!\!\!
     e^{\alpha B_{10}^*(U_{\wp'},V_{\wp'})}
     \Bigg\{ 
     \hbeta^{(1)}_0\left(U_{\wp'},V_{\wp'}\right)
     \prod_{i=1}^n \; j_{\{\wp'_i\}}^{\mu_i}(k_i)\; 
     +\sum_{j=1}^n 
     \hbeta^{(1)\mu_j}_{1\{\wp'_j\}}\left(U_{\wp'},V_{\wp'},k_j\right)
     \prod_{i\neq j}\; j_{\{\wp'_i\}}^{\mu_i}(k_i)\;
    \Bigg\}^*,
  \end{aligned}
\end{equation}
%%------------
where $U_\wp=p_{cd}+\sum_{\wp_i=D_1} k_{i}$ 
  and $V_\wp=p_{ef}+\sum_{\wp_i=D_2} k_{i}$.

The consistent method of omitting all of the remaining QED NFI interferences
between the production and two decays requires
omitting from the double sum over 
photon assignments all non-diagonal terms, $\wp\neq\wp'$,
and the interference terms in $B_{10}$.
After doing that the above omission the sum over photons 
can be reorganised into a product of three separate
sums, one for the production and two for the decays.
In this way we get the following EEX expression:
%///////////////////////////////////////////////////
\begin{equation}
  \label{eq:YFS3raw}
  \begin{aligned}
 &\sigma 
 = \frac{1}{flux}\; \sum_{n=0}^\infty\; \frac{1}{n!}\; 
   \int dLips_{4+n}(p_a+p_b;p_c,p_d,p_e,p_f,k_{1}...k_{n})
\\&\times
   \!\!\sum_{\wp\in\{P,D_1,D_2\}^n} 
    e^{2\alpha \Re B_{PDD}(U_\wp,V_\wp)}
     \prod_{i=1}^n \; |j_{\{\wp_i\}}^{\mu_i}(k_i)|^2\; 
     \Bigg\{ 
     |\hbeta^{(1)}_0\left(U_\wp,V_\wp \right)|^2
\\&+
\sum_{j=1}^{n} \left(  
                 2\Re \big(  \hbeta^{(1)}_{1\{\wp_j\}}(U_\wp,V_\wp,k_j)\cdot j_{\{\wp_j\}}(k_{j})^* \big)
                 + |\hbeta^{(1)}_{1\{\wp_j\}}(U_\wp,V_\wp,k_j)|^2 
                 \right)
                 |j_{\{\wp_j\}}(k_{j})|^{-2}\;
    \Bigg\}.
  \end{aligned}
\end{equation}
%%------------
In the above expression the YFS form-factor $e^{2\alpha \Re B_{10}}$
factorises into the product of independent form-factors 
for the production and two decay processes: 
\begin{equation}
  \label{eq:ReBPDD}
e^{2\alpha \Re B_{PDD}}=e^{2\alpha \Re B_P} e^{2\alpha \Re B_{D_1}} e^{2\alpha \Re B_{D_2}}.
\end{equation}
Eq.~(\ref{eq:YFS3raw}) can be rewritten in a more traditional
EEX notation as follows:
%///////////////////////////////////////////////////
\begin{equation}
  \label{eq:YFS3trad}
  \begin{aligned}
 &\sigma 
 = \frac{1}{flux}\; \sum_{n=0}^\infty\; \frac{1}{n!}\; 
   \int dLips_{4+n}(p_a+p_b;p_c,p_d,p_e,p_f,k_{1}...k_{n})
   \!\!\sum_{\wp\in\{P,D_1,D_2\}^n} 
\\&
    e^{2\alpha \Re B_{PDD}(U_\wp,V_\wp)}
     \prod_{i=1}^n \; \tilde{S}_{\{\wp_i\}}(k_i)\; 
     \Bigg\{ 
      \bbeta^{(1)}_0\left(U_\wp,V_\wp \right)
+\sum_{j=1}^{n}  \bbeta^{(1)}_{1\{\wp_j\}}(U_\wp,V_\wp,k_j)
                 \big[\tilde{S}_{\{\wp_j\}}(k_{j})\big]^{-1}\;
    \Bigg\},
  \end{aligned}
\end{equation}
%%------------
where
\begin{equation}
  \tilde{S}_X(k)= |j_X^\mu(k)|^2,\; X=P,D_1,D_2.
\end{equation}
Note that in the above expression for each photon assignment
we perfectly know the four momentum in each $W$ propagator 
-- simply because each photon is associated with the production or one of the two decays.

In fact eq.~(\ref{eq:YFS3raw}) looks like three separate EEX exponentiation schemes
for the three subprocesses.
They talk to each other only through total energy conservation and spin correlations%
\footnote{Connecting the production and the decays through the spin-density matrix formalism
   is the logical solution in the EEX case, as for the $\tau$-pair production and decay
   in \koralz.}.
This can be seen manifestly  even more clearly when,
for the purpose of the MC implementation, eq.~(\ref{eq:YFS3trad}) is transformed
into the following form in which the $\tilde{S}$-factors
for the production and the decays are factorised.
For $n$ photons in the overall sum over $3^n$ assignments of the photons $\{P,D_1,D_2\}^n$
there are groups (partitions) of $\frac{n!}{n_0!n_1!n_2!}$ choices,
with $n_0$ photons in the production, $n_1$ photons in the first decay and $n_2$
photons in the second decay, $n_0+n_1+n_2=n$.
The assignments in each partition are related by the permutation of the photons
within the partition.
We may replace in eq.~(\ref{eq:YFS3trad}) the whole such a partition
just by one permutation member, getting the following expression:
%///////////////////////////////////////////////////
\begin{equation}
  \label{eq:YFS3mc}
  \begin{aligned}
 &\sigma 
 = \sum_{n_0=0}^\infty \sum_{n_1=0}^\infty \sum_{n_2=0}^\infty
   \int dLips_{4+n_0+n_1+n_2}(p_a+p_b;p_c,p_d,p_e,p_f,k_{1}...k_{n_2})
\\&\times
   \frac{1}{n_0!} \prod_{i_1=0}^{n_0} \; \tilde{S}_{P}(k_{i_0})\; 
   \frac{1}{n_1!} \prod_{i_1=1}^{n_1} \; \tilde{S}_{D_1}(k_{i_1})\; 
   \frac{1}{n_2!} \prod_{i_2=1}^{n_2} \; \tilde{S}_{D_2}(k_{i_2})\; 
\\&\times
     e^{2\alpha \Re B_{PDD}(U_1, V_2)}
     \Bigg\{ 
     \bbeta^{(1)}_0\left(U_1, V_2 \right)
    +\sum_{j=1}^{n_0} \bbeta^{(1)}_{1\{P\}}(U_1,V_2,k_j)\;   \tilde{S}_{P}(k_{j})^{-1}\;
\\&
    +\sum_{j=1}^{n_1} \bbeta^{(1)}_{1\{D_1\}}(U_1,V_2,k_j)\; \tilde{S}_{D_1}(k_{j})^{-1}
    +\sum_{j=1}^{n_2} \bbeta^{(1)}_{1\{D_2\}}(U_1,V_2,k_j)\; \tilde{S}_{D_2}(k_{j})^{-1}
    \Bigg\},
  \end{aligned}
\end{equation}
where $U_1=p_{cd}+\sum_{i_1=0}^{n_1} k_{i_1}$ and $V_2=p_{ef}+\sum_{i_2=0}^{n_2} k_{i_2}$.
One can always come back to the configuration of eq.~(\ref{eq:YFS3trad})
by means of symmetrisation over photons.
In MC computations, the sum over photons is ``randomised'' in a natural way and only one 
partition member is generated at a time,
(using effectively eq.~(\ref{eq:YFS3mc}))
so the fact that the basic distribution
for EEX$_{R}$ is that of eq.~(\ref{eq:YFS3trad}) can be easily overlooked,
see also discussion in \cite{Jadach:1999vf}.

From the above algebra we see in a detail how EEX$_R$ can be
embedded in a natural way in the full CEEX$_R$,  defined in the previous section.

%%%%%%%%%%%%%%%%%%%%%%%%%%%%%%%%%%%%%%%%%%%%%%%%%%%%%%%%%%%%%%%%%%%%%%%%%%%
\subsection{Last step towards EEX$_R$ of \yfsww3}

The EEX of eq.~(\ref{eq:YFS3mc}) is not exactly that of EEX of \yfsww3\
and \kandy, as described in refs.~\cite{Jadach:2001uu,Jadach:2001mp}.
Let us discuss the remaining differences.
The most important difference is that QED matrix element for the $W$-boson decay in \yfsww3\
is implemented using the \photos\ program whose 
has matrix element is not in the EEX scheme, although very close to it.
At the precision of the LEP experiments this was the acceptable and economic solution.
There would be no problems with replacing \photos\ with the true EEX
implementation for the $W$ decays because such an implementation is
already available in the \winhac\ program developed for the single-$W$ production
at hadron colliders~\cite{Placzek:2003zg}.

The implementation of the EEX matrix elements for the production process in \yfsww3\
is described in fine detail in ref.~\cite{Jadach:2001uu}.
It is based on the {\tt YFS3} event generator~\cite{Jadach:1991dm}
for the $e^+e^-\to 2f$ process in which the final-state massive fermions are replaced with $W$'s.
The  {\tt YFS3} program does not include the QED initial-final state
interferences (IFI) between initial $e^\pm$ and final particles. 
Such interferences (present in EEX of eq.~(\ref{eq:YFS3mc})
were also added in \yfsww3\ using the reweighting technique of the
\bhwide\ program~\cite{Jadach:1995nk}.

%%%%%%%%%%%%%%%%%%%%%%%%%%%%%%%%%%%%%%%%%%%%%%%%%%%%%%%%%%%%%%%%%%%%%%%%%%%
\subsection{From EEX$_R$ to CEEX$_R$ in MC implementation}

The upgrade from EEX of eq.~(\ref{eq:YFS3mc}) to CEEX 
in the MC implementation is feasible and well defined.
In the Monte Carlo program implementing EEX, one usually generates MC events
according to some baseline distribution%
\footnote{The baseline distribution has to include all the soft and collinear
  singularities of the EEX distributions.}
and the final correcting weight introduces fine details of the EEX matrix element.
The CEEX matrix elements can be implemented by reweighting
events generated according to the same baseline distributions as in the EEX case,
just by replacing the EEX final MC correcting weight with that of CEEX,
without any changes in the baseline MC.
This kind of flexible and economic solution was already applied
in the \kkmc\ program~\cite{Jadach:1999vf}.
Similarly as in \kkmc,
the MC weight correcting from EEX to CEEX will be not bound from the above.
There are several solutions for this purely technical problem.

%%%%%%%%%%%%%%%%%%%%%%%%%%%%%%%%%%%%%%%%%%%%%%%%%%%%%%%%%%%%%%%%%%%%%%%%%%%%%%%%%
\subsection{Photon distributions around  $E_\gamma\sim\Gamma$}

Let us finally comments on two apparent deficiencies of the EEX$_{R}$ scheme:
\begin{itemize}
\item lack of transmutation of photon distributions around  $E_\gamma\sim\Gamma$,
\item excess of photon-multiplicity for very soft photons, $E_\gamma\le\Gamma$.
\end{itemize}

The phenomenon of ``transmutation of photon distributions'' occurs when photon energy changes
from the ``semisoft region'' $\Gamma < E_\gamma\ll E_{beam}$ down to ``true soft region''
$E_\gamma < \Gamma$.
In the true-soft region photon distributions do not reflect the existence of the the single
charged object, the resonance -- they reflect, instead, momenta and charges of all its decay products.
For these long range photons, the resonance itself just lives too shortly to be ``felt''.
On the other hand, the semi-soft photons with shorter wavelength can see the resonance
as a distinct object -- its presence is imprinted in the distributions of photon energy and angles.
In fact, it is the interference between the production-current $j^\mu_{P}$
and the decay-current $j^\mu_{D1,D2}$ which enforces the transition in the photon distributions.
This effect can be also seen explicitly in
the instrumental identity of eq.~(\ref{eq:ident1phot}),
or in the explicit one-photon emission amplitude of eq.~(\ref{eq:real-single}).
The absence of this interference in EEX, where all the NFI interferences are neglected,
causes that in EEX (of \yfsww3)
the above beautiful transmutation phenomenon cannot be present%
\footnote{The transition between these two situations is modeled in our new CEEX
  in a completely realistic way. It is continuous in the photon energy.}.

The lack of the above interferences causes also certain unphysical effect for very soft photons. 
As we know, in the real world (and in CEEX) there is no IR singularity
(neither real nor virtual) for the photon emission from the internal $W$ line, see eq.~(\ref{eq:real-single}),
while in EEX there is such (real and virtual), 
as seen explicitly in eq.~(\ref{eq:YFS3mc}). 
How to explain this paradox? Is this something dangerous?
The artificial IR divergence in EEX is not dangerous as long as we are
at the \Order{\frac{\alpha}{\pi}\frac{\Gamma}{M}} precision level for the distributions
which are inclusive enough, such that we do not examine multiplicities and angular spectra
of the photons  with $E_\gamma < \Gamma$.
Extra unphysical photons in this energy range do not contribute to integrated cross section,
because their contribution is countered immediately by the virtual form-factor.
They will however affect multiplicity of such very soft photons.

The good agreement of the soft photon spectra between \yfsww3 and \racoon\ confirms
that the effect is not sizeable.
The numerical estimates of  ref.~\cite{Fadin:1993dz} also suggest
that this effect is small, negligible for LEP2.
On the other hand, in the future high-statistics experiments it is worth
to examine the above effects for the photons with $E_\gamma \sim \Gamma_W$.
It was proposed in ref.~\cite{Fadin:1993dz} that it may even provide 
an independent relatively precise measurement of $\Gamma_W$.

Summarising, the presence of the extra unphysical soft photons with $E_\gamma < \Gamma$
in EEX (and its version implemented in \yfsww3) due to setting to zero all QED
interference effects between the production and decay processes
is not harmful at the precision level of \Order{\frac{\alpha}{\pi}\frac{\Gamma}{M}}.
For the higher-precision requirements, like that in FCC-ee,
one should go back to CEEX$_R$,
from which EEX$_{R}$ is derived, 
and get back for $E_\gamma \sim \Gamma$ fully exclusive realistic photon distributions.

%%%%%%%%%%%%%%%%%%%%%%%%%%%%%%%%%%%%%%%%%%%%%%%%%%%%%%%%%%%%%%%%%%%%%%%%%%%%%
\section{Summary and outlook}

In the present paper we have proposed a solution to the long-standing problem
of the systematic treatment of the soft and hard photon emission from
the unstable charged particles and the interferences between production
and decay parts of the process,
at any perturbative order.
This is of practical importance for high-precision measurements of
$W^+W^-$-pair production at electron--positron colliders, such as FCC-ee/ILC/CLIC, 
and for single-$W$ production at hadron colliders, such as LHC/FCC-hh, 
as well as in many other processes with 
production and decay of charged unstable particles of any spin.
So far there is no practical implementation of the full-scale
calculation in the proposed scheme.
However, it has been outlined how to accomplish it in the framework of some existing Monte Carlo (MC) event generators.

Our study has been focused on the process $e^+e^- \to W^+W^- \to 4f$ which is to be used e.g.\ 
for the high-precision $W$-boson mass and width measurements in the planned electron--positron colliders, particularly FCC-ee.
We have argued that the most economical (and perhaps the only feasible) way to achieve the required accuracy of
theoretical prediction for this process it to apply the so-called pole expansion (POE) to the general process of $e^+e^- \to 4f$,
and then to calculate the electroweak (EW) radiative corrections separately for 
each term of such an expansion to an appropriate order in
the coupling constant $\alpha$.
More specifically, for the leading term in POE, i.e.\ the so-called double-pole contribution 
which comprises two resonant $W$-bosons, one would need to include the fixed-order EW corrections 
up to \order{\alpha^2} for the on-shell-like $W$-pair production and $W$ decay processes,
while for the non-leading terms, i.e.\ the single-pole and non-pole contributions, the EW corrections at \order{\alpha^1} would be sufficient.
The calculations of the \order{\alpha^1} EW corrections are already available for the whole $e^+e^- \to 4f$ process,
while the \order{\alpha^2} ones for the double-resonant contribution do not exist yet, however they are feasible, in our opinion,
 by the time of the planned FCC-ee physics run.

In addition to the above fixed-order radiative corrections, in order to reach 
the requisite theoretical precision for the above process, 
one needs to include higher-order QED corrections corresponding to multiphoton emission from the initial- and final-state leptons 
as well as from the intermediate $W$-bosons.
We have argued that the best framework in which all this can be accomplished is
the so-called coherent exclusive exponentiation (CEEX) scheme.
Its main advantage over the traditional YFS exclusive exponentiation (EEX) method 
is that it operates directly at the level of spin amplitudes. 
Because of that, all multiphoton effects related to radiation from the resonant $W$-bosons
and to non-factorisable interferences can be accounted for in a straightforward way.
So far, the CEEX methodology was applied to  $e^+e^- \to 2f$  in the \kkmc\ event generator
and proved to be crucial in providing  precision theoretical predictions for this process necessary for the LEP experiments. 

We have provided the respective general cross-section formulae for the double-pole, single-pole and non-pole contributions
to the charged-current $e^+e^- \to 4f$ process which can be a basis for an appropriate MC implementation. 
An important ingredient in that is resummation of real-photon emissions including radiation from the intermediate $W$-bosons
and derivation of the corresponding virtual-photon form-factor, done explicitly in Appendices A, B and C. 
Our approach exploited the similarity between the virtual- and real-emission QED form-factors 
guaranteed by the infra-red cancellations. 

We have also discussed the relation of the above CEEX realisation to the existing EEX implementation
in terms of the hybrid MC program called \kandy, being the combination of two MC event generators: \koralw\ and \yfsww3.
In this implementation, the \order{\alpha^3} YFS exponentiation for initial-state radiation in the process $e^+e^- \to 4f$
was combined with the fixed-order \order{\alpha^1} EW corrections in the $W$-pair production and multiphoton
radiation in the $W$-decays generated by the \photos\ program,  while all the non-factorisable interferences were neglected.
Such a solution proved to be good enough for the LEP2 accuracy, but for the expected precision of the FCC-ee
experiments it will not suffice. 
However, it can constitute a good starting point and a MC platform for development and implementation of the CEEX scheme
described in this paper. In parallel, one can also develop an EEX approximation of the full-scale CEEX solution which will be
important for its numerical cross-checks. For this, the implementation of EEX for the $W$-boson decays
in the \winhac\ program can be used to replace the corresponding \photos\ radiation in \kandy.

%We have not discussed the problem of the Coulomb correction in the process of 
%unstable $W$-pair production.
%In our opinion, at the beginning it can be taken into account in the CEEX 
%double-pole contribution 
%in a similar way as it was done in \yfsww3 and then further refinements can be 
%added, if necessary.
%We have also skipped the issues related to the definition of 
%the mass and width of the resonance or to the UV renormalisation, 
%but this is beyond the scope of this paper.

%%%%%%%%%%%%%%%%%%%%%%%%%%%%%%%%%%%%%%%%%%%%%%%%%%%%%%%%%%%%%%%%%%%%%%%%%%%%%
%%%%%%%%%%%%%%%%%%%%%%%%%%%%%%%%%%%%%%%%%%%%%%%%%%%%%%%%%%%%%%%%%%%%%%%%%%%%%
%\vspace{3mm}
\section*{Acknowledgments}

Useful discussions with B.F.L~Ward and Z.~W\c{a}s are acknowledged.

\appendix
\newpage
\vspace{3mm}
\noindent{\bf\LARGE Appendices}
%%%%%%%%%%%%%%%%%%%%%%%%%%%%%%%%%%%%%%%%%%%%%%%%%%%%%%%%%%%%%%%%%%%%%%%%%%%
\section{Factoring photon-emission from $W$}
\label{appendixA}

The following considerations are valid for a charged unstable particle of any 
spin, 
eg. $W^\pm$, $\tau^\pm$ or $t$-quark.
Let us start with a simple identity for two propagators related to
single photon emission from an internal charged particle line
\begin{equation}
  \frac{1}{(Q_0^2-M^2)(Q_1^2-M^2)} 
  = \frac{1}{(Q_0^2-Q_1^2)(Q_1^2-M^2)}-\frac{1}{(Q_0^2-Q_1^2)(Q_0^2-M^2)}
\end{equation}
where $M^2=M_W^2+iM_W\Gamma_W$.

\begin{figure}[!ht]
\centering
\setlength{\unitlength}{0.1mm}
\includegraphics[height=15mm,width=60mm]{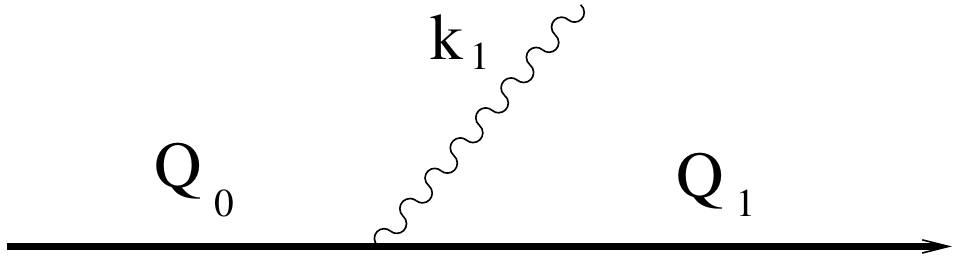} % was 40 mm
\caption{\small\sf Single emission from the internal W line.}
\label{fig:Wgam1}
\end{figure}

The kinematics is depicted in fig~\ref{fig:Wgam1}.
%/////////////////////////////////////////////////////////////////////////////////////////
%-----------------------------------------------------------------------------------------
Noticing that $Q_0^2-Q_1^2 =2k_1Q_0-k_1^2=2k_1Q_1+k_1^2$,
we may rewrite the above as follows:
\begin{equation}
  \label{eq:ident1phot}
  \frac{1}{(Q_0^2-M^2)(Q_1^2-M^2)} 
  = \frac{1}{(2k_1Q_1+k_1^2)(Q_1^2-M^2)}+\frac{1}{(-2k_1Q_0+k_1^2)(Q_0^2-M^2)}.
\end{equation}
The reader will recognise the first term as representing a photon (eikonal) emission factor
in the {\em production}  part of the process times a resonance propagator 
(with the reduced four momentum $Q_1=Q_0-k_1$)
and the second term as the analogous emission factor
in the {\em decay} process times the resonance propagator 
(with the four-momentum $Q_0=Q_1+k_1$).
Each of the two terms look IR-divergent, however
the two IR divergences cancel -- the difference is finite.
In the original expression it was the resonance width $\Gamma_W$
which was providing an infrared regulator for a photon with the momentum $k_1=Q_1-Q_2$.

%/////////////////////////////////////////////////////////////////////////////////////////
\begin{figure}[!ht]
\centering
\setlength{\unitlength}{0.1mm}
\includegraphics[height=15mm,width=140mm]{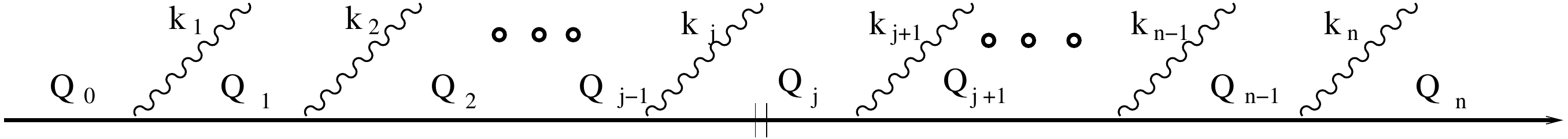} % was 40 mm
\caption{\small\sf Multiple emission from the internal W line.}
\label{fig:WgamM}
\end{figure}
Let us now consider the general case of the $n$-photon emissions from the internal 
charged particle line, depicted in fig.~\ref{fig:WgamM}, in the soft-photon approximation.
The reorganisation of the product of the propagators starts with the following identity:
%(partial fraction with respect to the $M^2$)
%-----------------------------------------------------------------------------------------
\begin{equation}
  \label{eq:intermediate}
  \begin{aligned}
  &\frac{1}{(Q_0^2-M^2)(Q_1^2-M^2)\dots(Q_n^2-M^2) }\\
  & \qquad=\sum_{j=0}^{n}  
    \frac{1}{ \prod_{i=0}^{j-1} (Q_i^2-Q_j^2) \; \;(Q_j^2-M^2) \prod_{i=1}^{n-j}(Q_{j+i}^2-Q_j^2)  }.
  \end{aligned}
%\label{eq:identQ}
\end{equation}
%-----------------------------------------------------------------------------------------
It can be proven using the mathematical induction method.
Assuming that the identity is true for $n$, let us prove it for $n+1$.
Using a short-hand notation $y_i= Q_i^2-M^2$, one obtains%
\footnote{ The identity $\sum_{j=0}^n \prod_{i=0,i\neq j}^n \frac{1}{x_i-x_j}=0$ 
           is used in the last step.}
\begin{equation}
\begin{split}
&\prod\limits_{i=0}^{n+1} \frac{1}{y_i}
=  \frac{1}{ y_{n+1}}\; \sum_{j=0}^{n}  \frac{1}{y_j }\;
 \prod\limits_{i\neq j}^n \frac{1}{(y_i-y_j)}\;
=\sum_{j=0}^{n}\;  \frac{1}{(y_{n+1}-y_j) }
 \Big( \frac{1}{y_j } - \frac{1}{y_{n+1} }\Big)
 \prod\limits_{i\neq j}^n \frac{1}{(y_i-y_j)}
%\\&
%=\sum_{j=0}^{n}\; \frac{1}{y_j }   \frac{1}{(y_{n+1}-y_j) }
% \prod\limits_{i\neq j}^n \frac{1}{(y_i-y_j)}
%-\sum_{j=0}^{n}\; \frac{1}{y_{n+1}} \frac{1}{(y_{n+1}-y_j) }
% \prod\limits_{i\neq j}^n \frac{1}{(y_i-y_j)}
\\&
=\sum_{j=0}^{n}\; \frac{1}{y_j } 
 \prod\limits_{i\neq j}^{n+1} \frac{1}{(y_i-y_j)}
- \frac{1}{y_{n+1}} \sum_{j=0}^{n}\;
 \prod\limits_{i\neq j}^{n+1} \frac{1}{(y_i-y_j)}
\\&
=\sum_{j=0}^{n+1}\; \frac{1}{y_j } 
 \prod\limits_{i\neq j}^{n+1} \frac{1}{(y_i-y_j)}
-\frac{1}{y_{n+1} } 
 \prod\limits_{i\neq n+1}^{n+1} \frac{1}{(y_i-y_{n+1})}
- \frac{1}{y_{n+1}} \sum_{j=0}^{n}\;
 \prod\limits_{i\neq j}^{n+1} \frac{1}{(y_i-y_j)}
\\&
=\sum_{j=0}^{n+1}\; \frac{1}{y_j } 
 \prod\limits_{i\neq j}^{n+1} \frac{1}{(y_i-y_j)}
- \frac{1}{y_{n+1}} \sum_{j=0}^{n+1}\;
 \prod\limits_{i\neq j}^{n+1} \frac{1}{(y_i-y_j)}
=\sum_{j=0}^{n+1}\; \frac{1}{y_j } 
 \prod\limits_{i\neq j}^{n+1} \frac{1}{(y_i-y_j)}.
\end{split}
\end{equation}
Alternatively, one can prove it with the help of partial fractioning with
respect to $M^2$:
\begin{equation}
  \label{eq:partfract}
  \begin{aligned}
  &\frac{1}{(Q_0^2-M^2)(Q_1^2-M^2)\dots(Q_n^2-M^2) } 
     =\sum_{j=0}^{n} \frac{A_j}{Q_j^2-M^2}.
  \end{aligned}
%\label{eq:identQ}
\end{equation}
Multiplying eq.~(\ref{eq:partfract}) in a standard way by $Q_j^2-M^2$ and
substituting $Q_j^2=M^2$ we obtain
\begin{equation}
  \label{eq:partfract2}
  \begin{aligned}
  A_j & =
    \frac{1}{ \prod_{i=0}^{j-1} (Q_i^2-Q_j^2)
\prod_{i=1}^{n-j}(Q_{j+i}^2-Q_j^2)  }.
  \end{aligned}
%\label{eq:identQ}
\end{equation}

Let us now examine the soft-photon limit in eq.\ref{eq:intermediate}.
Taking the $j$-th term, we may identify
\begin{equation}
  \label{eq:appr1}
  \begin{aligned}
  Q_0^2-Q_j^2     &\simeq(2k_jQ_j+k_j^2)+\dots+(2k_2Q_j+k_2^2) +(2k_1Q_j+k_1^2),\\
  Q_1^2-Q_j^2     &\simeq(2k_jQ_j+k_j^2)+\dots+(2k_2Q_j+k_2^2), \\
  Q_{j-1}^2-Q_j^2 &=     (2k_jQ_j+k_j^2).
  \end{aligned}
\end{equation}
and 
\begin{equation}
  \label{eq:appr2}
  \begin{aligned}
  Q_{j+1}^2-Q_j^2 & =     (-2k_{j+1}Q_j+k_{j+1}^2)\\
  Q_{n-1}^2-Q_j^2 & \simeq(-2k_{j+1}Q_j+k_{j+1}^2)+\dots+(-2k_{n-1}Q_j+k_{n-1}^2) \\
  Q_{n}^2  -Q_j^2 & \simeq(-2k_{j+1}Q_j+k_{j+1}^2)+\dots+(-2k_{n-1}Q_j+k_{n-1}^2) +(-2k_{n}Q_j+k_{n}^2)
  \end{aligned}
\end{equation}
In the above equations we have neglected the subleading products $k_ik_j$. 
This is allowed in the soft-photon approximation.
On the other hand, terms $k_i^2$
could also be omitted in the soft-photon approximation,
but they are kept because they render virtual photon integrals UV-finite.

In the next step we perform the usual sum over permutation over all photons. 
This will lead to a ``Poissonian'' emission formula,
separately for the resonance production and decay stages of the entire process,
with the explicit sum over the assignments of photons
to the production, denoted by the index $P$, and to the decay, denoted by the index $D$.
We start from eq.~(\ref{eq:intermediate}) switching to a more compact notation:
%-----------------------------------------------------------------------------------------
\begin{equation}
  \begin{aligned}
    R(Q^2_i)&= Q_i^2-M^2,\\
    N^{P}_i &= 2k_{i}Q_j+k_{i}^2=(Q_j+k_{i})^2-Q_j^2,\\
    N^{D}_i &=-2k_{i}Q_j+k_{i}^2=(Q_j-k_{i})^2-Q_j^2.
  \end{aligned}
\end{equation}
Inserting the relations of eqs.~(\ref{eq:appr1}) and (\ref{eq:appr2})
into eq.~(\ref{eq:intermediate})
and summing over permutations we obtain
%-----------------------------------------------------------------------------------------
\begin{equation}
  \label{eq:identity}
  \begin{aligned}
  &\sum_{permut.} \frac{1}{R(Q_0^2)R(Q_1^2)\dots R(Q_n^2) }\\
  &=\sum_{permut.}
    \sum_{j=0}^{n} \Bigl[  
    \frac{1}{N^{P}_1+N^{P}_2+N^{P}_3+\dots N^{P}_j}\;
    \frac{1}{N^{P}_2+N^{P}_3+\dots N^{P}_j} \dots \frac{1}{N^{P}_j}\Bigr]\\
   &\qquad\qquad\qquad
    \times
    \frac{1}{R(Q^2_j)} \times \Bigl[
    \frac{1}{N^{D}_{j+1}}\;
    \frac{1}{N^{D}_{j+1}+N^{D}_{j+2}} \dots 
    \frac{1}{N^{D}_{j+1}+N^{D}_{j+2}\dots N^{D}_{n}} \Bigr],\\
  \end{aligned}
\end{equation}
where for $j=0$ and $j=n$, respectively, the term in the first/second square-bracket pair should read as $1$.
Next, for each $j$-th term we split the sum over all permutations
of $(1,2,3,\dots,n)$ into two separate sums: 
one over permutations of $(1,2,3,\dots,j)$
and another over permutations of $(j+1,j+2,\dots,n)$.
These two sums are performed%
\footnote{ \label{foot:23}
  Here we use twice the well-known identity
  $\sum\limits_{perm.} 
      \frac{1}{a_1(a_1+a_2)(a_1+a_2+a_3)\dots (a_1+a_2+\dots a_n)}
                  =\frac{1}{a_1a_2\dots a_n}$, 
where the sum is over all permutations of $(1,2,3,\dots ,n)$.}.
The sum over $\big( {n \atop j} \big)$ assignments of photons to
production and decay remains.
Alternatively, the entire remaining sum can be represented as a sum over
$\sum_j \big( {n \atop j} \big) = (1+1)^n= 2^n$ terms 
(photon assignments) as follows
%-----------------------------------------------------------------------------------------
\begin{equation}
  \label{eq:principal-ident}
\begin{aligned}
   &\sum_{permut.} \frac{1}{R(Q_0^2)R(Q_1^2)\dots R(Q_n^2) }\\
   &\qquad\qquad =
    \sum_{\wp=(P,D)^n} 
    \prod_{\wp_i=P} \frac{1}{(Q_\wp +k_{i})^2 -Q^2_\wp}\times
    \frac{1}{R(Q^2_\wp)}\times
    \prod_{\wp_i=D} \frac{1}{(Q_\wp -k_{i})^2 -Q^2_\wp},
\end{aligned}
\end{equation}
where
\begin{equation}
  Q\wp = Q_0 - \sum_{\wp_i=P} k_i= Q_n + \sum_{\wp_i=D} k_i.
\end{equation}
The vectors $\wp=(\wp_1,\wp_2, \dots \wp_n)$ of the photon assignments whose
components have values equal to $P$ or $D$, while $\sum_{\wp_i=P}$ 
($\prod_{\wp_i=P}$)
denotes the sum over (product of) all $i$ for which $\wp_i=P$, i.e.
all photon which belong to the production stage of the process.

Main features of eq.~(\ref{eq:principal-ident}),
the principal result of this Appendix, are the following:
\begin{itemize}
\item
Its left-hand side represents ``raw'' Feynman diagrams for multiple-photon 
emission from the charged-particle internal line.
\item
Its right-hand side includes two photon emission factors: one for
the production part of the process (resonance formation)
and the second one for the decay part of the process (resonance decay).
\item
It includes the single-resonance propagator of the standard form, with the complex mass $M$,
and the four-momentum $Q_\wp$, which comprises momenta of all photons assigned
to the resonance decay.
\item
It is rather striking that all photon-emission factors look as
if photons were emitted by the charged particle of the mass $Q_\wp^2$!
This is, of course, intuitively well justified and quite appealing.
\item
The fact that the {\em coherent} sum is performed over all the photon assignments
to the production and the decay reflects the QED gauge invariance and the Bose--Einstein statistics.
\item
It holds both for the virtual and real photons (this is why we have kept $k_i^2$).
\end{itemize}

%%%%%%%%%%%%%%%%%%%%%%%%%%%%%%%%%%%%%%%%%%%%%%%%%%%%%%%%%%%%%%%%%%%%%%%%%%%
%\newpage
\section{Resummation of real-photon emissions}
\label{app:b}
%%%%%%%%%%%%%%%%%%%%%%%%%%%%%%%%%%%%%%%%%%%%%%%%%%%%%%%%%%%%%%%%%%%%%%%%%%%
In this Appendix we show how to do the resummation of the amplitude of the
multiple-real-photon emission.
We expect that because of IR cancellations the basic algebraic structure
of our derivation holds for the integrands of multiloop corrections.

Let us begin with a short summary of the YFS method performed
in a combinatorial way. The process under consideration is 
$$e(p_a)\bar\nu_e(p_b)\longrightarrow W \longrightarrow \mu(p_c)\bar\nu_\mu(p_d).$$
At first, we consider
the standard YFS scheme without radiation nor recoil from $W$. 
As proven by YFS \cite{Yennie:1961ad}, the 
IR radiation comes entirely from the charged external legs ($e$ and $\mu$) and 
has a form of soft currents. The sum of graphs with $N$ real emissions is the 
following:
\begin{equation}
  \label{eq:multi-real1}
  \begin{aligned}
   &\!\!\!\!\!\!{\Meu^{(0)}_N}^{\mu_1,\dots,\mu_N}(k_1,\dots,k_N) \simeq
  \\ \simeq &
\sum_{l=0}^N \sum_{\pi}^{N!}
  \biggl(\frac{2p_a^{\mu_1}}{2p_a k_{\pi_1}}
         \frac{2p_a^{\mu_2}}{2p_a k_{\pi_1}+2p_a k_{\pi_2}}
         \dots
         \frac{2p_a^{\mu_l}}{2p_a k_{\pi_1}+2p_a k_{\pi_2}+\dots+2p_a k_{\pi_l}}
  \biggr)
  \\ & \times 
  \biggl(\frac{-2p_c^{\mu_{l+1}}}{2p_c k_{\pi_{l+1}}}
         \frac{-2p_c^{\mu_{l+2}}}{2p_c k_{\pi_{l+1}}+2p_c k_{\pi_{l+2}}}
         \dots
         \frac{-2p_c^{\mu_{N}}}{2p_c k_{\pi_{l+1}} +2p_c k_{\pi_{l+2}}
               +\dots +2p_ck_{\pi_N}}
  \biggr)
  \\ & \times
  \frac{1}{p_{ab}^2-M^2}.
  \end{aligned}
\end{equation}
We execute now the sum over permutations of photons within the $a$ and $c$
sub-groups according to the formula of footnote \ref{foot:23}. This turns
complicated sums into simple products:
\begin{equation}
  \label{eq:multi-real2}
  \begin{aligned}
   &\!\!\!\!\!\!{\Meu^{(0)}_N}^{\mu_1,\dots,\mu_N}(k_1,\dots,k_N) \simeq
%  \\ \simeq &
  \frac{1}{p_{ab}^2-M^2}
  \sum_{l=0}^N \sum_{\pi/\pi_l/\pi_{N-l}}^{N!/l!/(n-l)!}
  \biggl(\prod_{i=1}^l \frac{2p_a^{\mu_i}}{2p_a k_{\pi_i}}
  \biggr)
  \biggl(\prod_{i=1}^{N-l} \frac{-2p_c^{\mu_{l+i}}}{2p_c k_{\pi_{l+i}}}
  \biggr).
  \end{aligned}
\end{equation}

It takes now a few moments to realise that the combinatorial sum over permutations
can be replaced by the sum over partitions (cf.\ eqs.\
(\ref{eq:identity}) and (\ref{eq:principal-ident}))%
\footnote{
\label{foot:13}
Note that the identity (\ref{eq:sumpart}) generalises to more than two
particles, for example:
\begin{equation}
  \label{eq:sumpart3}
  \sum_{{l_a,l_c,l_e=0}\atop{l_a+l_c+l_e=N}}^N
\sum_{\pi/\pi_a/\pi_c/\pi_e}^{N!/l_a!/l_c!/l_e!}
  =
    \sum_{\wp=(a,c,e)^N}^{3^N}.
\end{equation}
}
\begin{equation}
  \label{eq:sumpart}
  \sum_{l=0}^N \sum_{\pi/\pi_a/\pi_c}^{N!/l!/(N-l)!}
  =
    \sum_{\wp=(a,c)^N}^{2^N}.
\end{equation}
Consequently we get
\begin{equation}
  \label{eq:multi-real3}
  \begin{aligned}
   &\!\!\!\!\!\!{\Meu^{(0)}_N}^{\mu_1,\dots,\mu_N}(k_1,\dots,k_N) \simeq
%  \\ \simeq &
  \frac{1}{p_{ab}^2-M^2}
  \sum_{\wp=(a,c)^N}^{2^N}
  \biggl(
     \prod_{i=1}^N \frac{2\theta_{\wp_i}p_{\wp_i}^{\mu_i}}{2p_{\wp_i}k_{i}}
  \biggr),
  \end{aligned}
\end{equation}
where $\theta$ equals $+1$ for initial state and $-1$ for final state.
Finally, we notice that the sum over partitions in eq.\ (\ref{eq:multi-real3}) can be
rewritten in a compact form as%
\footnote{
For instance, for N=2 we have four partitions:
\begin{equation}
\frac{2p_{a}^{\mu_1}}{2p_{a} k_{1}}
\frac{2p_{a}^{\mu_2}}{2p_{a} k_{2}}
-
\frac{2p_{a}^{\mu_1}}{2p_{a} k_{1}}
\frac{2p_{c}^{\mu_2}}{2p_{c} k_{2}}
-
\frac{2p_{c}^{\mu_1}}{2p_{c} k_{1}}
\frac{2p_{a}^{\mu_2}}{2p_{a} k_{2}}
+
\frac{2p_{c}^{\mu_1}}{2p_{c} k_{1}}
\frac{2p_{c}^{\mu_2}}{2p_{c} k_{2}}
=
  \biggl(
      \frac{2p_{a}^{\mu_1}}{2p_{a} k_{1}}
     -\frac{2p_{c}^{\mu_1}}{2p_{c} k_{1}}
  \biggr)
  \biggl(
      \frac{2p_{a}^{\mu_2}}{2p_{a} k_{2}}
     -\frac{2p_{c}^{\mu_2}}{2p_{c} k_{2}}
  \biggr).
\notag
\end{equation}
}
\begin{equation}
  \label{eq:multi-real4}
  \begin{aligned}
   &\!\!\!\!\!\!{\Meu^{(0)}_N}^{\mu_1,\dots,\mu_N}(k_1,\dots,k_N) \simeq
%  \\ \simeq &
  \frac{1}{p_{ab}^2-M^2}
  \prod_{i=1}^N \biggl(
      \frac{2p_{a}^{\mu_i}}{2p_{a} k_{i}}
     -\frac{2p_{c}^{\mu_i}}{2p_{c} k_{i}}
  \biggr).
  \end{aligned}
\end{equation}

Let us now allow for the radiation from the $W$-boson. We begin by analysing the numerator of
the multiple-emission graph of fig.\ \ref{fig:WgamM}, i.e.\ of 
LHS of eq.\ (\ref{eq:intermediate}). The numerator of the single photon emission
with two accompanying $W$ propagators (in the small-photon-momentum limit)
 looks as follows:
\begin{equation}
  \label{eq:single-phot}
  \begin{aligned}
 & \bigl(-g^{\lambda\lambda'}+p^\lambda p^{\lambda'}/M_W^2\bigr)
   V(p,k,p-k)_{\lambda'\rho\sigma'}
   \bigl(-g^{\sigma'\sigma}+(p-k)^{\sigma'} (p-k)^{\sigma}/M_W^2\bigr)
\\ &
\eqop^{k\to 0}
   \bigl(-g_{\lambda\sigma}+ p_\lambda p_\sigma/M_W^2\bigr) (-2p_\rho) 
   + g_{\lambda\rho} p_\sigma(p^2-M_W^2)
   + g_{\sigma\rho}p_\lambda(p^2-M_W^2),
  \end{aligned}
\end{equation}
where $V(p,k,p-k)_{\lambda'\rho\sigma'}$ is the $W\gamma W$ vertex. Dropping
also the terms proportional to $p^2-M_W^2$ (i.e.\ putting $p$ on-shell) we
obtain a self-repeating structure and the numerator of the whole line becomes
\begin{equation}
  \label{eq:multi-phot}
   \bigl(-g_{\lambda\sigma}+ Q_{j\lambda} Q_{j\sigma}/M_W^2\bigr) 
   \prod_{i=1}^n (-2p_{\mu_i}).
\end{equation}
Now inclusion of the radiation from $W$ into eq.\ (\ref{eq:multi-real1})
amounts to the following modification:
\begin{equation}
  \label{eq:multi-W-real1}
  \begin{aligned}
   &\!\!\!\!\!\!{\Meu^{(0)}_N}^{\mu_1,\dots,\mu_N}(k_1,\dots,k_N) \simeq
  \\ \simeq &
\sum_{{l_a,l_c,n=0}\atop{l_a+l_c+n=N}}^N \sum_{\pi}^{N!}
\biggl[
  \biggl(\frac{2p_a^{\mu_{\pi_1}}}{2p_a k_{\pi_1}}
         \frac{2p_a^{\mu_{\pi_2}}}{2p_a k_{\pi_1}+2p_a k_{\pi_2}}
         \dots
         \frac{2p_a^{\mu_{\pi_{l_a}}}}
              {2p_a k_{\pi_1}+2p_a k_{\pi_2}+\dots+2p_a k_{\pi_{l_a}}}
  \biggr)
  \\ &
  \frac{-g^{\lambda\sigma}+Q_{\pi_0}^\lambda Q_{\pi_0}^{\sigma}/M_W^2}
  {Q_{\pi_0}^2-M_W^2}
  \prod_{i=1}^n
   \frac{(-2Q_{\pi_0}^{\mu_{l_a+i}})}{Q_{\pi_{l_a+i}}^2-M_W^2}
  \\&
  \biggl(\frac{-2p_c^{\mu_{\pi_{l_a+n+1}}}}{2p_c k_{\pi_{l_a+n+1}}}
  \frac{-2p_c^{\mu_{\pi_{l_a+n+2}}}}
       { 2p_c k_{\pi_{l_a+n+1}}+2p_c k_{\pi_{l_a+n+2}}}
         \dots
         \frac{-2p_c^{\mu_{\pi_{l_a+n+l_c}}}}
       {2p_c k_{\pi_{l_a+n+1}} +2p_c k_{\pi_{l_a+n+2}} +\dots +2p_c
                                                         k_{\pi_{l_a+n+l_c}}}
  \biggr)
\biggr],
  \end{aligned}
\end{equation}
where we temporarily choose $Q_0$ as the four-momentum in the numerators. The unmatched
indices $\lambda\sigma$ in the $W$ propagator are to be treated ``symbolically'' as we do not write a complete expression for $\Meu$. At this
moment we plug in the formula (\ref{eq:identity}) 
% and then (\ref{eq:principal-ident})
\begin{equation}
  \label{eq:multi-W-real2}
  \begin{aligned}
   &\!\!\!\!\!\!{\Meu^{(0)}_N}^{\mu_1,\dots,\mu_N}(k_1,\dots,k_N) \simeq
%----------
  \\ \simeq &
\sum_{{l_a,l_c,n=0}\atop{l_a+l_c+n=N}}^N 
\sum_{{l_g,l_h=0}\atop{l_g+l_h=n}}^n \sum_{\pi}^{N!}
\biggl[
  \biggl(\frac{2p_a^{\mu_{\pi_1}}}{2p_a k_{\pi_1}}
         \frac{2p_a^{\mu_{\pi_2}}}{2p_a k_{\pi_1}+2p_a k_{\pi_2}}
         \dots
         \frac{2p_a^{\mu_{\pi_{l_a}}}}
              {2p_a k_{\pi_1}+2p_a k_{\pi_2}+\dots+2p_a k_{\pi_{l_a}}}
  \biggr)
%----------
  \\&
  \biggl(\frac{-2Q_{\pi_0}^{\mu_{\pi_{l_a+1}}}}{2Q_{\pi_{l_g}} k_{\pi_{l_a+1}}}
  \frac{-2Q_{\pi_0}^{\mu_{\pi_{l_a+2}}}}
       {2Q_{\pi_{l_g}} k_{\pi_{l_a+1}}+2Q_{\pi_{l_g}} k_{\pi_{l_a+2}}}
         \dots
         \frac{-2Q_{\pi_0}^{\mu_{\pi_{l_a+l_g}}}}
    {2Q_{\pi_{l_g}} k_{\pi_{l_a+1}} +2Q_{\pi_{l_g}} k_{\pi_{l_a+2}} +\dots
                +2Q_{\pi_{l_g}} k_{\pi_{l_a+l_g}}}
  \biggr)
%----------
  \\ &
  \frac{-g^{\lambda\sigma}+Q_{\pi_{l_g}}^\lambda Q_{\pi_{l_g}}^{\sigma}/M_W^2}
  {Q_{\pi_{l_g}}^2-M_W^2}
%  \prod_{i=1}^n
%   \frac{(-2Q_{\pi_0}^{\mu_{l_a+i}})}{Q_{\pi_{l_a+i}}^2-M_W^2}
%---------
  \\&
  \biggl(
    \frac{2Q_{\pi_0}^{\mu_{\pi_{l_a+l_g+1}}}}
         {2Q_{\pi_{l_g}} k_{\pi_{l_a+l_g+1}}}
  \frac{2Q_{\pi_0}^{\mu_{\pi_{l_a+l_g+2}}}}
      {2Q_{\pi_{l_g}} k_{\pi_{l_a+l_g+1}}+2Q_{\pi_{l_g}} k_{\pi_{l_a+l_g+2}}}
%--------
  \\&
     ~~~~~~~~~~~~~~~~~~~    \dots
         \frac{2Q_{\pi_0}^{\mu_{\pi_{l_a+l_g+l_h}}}}
    {2Q_{\pi_{l_g}} k_{\pi_{l_a+l_g+1}} +2Q_{\pi_{l_g}} k_{\pi_{l_a+l_g+2}} 
       +\dots +2Q_{\pi_{l_g}} k_{\pi_{l_a+l_g+l_h}}}
  \biggr)
%-------
  \\&
  \biggl(\frac{-2p_c^{\mu_{\pi_{l_a+n+1}}}}{2p_c k_{\pi_{l_a+n+1}}}
  \frac{-2p_c^{\mu_{\pi_{l_a+n+2}}}}
       {2p_c k_{\pi_{l_a+n+1}}+2p_c k_{\pi_{l_a+n+2}}}
         \dots
         \frac{-2p_c^{\mu_{\pi_{l_a+n+l_c}}}}
       {2p_c k_{\pi_{l_a+n+1}} +2p_c k_{\pi_{l_a+n+2}} +\dots +2p_c
                                                         k_{\pi_{l_a+n+l_c}}}
  \biggr)
\biggr].
  \end{aligned}
\end{equation}
The double sum can be converted into a single one
\begin{equation}
\sum_{{l_a,l_c,n=0}\atop{l_a+l_c+n=N}}^N 
\sum_{{l_g,l_h=0}\atop{l_g+l_h=n}}^n 
=
\sum_{{l_a,l_c,l_g,l_h=0}\atop{l_a+l_c+l_g+l_h=N}}^N 
\end{equation}
and the four groups of permutations can be executed as in 
eq.\ (\ref{eq:multi-real2}):
\begin{equation}
  \label{eq:multi-W-real3}
  \begin{aligned}
   &{\Meu^{(0)}_N}^{\mu_1,\dots,\mu_N}(k_1,\dots,k_N)
  \simeq
  \sum_{{l_a,l_c,l_g,l_h=0}\atop{l_a+l_c+l_g+l_h=N}}^N 
  \sum_{\pi/\pi_{l_a}/\pi_{l_g}/\pi_{l_h}/\pi_{l_c}}
      ^{N!/l_a!/l_g!/l_h!/l_c!}
%  \\ &
  \frac{-g^{\lambda\sigma}+Q_{\pi_{l_g}}^\lambda Q_{\pi_{l_g}}^{\sigma}/M_W^2}
  {Q_{\pi_{l_g}}^2-M_W^2}
\\ &
  \biggl(\prod_{i=1}^{l_a} \frac{2p_a^{\mu_{\pi_i}}}{2p_a k_{\pi_i}}
  \biggr)
  \biggl(\prod_{i=1}^{l_g} 
     \frac{-2Q_{\pi_0}^{\mu_{\pi_{l_a+i}}}}{2Q_{\pi_{l_g}} k_{\pi_{l_a+i}}}
  \biggr)
  \biggl(\prod_{i=1}^{l_h} 
     \frac{2Q_{\pi_0}^{\mu_{\pi_{l_a+l_g+i}}}}
          {2Q_{\pi_{l_g}} k_{\pi_{l_a+l_g+i}}}
  \biggr)
  \biggl(\prod_{i=1}^{l_c} 
         \frac{-2p_c^{\mu_{\pi_{l_a+l_g+l_h+i}}}}{2p_c k_{\pi_{l_a+l_g+l_h+i}}}
  \biggr).
  \end{aligned}
\end{equation}
The first two terms in the brackets describe the emission from the production
part (from the lines $a$ and $g$). $Q_{\pi_{l_g}}\equiv Q_g$ 
is defined there as 
$Q_g=p_{ab} -K_P,~~K_P=K_a +K_g$ and is the same for all terms of the product 
for given (fixed) $K_P$, regardless of the choice of $K_a$ and 
$K_g$. Therefore, these
two products can be combined into one as in eqs.\  (\ref{eq:multi-real3})
and (\ref{eq:multi-real4}). The same holds for the last two terms which describe
emission from the decays with $Q_g=p_{cd} +K_D,~~K_D=K_c +K_h$:
\begin{equation}
  \label{eq:multi-W-real4}
  \begin{aligned}
   &{\Meu^{(0)}_N}^{\mu_1,\dots,\mu_N}(k_1,\dots,k_N)
  \simeq
  \sum_{{l_P,l_D=0}\atop{l_P+l_D=N}}^N 
  \sum_{\pi/\pi_{l_P}/\pi_{l_D}}
      ^{N!/l_P!/l_D!}
%  \\ &
  \frac{-g^{\lambda\sigma}+Q_{\pi_{l_g}}^\lambda Q_{\pi_{l_g}}^{\sigma}/M_W^2}
  {Q_{\pi_{l_g}}^2-M_W^2}
\\ &
  \prod_{i=1}^{l_P}\biggl( \frac{2p_a^{\mu_{\pi_i}}}{2p_a k_{\pi_i}}
%  \biggr)
%  \biggl(\prod_{i=1}^{l_g} 
     -\frac{2Q_{\pi_0}^{\mu_{\pi_{i}}}}{2Q_{\pi_{l_g}} k_{\pi_{i}}}
  \biggr)
  \prod_{i=1}^{l_D} \biggl(
     \frac{2Q_{\pi_0}^{\mu_{\pi_{l_P+i}}}}
          {2Q_{\pi_{l_g}} k_{\pi_{l_P+i}}}
%  \biggr)
%  \biggl(\prod_{i=1}^{l_c} 
         -\frac{2p_c^{\mu_{\pi_{l_P+i}}}}{2p_c k_{\pi_{l_P+i}}}
  \biggr),
  \end{aligned}
\end{equation}
where $l_a+l_g=l_P$ and $l_h+l_c=l_D$. The sum over permutations can be once
more replaced by the sum over partitions (cf.\ eq.\ (\ref{eq:sumpart})):
\begin{equation}
  \label{eq:multi-W-real5}
  \begin{aligned}
   &{\Meu^{(0)}_N}^{\mu_1,\dots,\mu_N}(k_1,\dots,k_N)
  \simeq
  \sum_{\wp=(P,D)^N}^{2^N}
%  \\ &
  \frac{-g^{\lambda\sigma}+Q_{g}^\lambda Q_{g}^{\sigma}/M_W^2}
  {Q_{g}^2-M_W^2}
%\\ &
  \prod_{i=1}^{N} 
  \biggl(\frac{2\theta_{\wp_i}p_{\wp_i}^{\mu_{i}}}{2p_{\wp_i} k_i}
%  \biggr)
%  \biggl(\prod_{i=1}^{l_g} 
     -\frac{2\theta_{\wp_i}Q_{\wp_i}^{\mu_{i}}}
           {2Q_{\wp_i} k_i}
  \biggr)
\\&
  =
  \sum_{\wp=(P,D)^N}^{2^N}
  \frac{-g^{\lambda\sigma}+Q_{g}^\lambda Q_{g}^{\sigma}/M_W^2}
  {Q_{g}^2-M_W^2}
%\\ &
  \prod_{i=1}^{N}
  j_{\wp_i}^{\mu_i},
  ~~~j_{P}^{\mu_i}=
  \frac{2 p_{a}^{\mu_i}}{2p_{a} k_i}
     -\frac{2 Q_{g}^{\mu_{i}}}
           {2Q_{g} k_i},
   ~~~j_{D}^{\mu_i}=
  \frac{2 Q_{g}^{\mu_{i}}}
           {2Q_{g} k_i}
     -\frac{2 p_{c}^{\mu_i}}{2p_{c} k_i}.
  \end{aligned}
\end{equation}
In eq.\ (\ref{eq:multi-W-real5}) we have used a freedom of defining $Q_{\pi_0}$ to
replace it with $Q_{\wp_i}\equiv Q_g$. Note that, contrary to $p_X$, the
vectors $Q_X$ depend on the choice of partitions, i.e.\ vary from a partition to partition. 
This prevents us from collapsing the remaining sum over partitions,
quite analogously as in the case of the neutral resonance.

%The $Q_X$-es could be replaced by
%partition independent $q_X$-es defined as $q_P= p_a+p_b$ and $q_D= p_c+p_d$. 
%In such a case it would be tempting to rewrite the sum over partitions
%$\wp=(P,D)$ in a compact form as in eq.\ (\ref{eq:multi-real4})
%\begin{equation}
%  \label{eq:multi-W-real6}
%  \begin{aligned}
%   &{\Meu^{(0)}_N}^{\mu_1,\dots,\mu_N}(k_1,\dots,k_N) \simeq
%%  \\ \simeq &
%  \prod_{i=1}^N \biggl(
%      \frac{2p_{a}^{\mu_i}}{2p_{a} k_{i}}
%     -\frac{2q_{P}^{\mu_i}}{2q_{P} k_{i}}
%     +\frac{2q_{D}^{\mu_i}}{2q_{D} k_{i}}
%     -\frac{2p_{c}^{\mu_i}}{2p_{c} k_{i}}
%  \biggr)
%  \frac{-g^{\lambda\sigma}+Q_{g}^\lambda Q_{g}^{\sigma}/M_W^2}
%  {Q_{g}^2-M_W^2}
%  \end{aligned}
%\end{equation}
%Unfortunately, the above equation is incorrect, because the $Q_g$ has to be
%decoded separately for each term {\em after} the product of currents is
%executed and partitions are known. Only then one knows the value of $K_P$ or 
%$K_D$ necessary to construct the vector $Q_g$.

%%%%%%%%%%%%%%%%%%%%%%%%%%%%%%%%%%%%%%%%%%%%%%%%%%%%%%%%%%%%%%%%%%%%%%%%%%%
%\newpage
\section{Details of virtual form-factor}

In the following we are going to generalise the YFS \cite{Yennie:1961ad}
virtual form-factor function $\alpha B$ to the general case
with charged intermediate resonances.
%/////////////////////////////////////////////////////////////////////////////////////////
\begin{figure}[!ht]
\centering
\setlength{\unitlength}{0.1mm}
\includegraphics[height=80mm,width=80mm]{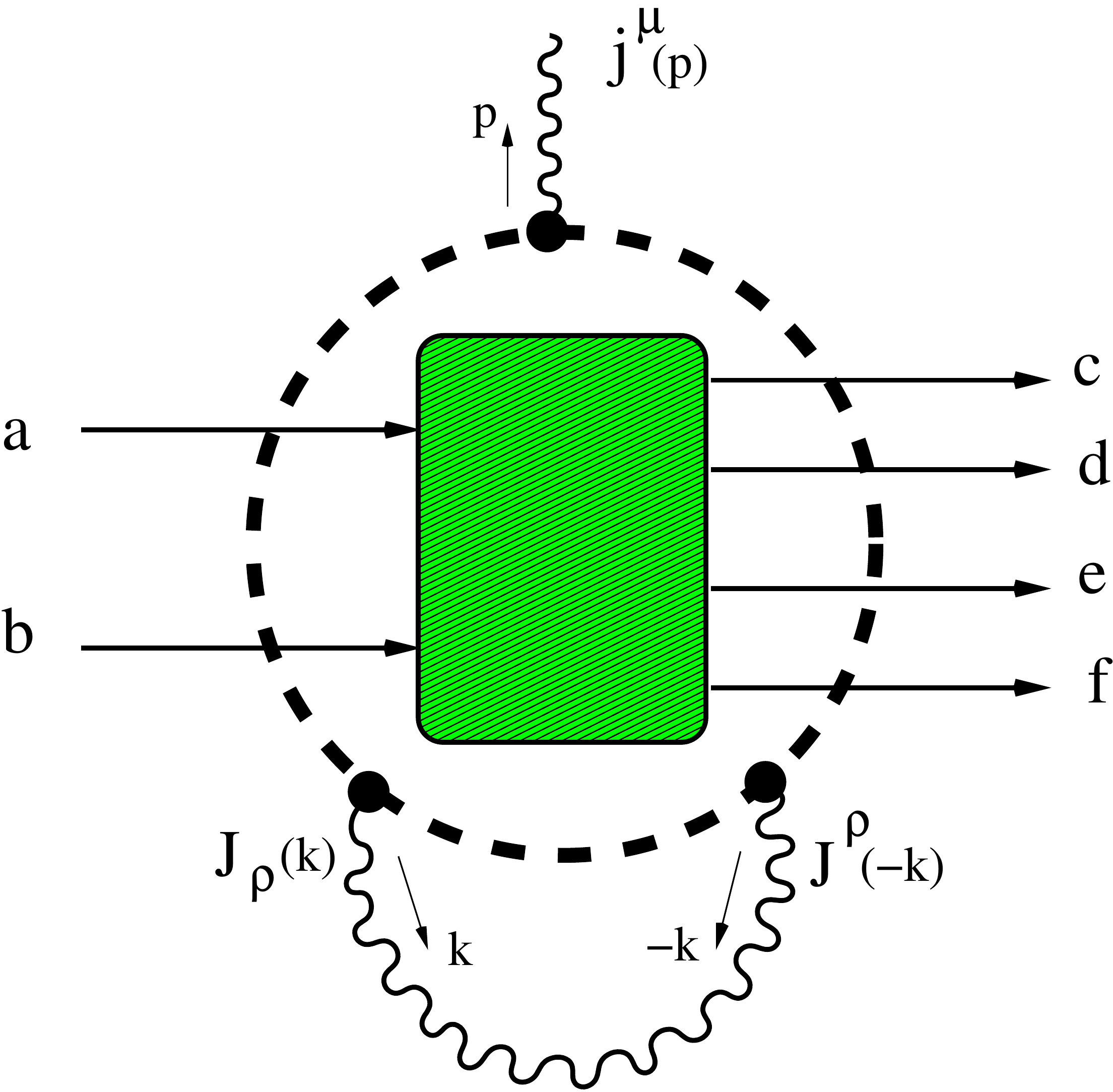} % was 40 mm
\caption{\small\sf
  Example of one real and one virtual photon emissions.
  The electric current is a sum of contributions from all external particles.
  This is why it is attached to the dashed line which crosses all relevant external lines.
  The rest of the Feynman diagram is visualised as the internal dark box.
}
\label{fig:current}
\end{figure}
%-----------------------------------------------------------------------------------------
In order to introduce the notation, let us first write down explicitly the
emission factor for a single-real photon
%%%%%%%%%%%%%%%%%%%%%%%%%%%%
\begin{equation}
j^\mu(k)= ie \sum_{X=a,b,c,d,e,f}\!\!\! Q_X \theta_X\;
         \frac{ 2p_X^\mu}{2p_X k},
\end{equation}
where $\theta_X=+1,-1$ for particles in the initial and final state, respectively,
$Q_X$ is the charge of the particle (in the units of the positron charge $e$)
and the single-virtual photon current reads
%%%%%%%%%%%%%%%%%%%%%%%%%%%%
\begin{equation}
J^\mu(k)= \sum_{X=a,b,c,d,e,f} \hat{J}_{X}^\mu(k),\qquad
\hat{J}_{X}^\mu(k) \equiv
 Q_X \theta_X\; \frac{ 2p_X^\mu\theta_X -k^\mu}
                     {k^2- 2p_X k\theta_X +i\varepsilon},
\end{equation}
see fig.~\ref{fig:current}.
For the virtual corrections we always have an even number of the $J^\mu(k)$ currents
paired in the so-called virtual $S$-factor
%%%%%%%%%%%%%%%%%%%%%%%%%%%%
\begin{equation}
S(k)= J(k) \circ J(k) =\sum_{ {X=a,b,c,d,e,f}\atop{Y=a,b,c,d,e,f} } J_X(k) \circ J_Y(k),
\end{equation}
where
%%%%%%%%%%%%%%%%%%%%%%%%%%%%
\begin{equation}
  J_X(k) \circ J_Y(k) = J_X(k) \cdot J_Y(-k),\; {\rm for}\; X\neq Y,\quad 
  J_X(k) \circ J_X(k) = J_X(k) \cdot J_X(k).
\end{equation}
In fig.~\ref{fig:current} we illustrate all that in a visual way.
The contribution $J_X(k) \cdot J_X(k)$ looks diagrammatically like the self-energy, but
in fact it comes from the charge renormalisation, see the discussion
in refs.~\cite{Yennie:1961ad,weinberg95-book}.

In the derivation of  $\exp(\alpha B)$ of ref.~\cite{Yennie:1961ad}
(taking as an example the four-fermion production process)
we arrive at a certain stage where the contributions from all the real and virtual photons
are factorised.
%/////////////////////////////////////////////////////////////////////////////////////////
\begin{figure}[!ht]
\centering
\setlength{\unitlength}{0.1mm}
\includegraphics[height=65mm,width=160mm]{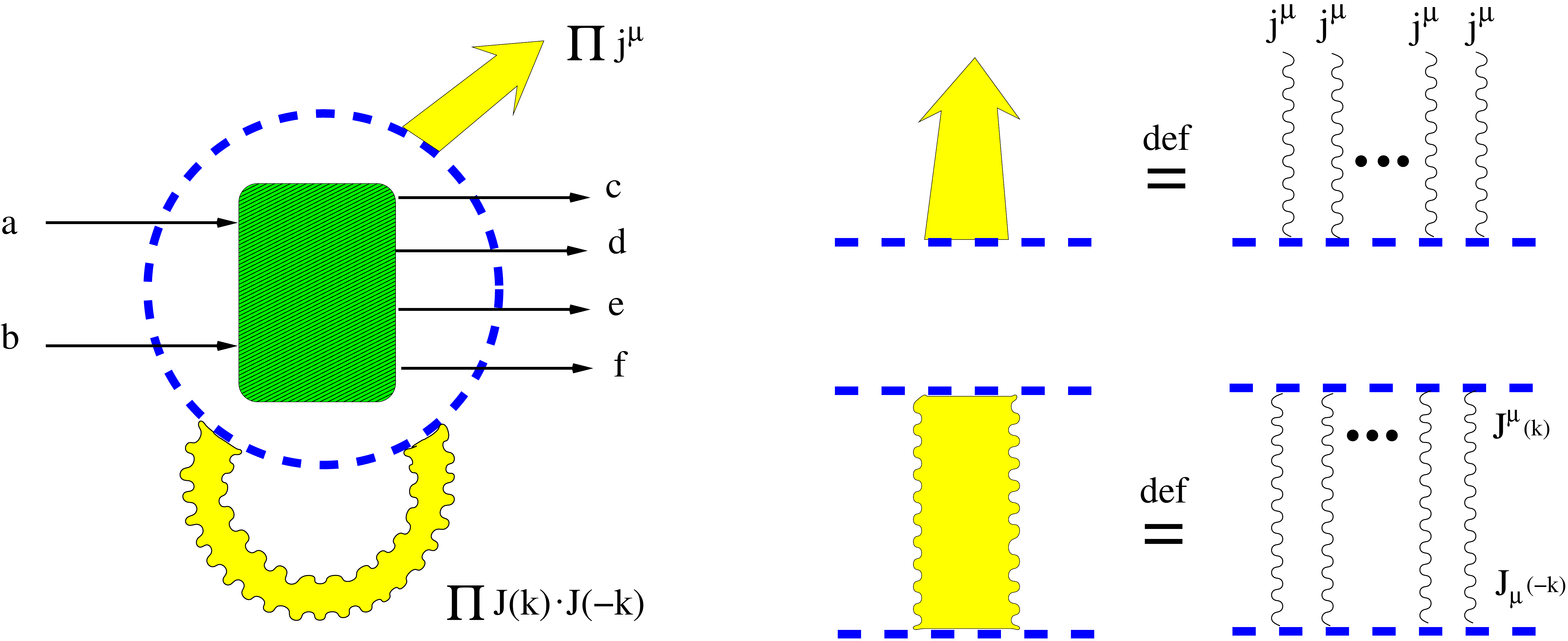} % was 40 mm
\caption{\small\sf
  The EEX amplitude for four-fermion production.
  The electric current is a sum of contributions from all external particles.
  This is why it is attached to the dashed line which crosses all relevant external lines.
  We use the collective notation for multiple-photon lines which is defined in the plot.
}
\label{fig:yfs61amp}
\end{figure}
%-----------------------------------------------------------------------------------------
The corresponding scattering amplitude with $m$ real and any number of virtual
photons
taken in the soft-photon approximation, visualised in fig.~\ref{fig:yfs61amp},
reads
%//////////////////////////////////////////////////
\begin{equation}
  \begin{aligned}
   &M^{\mu_1\mu_2...\mu_m}(k_1,k_2,...,k_m)=
   \\&=\Meu\;
    \prod_{l=1}^m j^\mu(k_l) \;
    \sum_{n=0}^\infty
    \frac{1}{n!}
    \prod_{i=1}^n 
            \alpha \int  \frac{i}{(2\pi)^3}\; 
            \frac{d^4 k_i }{k_i^2-\lambda^2+i\varepsilon}\; 
                J^\mu(k_i) \circ J_\mu(k_i).
  \end{aligned}
\end{equation}
The sum over virtual photons is done  trivially, resulting in the exponential form-factor:
%//////////////////////////////////////////////////
\begin{equation}
  \begin{aligned}
   &M^{\mu_1\mu_2...\mu_m}(k_1,k_2,...,k_m)= \Meu\;
    \prod_{l=1}^m j^\mu(k_l) \; e^{\alpha B_{6}},
\\
   &B_{6}=
            \int  \frac{i}{(2\pi)^3}\; 
            \frac{d^4 k }{k^2-\lambda^2+i\varepsilon}\; J(k) \circ J(k).
  \end{aligned}
\end{equation}
Note that in the residual function $\Meu$ there is no ``recoil'' dependence on photon momenta,
we are therefore limited to very soft photons ($E_\gamma\ll \Gamma_W$) 
in the process of our interest.

%/////////////////////////////////////////////////////////////////////////////////////////
\begin{figure}[!ht]
\centering
\setlength{\unitlength}{0.1mm}
\includegraphics[height=80mm,width=100mm]{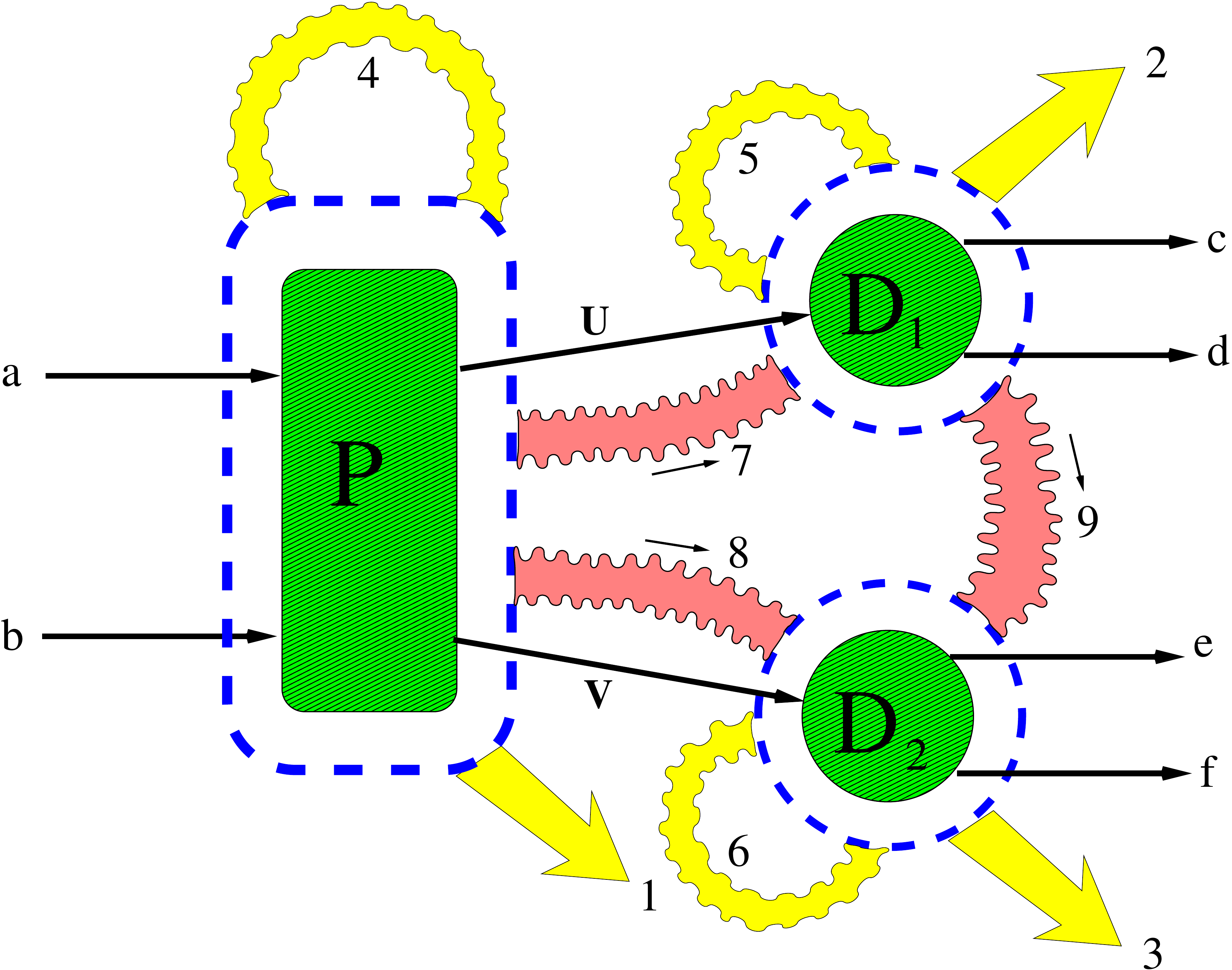} % was 40 mm
\caption{\small\sf
  The CEEX amplitude for $WW$ production and two decays in the soft-photon approximation.
  Visualised are all classes of virtual and real photon emissions.
}
\label{fig:ceex}
\end{figure}
%-----------------------------------------------------------------------------------------

Let us now take into account the double-resonant character of the process, 
see fig. \ref{fig:ceex}.
After factorising all the real and virtual soft photons, 
and introducing a new source of emission from the intermediate resonant $W$'s, 
we use the identity of eq.~(\ref{eq:principal-ident})
to arrive at the amplitude depicted in fig.~\ref{fig:ceex},
which can be written explicitly as follows:
%//////////////////////////////////////////////////
\begin{equation}
  \begin{aligned}
   &M^{\mu_{_{1_1}}...\mu_{_{3_{n_3}}}}_{n_1 n_2 n_3}(k_{1_1},...,k_{3_{n_3}})=
    \Meu_0\;
    \prod_{i_1={1}}^{n_1} j^{\mu_{i_1}}_{P}(k_{i_1}) \;
    \prod_{i_2={1}}^{n_2} j^{\mu_{i_2}}_{D_1}(k_{i_2}) \;
    \prod_{i_3={1}}^{n_3} j^{\mu_{i_3}}_{D_2}(k_{i_3}) \;
   \\&\quad
    \sum_{n_4=0}^\infty \frac{1}{n_4!} 
        \prod_{i_4=1}^{n_4} \alpha \int \frac{i}{(2\pi)^3}\; 
            \frac{d^4 k_{i_4} }{k_{i_4}^2-m_\gamma^2+i\varepsilon}\;
        J_P(k_{i_4})\circ J_P(k_{i_4})\;
   \\&\quad
    \sum_{n_5=0}^\infty \frac{1}{n_5!} 
        \prod_{i_5=1}^{n_5} \alpha \int  \frac{i}{(2\pi)^3}\; 
            \frac{d^4 k_{i_5} }{k_{i_5}^2-m_\gamma^2+i\varepsilon}\;
        J_{D_1}(k_{i_5})\circ J_{D_1}(k_{i_5})
   \\&\quad
    \sum_{n_6=0}^\infty \frac{1}{n_6!} 
        \prod_{i_6=1}^{n_6} \alpha \int  \frac{i}{(2\pi)^3}\; 
            \frac{d^4 k_{i_6} }{k_{i_6}^2-m_\gamma^2+i\varepsilon}\; 
        J_{D_2}(k_{i_6})\circ J_{D_2}(k_{i_6})
   \\&\quad
    \sum_{n_7=0}^\infty \frac{1}{n_7!} 
        \prod_{i_7=1}^{n_7} 2\alpha \int  \frac{i}{(2\pi)^3}\; 
            \frac{d^4 k_{i_7} }{k_{i_7}^2-m_\gamma^2+i\varepsilon}\; 
        J_{P}(k_{i_7})\circ J_{D_1}(k_{i_7})
   \\&\quad
    \sum_{n_8=0}^\infty \frac{1}{n_8!} 
        \prod_{i_8=1}^{n_8} 2\alpha \int  \frac{i}{(2\pi)^3}\; 
            \frac{d^4 k_{i_8} }{k_{i_8}^2-m_\gamma^2+i\varepsilon}\; 
        J_{P}(k_{i_8})\circ J_{D_2}(k_{i_8})
   \\&\quad
    \sum_{n_9=0}^\infty \frac{1}{n_9!} 
        \prod_{i_9=1}^{n_9} 2\alpha \int  \frac{i}{(2\pi)^3}\; 
            \frac{d^4 k_{i_9} }{k_{i_9}^2-m_\gamma^2+i\varepsilon}\; 
        J_{D_1}(k_{i_9})\circ J_{D_2}(k_{i_9})
   \\&\quad
    \frac{1}{(p_{cd}+K_2-K_7+K_9)^2-M^2 }\; \frac{1}{(p_{ef}+K_3-K_8-K_9)-M^2 },
  \end{aligned}
\end{equation}
where $K_l=\sum_{i_l=1}^{n_l} k_{i_l}$. We have defined {\it nine} groups of 
photons (labeled by $l$); $l=1,2,3$ corresponds, respectively, to the real emission 
from the $W$-pair production and their decays, the virtual photons corresponding to the same 
sources are denoted as the group (4,5,6), while (7,8) corresponds to the virtual photons 
attached  to the production and the decay. Finally, the virtual photons denoted as the group (9) 
connect the decays of the first and the second $W$-boson.

The most interesting part in the above expression is that
the product of two resonance propagators 
includes all the relevant recoil dependence on the real and virtual
photon momenta. 
This dependence can be read easily from fig.~\ref{fig:ceex}.
We are now not limited by $E_\gamma\ll \Gamma_W$, but rather by 
$E_\gamma\ll \sqrt{s}$.
One important feature is that propagators do not depend on $K_4$,  $K_5$ and 
$K_6$
-- this is why the sums over relevant photons can be  immediately folded into
three standard YFS form-factor $e^{\alpha B}$, for the production and decay processes. 
This, however, cannot be done for the three virtual-interference contributions
because the propagators do depend on $K_7$,  $K_8$ and  $K_9$.
The dependence on the real photons $K_2$ and  $K_3$ is not harmful for our task
of summing up {\it virtual} contributions to the infinite order:
%//////////////////////////////////////////////////
\begin{equation}
  \begin{aligned}
   &M^{\mu_{_{1_1}}...\mu_{_{3_{n_3}}}}_{n_1 n_2 n_3}(k_{1_1},...,k_{3_{n_3}})=
    \Meu_0\;
    \prod_{i_1={1}}^{n_1} j^{\mu_{i_1}}_{P}(k_{i_1}) \;
    \prod_{i_2={1}}^{n_2} j^{\mu_{i_2}}_{D_1}(k_{i_2}) \;
    \prod_{i_3={1}}^{n_3} j^{\mu_{i_3}}_{D_2}(k_{i_3}) \;
   \\&\quad
    e^{\alpha B_P}\; e^{\alpha B_{D_1}}\; e^{\alpha B_{D_2}}\;
    \sum_{n_7=0}^\infty \frac{1}{n_7!} 
         \prod_{i_7=1}^{n_7} 2\alpha  \int  \frac{i}{(2\pi)^3}\; 
            \frac{d^4 k_{i_7} }{k_{i_7}^2-m_\gamma^2}\; 
         J_P(k_{i_7})\circ J_{D_1}(k_{i_7})\;
   \\&\quad
    \sum_{n_8=0}^\infty \frac{1}{n_8!} 
         \prod_{i_8=1}^{n_8} 2\alpha  \int  \frac{i}{(2\pi)^3}\; 
            \frac{d^4 k_{i_8} }{k_{i_8}^2-m_\gamma^2}\;
         J_{P}(k_{i_8})\circ J_{D_2}(k_{i_8})
   \\&\quad
    \sum_{n_9=0}^\infty \frac{1}{n_9!} 
         \prod_{i_9=1}^{n_9} 2\alpha  \int  \frac{i}{(2\pi)^3}\; 
            \frac{d^4 k_{i_9} }{k_{i_9}^2-m_\gamma^2}\; 
         J_{D_1}(k_{i_9})\circ J_{D_2}(k_{i_9})
   \\&\quad
    \frac{1}{(U_2-K_7+K_9)^2-M^2 }\; \frac{1}{(V_3-K_8-K_9)-M^2 },
  \end{aligned}
\end{equation}
where $U_2=p_{cd}+K_2$ and $V_3=p_{ef}+K_3$, and
\begin{equation}
  \alpha B_X = \int  \frac{i}{(2\pi)^3}\; 
  \frac{d^4 k }{k^2-m_\gamma^2 +i\varepsilon}\; J_{X}(k)\circ J_{X}(k),\quad X=P,D_1,D_2.
\end{equation}
At this point we use the following  approximations (valid in the soft-photon limit):
\begin{equation}
  \begin{aligned}
 &\frac{1}{(U_2-K_7+K_9)^2-M^2 }\;
  \simeq \frac{1}{ U_2^2-M^2 -2U_2 K_7 +2U_2 K_9 }\;
\\&\quad
  = \frac{1}{U_2^2-M^2}\; 
    \frac{1}{ 1-\sum_{i_7} \frac{2U_2 k_{i_7}}{U_2^2-M^2} +\sum_{i_9} \frac{2U_2 k_{i_9}}{U_2^2-M^2} }\;
\\&\quad
  \simeq \frac{1}{U_2^2-M^2}\; 
    \prod_{i_7} \frac{1}{1 -\frac{2U_2 k_{i_7}}{U_2^2-M^2}}\; 
    \prod_{i_9} \frac{1}{1 +\frac{2U_2 k_{i_9}}{U_2^2-M^2}}\;
\\&\quad
  \simeq \frac{1}{U_2^2-M^2}\; 
    \prod_{i_7} \frac{U_2^2-M^2}{(U_2-k_{i_7})^2-M^2}\; 
    \prod_{i_9} \frac{U_2^2-M^2}{(U_2+k_{i_9})^2-M^2}\;,
  \end{aligned}
\end{equation}
which, together with the analogous approximation for the second propagator,
allows us to fold-in three remaining sums into exponents:
%//////////////////////////////////////////////////
\begin{equation}
  \label{eq:ceex-formfac}
  \begin{aligned}
   &M^{\mu_{_{1_1}}...\mu_{_{3_{n_3}}}}_{n_1 n_2 n_3}(k_{1_1},...,k_{3_{n_3}})=
    \Meu_0\;
    \prod_{i_1={1_1}}^{n_1} j^{\mu_{i_1}}_{P}(k_{i_1}) \;
    \prod_{i_2={1_2}}^{n_2} j^{\mu_{i_2}}_{D_1}(k_{i_2}) \;
    \prod_{i_3={1_3}}^{n_3} j^{\mu_{i_3}}_{D_2}(k_{i_3}) \;
   \\&\qquad\qquad
    e^{ \alpha B_{10}^{\rm CEEX}(p_{cd}+K_2,p_{ef}+K_3 )}\;
    \frac{1}{(p_{cd}+K_2)^2-M^2 }\; \frac{1}{(p_{ef}+K_3)-M^2 },
  \end{aligned}
\end{equation}
where
%//////////////////////////////////////////////////
\begin{equation}
  \begin{aligned}
   &\alpha B_{10}^{\rm CEEX}(U,V)= \alpha B_P+\alpha B_{D_1}+\alpha B_{D_2}
                         +2\alpha B_{P\otimes D_1}(U) 
                         +2\alpha B_{P\otimes D_2}(V)
                         +2\alpha B_{D_1\otimes D_2}(U,V),\;
   \\&
   \alpha B_{P\otimes D_1}(U)=
    \int  \frac{i}{(2\pi)^3}\; 
            \frac{d^4 k }{k^2-m_\gamma^2+i\varepsilon}\; J_P(k)\circ J_{D_1}(k)
            \frac{U^2-M^2}{(U-k)^2-M^2},
   \\&
    \alpha B_{P\otimes D_2}(V)=
    \int  \frac{i}{(2\pi)^3}\; 
            \frac{d^4 k}{k^2-m_\gamma^2+i\varepsilon}\; J_{P}(k)\circ J_{D_2}(k)
            \frac{V^2-M^2}{(V-k)^2-M^2},
   \\&
    \alpha B_{D_1\otimes D_2}(U,V)=
    \int  \frac{i}{(2\pi)^3}\; 
            \frac{d^4 k }{k^2-m_\gamma^2+i\varepsilon}\; J_{D_1}(k)\circ J_{D_2}(k)
            \frac{U^2-M^2}{(U+k)^2-M^2}\;
            \frac{V^2-M^2}{(V-k)^2-M^2}.
  \end{aligned}
\end{equation}
Let us note that in the no-recoil limit $U-k\to U$, $V-k\to V$, i.e.\ $k\ll
\Gamma_W$, 
the function $B_{10}^{\rm CEEX}(U,V)$ reduces to $B_6^{\rm CEEX}$, in an analogous way to
eq.~(\ref{eq:real-single}).

In eq.~(\ref{eq:ceex-formfac}) and in all previous steps the contributions of real
photon were taken as just one term (in which we know to which subprocess every real 
photon belongs) from the grand sum (as defined e.g.\ in 
eq.~(\ref{eq:ceex1b-resonanat})), over all 
$3^n$ photon assignments $(P,D_1,D_2)^n$,
in which we know to which subprocess every real photon belongs%
\footnote{The final form of the result took 
shape thanks to the use, at the earlier step, the identity of eq.~(\ref{eq:identity}).
}.
Let us restore this coherent sum over all 
photon assignments in the following compact final expression:
%//////////////////////////////////////////////////
\begin{equation}
  \label{eq:ceex-formfactor}
   M^{\mu_1...\mu_n}(k_1,k_2,...,k_n)=\!\!\!\!\!
    \sum_{\wp\in(P,D_1,D_2)^n}\!\!\!\!
    \Meu_0\; \prod_{i=1}^{n}  j^{\mu_i}_{\wp_i}(k_i)\;
    e^{ \alpha B_{10}^{\rm CEEX}(U_\wp,V_\wp )}\;
    \frac{1}{U_\wp^2-M^2 }\; \frac{1}{V_\wp^2-M^2 },
\end{equation}
where $U_\wp = p_{cd}+\sum\limits_{\wp_i=D_1} k_i$ and
      $V_\wp = p_{ef}+\sum\limits_{\wp_i=D_2} k_i$.

Eq.~(\ref{eq:ceex-formfactor}) is the principal result of this Appendix.
The CEEX form-factor of eq.~(\ref{eq:ceex-formfactor}) is valid for production
of a pair of any charged resonances of any spin.
The case of a single charged resonance, or more than two charged
resonances, can be treated in the same way.

The presented derivation of the virtual form-factor is based to a large extent on 
the analogy with the real-emission part (see  Appendix \ref{app:b})
and the cancellation between the real and virtual emissions. 
%Technical details  related to the definition 
%of the width and the mass of the $W$-boson
%, along with the renormalisation issues, 
%are to be addressed in future works. 

\newpage
%%%%%%%%%%%%%%%%%%%%%%%%%%%%%%%%%%%%%%%%%%%%%%%%%%%%%%%%%%%%%%%%%%%%%%%%%%%%
%%%%%%%%%%%%%%%%%%%%%%%%%%%%%%%%%%%%%%%%%%%%%%%%%%%%%%%%%%%%%%%%%%%%%%%%%%%%
%%%%%%%%%%%%%%%%%%%%%%%%%%%%%%%%%%%%%%%%%%%%%%%%%%%%%%%%%%%%%%%%%%%%%%%%%%%%
%\bibliographystyle{elsarticle-num}
%\bibliography{fccee}
%%%%%%%%%%%%%%%%%%%%%%%%%%%%%%%%%%%%%%%%%%%%%%%%%%%%%%%%%%%%%%%%%%%%%%%%%%%
%%%%%%%%%%%%%%%%%%%%%%%%%%%%%%%%%%%%%%%%%%%%%%%%%%%%%%%%%%%%%%%%%%%%%%%%%%%%

                                     %%%%%%%%%%%%%%%%%%
\end{document}